\documentclass[fleqn,usenatbib]{mnras}

\usepackage{eso-pic}

\AddToShipoutPictureBG*{%
  \AtPageUpperLeft{%
    \hspace{0.73\paperwidth}%
    \raisebox{-1.\baselineskip}{%
      \makebox[0pt][l]{\textnormal{DES-2021-0643}}
}}}%

\AddToShipoutPictureBG*{%
  \AtPageUpperLeft{%
    \hspace{0.73\paperwidth}%
    \raisebox{-2.\baselineskip}{%
      \makebox[0pt][l]{\textnormal{FERMILAB-PUB-21-497-AE}}
}}}%

\usepackage{mdframed}
\usepackage{graphicx}
\usepackage{subcaption}
\usepackage{multirow}
\usepackage{lineno}
\usepackage[T1]{fontenc}
\usepackage{ae,aecompl}
\usepackage{todonotes}

\usepackage{graphicx}	
\usepackage{newtxtext,newtxmath}
\usepackage{adjustbox}
\usepackage{cleveref}

\newcommand{\mcal}{\textsc{metacalibration}}
\newcommand{\Mcal}{\textsc{Metacalibration}}

\defcitealias{ZuntzY1}{ZS18}

\newcommand{\pkdgrav}{{\textsc{PkdGrav3}}}
\newcommand{\darkgrid}{{\textsc{DarkGridV1}}}

\defcitealias{kacprzak2016cosmology}{K16}	
\defcitealias{zurcher2021cosmological}{Z21}	

\definecolor{CLs}{HTML}{0173B2}
\definecolor{Peaks}{HTML}{DE8F05}
\definecolor{CLsPeaks}{HTML}{029E73}

\let\oldequation\equation
\let\oldendequation\endequation
\renewenvironment{equation}
  {\linenomathNonumbers\oldequation}
  {\oldendequation\endlinenomath}

\title[DES Y3 results: Cosmology with peaks]{Dark Energy Survey Year 3 results: Cosmology with peaks using an emulator approach}

\makeatletter
\def \blfootnote{\xdef\@thefnmark{}\@footnotetext}
\makeatother

\author[D. Z\"urcher et~al.]{
\parbox{\textwidth}{
\Large 
D. ~Z\"urcher$^{1}$\thanks{dominik.zuercher@phys.ethz.ch},
J. Fluri$^{1}$,
R. Sgier$^{1}$,
T. Kacprzak$^{1}$,
M. Gatti$^{2}$,
C. Doux$^{2}$,
L. Whiteway$^{3}$,
A. R{\'e}fr{\'e}gier$^{1}$,
C. Chang$^{4,5}$,
N. Jeffrey$^{3, 6}$
% ...
\begin{center} (DES Collaboration) \end{center}
}
}

\vspace{0.2cm}

\date{Accepted XXX. Received YYY; in original form ZZZ}

\pubyear{2021}

\begin{document}
\label{firstpage}
\pagerange{\pageref{firstpage}--\pageref{lastpage}}
\maketitle

\definecolor{pink}{rgb}{0.848, 0.188, 0.478}

\begin{abstract}
We constrain the matter density $\Omega_{\mathrm{m}}$ and the amplitude of density fluctuations $\sigma_8$ within the $\Lambda$CDM cosmological model with shear peak statistics and angular convergence power spectra using  mass maps constructed from the first three years of data of the Dark Energy Survey (DES Y3). 
We use tomographic shear peak statistics, including cross-peaks: peak counts calculated on maps created by taking a harmonic space product of the convergence of two tomographic redshift bins.
Our analysis follows a forward-modelling scheme to create a likelihood of these statistics using N-body simulations, using a Gaussian process emulator. 
We take into account the uncertainty from the remaining, largely unconstrained $\Lambda$CDM parameters ($\Omega_{\mathrm{b}}$, $n_{\mathrm{s}}$ and $h$). 
We include the following lensing systematics: multiplicative shear bias, photometric redshift uncertainty, and galaxy intrinsic alignment. Stringent scale cuts are applied to avoid biases from unmodelled baryonic physics. We find that the additional non-Gaussian information leads to a tightening of the constraints on the structure growth parameter yielding $S_8~\equiv~\sigma_8\sqrt{\Omega_{\mathrm{m}}/0.3}~=~0.797_{-0.013}^{+0.015}$ (68\% confidence limits), with a precision of 1.8\%, an improvement of ~38\% compared to the angular power spectra only case. The results obtained with the angular power spectra and peak counts are found to be in agreement with each other and no significant difference in $S_8$ is recorded. We find a mild tension of $1.5 \thinspace \sigma$ between our study and the results from Planck 2018, with our analysis yielding a lower $S_8$.
Furthermore, we observe that the combination of angular power spectra and tomographic peak counts breaks the degeneracy between galaxy intrinsic alignment $A_{\mathrm{IA}}$ and $S_8$, improving cosmological constraints.
We run a suite of tests concluding that our results are robust and consistent with the results from other studies using DES Y3 data.
\end{abstract}


\clearpage

\section{Introduction}
\label{sec:intro}

The large scale structure (LSS) of the Universe is a powerful probe for testing cosmological models \citep[see][for reviews]{Kilbinger2015review,Albrecht2006taskforce}.
Recent measurements from observational programs that map the LSS such as the Dark Energy Survey\footnote{\url{https://www.darkenergysurvey.org}} (DES), Kilo-Degree Survey\footnote{\url{http://kids.strw.leidenuniv.nl}} (KIDS), and Hyper-Suprime Cam\footnote{\url{https://hsc.mtk.nao.ac.jp/ssp/survey/}} (HSC) have delivered cosmological constraints on the matter density $\Omega_{\rm m}$ and amplitude of density fluctuations $\sigma_8$ with better than $5\%$ precision\footnote{We define $\sigma_8$ as the present-day root-mean-square amplitude of the matter fluctuations averaged in spheres of radius 8 $h^{-1}$ Mpc as computed from linear theory.}.
This has revealed evidence of moderate tensions between the LSS measurements and measurements using the cosmic microwave background (CMB) for these parameters \citep{Leauthaud2017low,DiValentino2020intertwined2,Lemos2020tensions}; these tensions may indicate that the Lambda Cold Dark Matter ($\Lambda$CDM) model is unable to explain these observations jointly, or that our understanding of systematic effects is insufficient.
Multiple modelling choices and analysis variants have been explored \citep*{Troxel2018consistency,Joudaki2020combined} in an attempt to understand this discrepancy. \\

\noindent The evolution of the cosmic web substructures of the LSS, consisting of halos, filaments, sheets, and voids \citep{Bond1996how,Foreroromero2009dynamical,Dietrich2012filament,Liebskind2017tracing},
carries information about the underlying cosmological parameters.
Weak gravitational lensing is one of the probes able to map the distribution of these structures directly \citep[see][for review]{Kilbinger2015review}. Not only does it enable us to map their spatial distribution, it also allows us to study their temporal evolution through tomography.
Weak lensing mass maps are created using small shape distortions in the images of background galaxies caused by the gravitational lensing due to the foreground LSS \citep[see][for reviews]{Bernsetin2020shapes,Bartelmann2001review}. 
The information contained in these maps is typically extracted using 2-point statistics such as the real space correlation function, the angular power spectrum, or the wavelet-like COSEBIs (complete orthogonal sets of E/B-integrals) \citep{Asgari2021kids}.
However, 2-point statistics do not capture all the information available in the highly non-Gaussian mass maps \citep{Springel2006lss,Yang2011information}.
Multiple approaches have been proposed to extract this information:
the peak count function, which counts local maxima of the mass maps and thus probes their highly non-linear parts
\citep[][hereafter \citetalias{kacprzak2016cosmology}]{jain2000statistics, Dietrich2010peaks, Shan2018peaks,Martinet2018peaks,harnois2020cosmic, des_sv_lfi, kacprzak2016cosmology},
3-point statistics, which analyze the configurations of triangles at various scales \citep{Takada2003threepoint,Semboloni2011space,Fu2014threepoint},
the higher-order moments of convergence maps \citep*{Patton2017onepoint,Gatti2020moments},
the Minkowski functionals, which analyze the topology of the maps \citep{Shirasaki2014minkowski,Petri2015minkowski,Parroni2020minkowski},
and machine learning approaches, which aim to detect features automatically \citep{Gupta2018deeplearning,Fluri2019kids, des_sv_lfi}.
Some of these statistics and their combinations increase the precision of the cosmological parameter measurement, as well as responding differently to systematic effects such as galaxy intrinsic alignments \cite[see][hereafter \citetalias{zurcher2021cosmological}, for example]{zurcher2021cosmological}.

\noindent These approaches increase the cosmological constraining power, but face a major difficulty in their application: the prediction from theory is more challenging than that of 2-point statistics.
A possible solution is to use numerical simulations to provide predictions for the statistics.
These simulations are computationally expensive, especially for high dimensional parameter spaces that include cosmological, astrophysical, and systematics parameters.
Major progress has been made in recent years in creating a fully simulation-based likelihood for cosmic shear measurements, with emerging emulators \citep{Lawrence2017cosmicemu,Knabenhans2020emulator,Angulo2020Bacco} and simulation grids \citep{DeRose2019aemulus,VillaescusaNavarro2020camels}.
Map-level implementations of intrinsic alignments \citep{Joachimi2013ia2} and baryonic effects \citep{Schneider2020baryon} have recently been used in cosmic shear measurements \citep{Fluri2018deeplearning}.
This enables a reliable measurement of the structure growth parameter $S_8 \equiv \sigma_8 \sqrt{\Omega_{\mathrm{m}}/0.3}$ (the quantity to which weak lensing measurements are the most sensitive) at large and intermediate scales \citep{weiss2019effects}.
These approaches can shed more light on the tension in this parameter between CMB and LSS measurements by providing cosmological information complementary to the 2-point statistics. \\

\noindent In this work, we infer cosmological parameter constraints using peak counts and the angular power spectra of the tomographic weak lensing mass maps from the first three years of data from the Dark Energy Survey (DES Y3) \citep{y3-gold}.
The DES Y3 mass maps were first presented in \cite*{y3-massmapping}.
We follow a forward-modelling scheme by building an emulator of the peak counts at different cosmologies. 
The emulator is trained on a suite of \pkdgrav\ N-Body simulations \citep{potter2017pkdgrav3} that we created, dubbed \darkgrid.
We measure the cosmological parameters $\sigma_8$ and $\Omega_{\mathrm{m}}$ in the $\Lambda$CDM model, as well as the galaxy intrinsic alignment amplitude $A_{\mathrm{IA}}$.
We do not infer the values of the remaining $\Lambda$CDM parameters (baryon density $\Omega_{\mathrm{b}}$, scalar spectral index $n_{\mathrm{s}}$ and dimensionless Hubble parameter $h$) as they are mostly unconstrained by weak lensing measurements, but we take into account their contribution to the measurement uncertainty.
Further, we incorporate the effects of photometric redshift uncertainty, shear calibration biases and the redshift dependence of galaxy intrinsic alignment into the analysis.
Stringent scale cuts are applied to ensure that the results are not sensitive to baryon modelling. 
The design of the analysis, combined with the different sensitivity of peak counts to intrinsic alignment modelling, provides an alternative and complementary measurement to the main cosmic shear analysis of the DES Y3 data \citep{des2021cosmic, secco2021dark}. \\

\noindent This analysis is done in a blinded way to avoid intentional or unintentional confirmation biases.
We formulate a number of criteria that need to be satisfied before unblinding. \\

\noindent This work starts by introducing the DES Y3 shape catalogue and the \darkgrid\ simulation suite in Section~\ref{sec:data}.
The various systematic effects affecting weak lensing measurements are discussed in Section~\ref{sec:systematics} and their treatment in this study is outlined.
Section~\ref{sec:method} explains the forward modelling of the DES Y3 mass maps and the derived statistics, which are then compared to the statistics measured from the DES Y3 mass maps. This section also explains the inference pipeline.
Section~\ref{sec:blinding} discusses the blinding procedure followed in this work as well as the different unblinding tests that were performed. 
The inferred cosmological constraints are presented, discussed, and compared to the results of other studies in Section~\ref{sec:results}.
We summarise our findings in Section~\ref{sec:summary}.

\section{Data}
\label{sec:data}

\subsection{Dark Energy Survey Year 3 shape catalogue}

\begin{figure}
\includegraphics[width=0.43\textwidth]{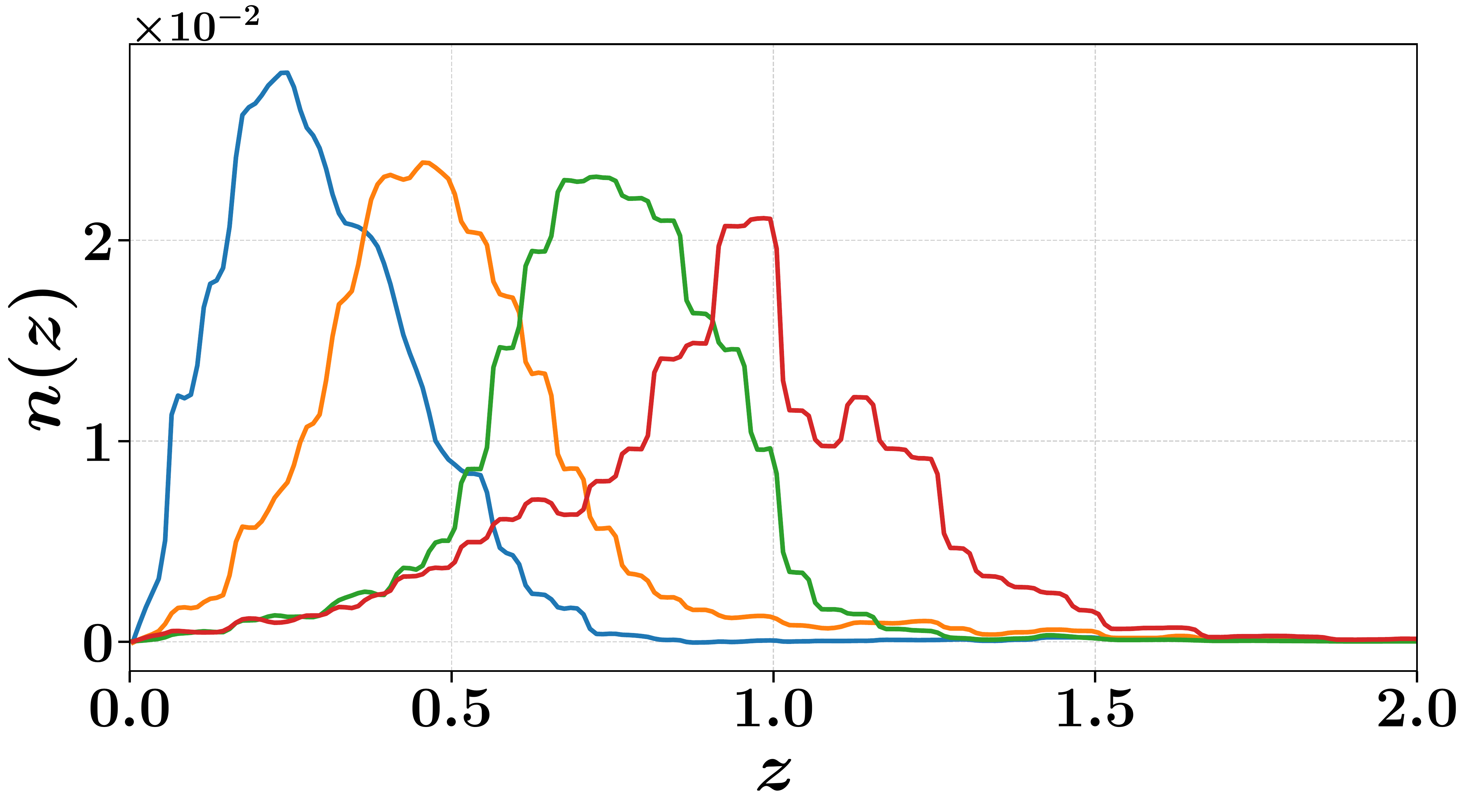}
\caption{Tomographic redshift distributions $n(z)$ of the galaxies in the DES Y3 shape catalogue. The calculation of the tomographic redshift distributions was performed using the SOMPZ method \citep*{Myles2020}. The distributions shown are normalised by the number of galaxies in each bin. 
The distributions span the redshift range $z=0\, -\, 3$.}
\label{fig:nz}
\end{figure}

This work uses data from the first three years of data (Y3) of the Dark Energy Survey (DES, \cite{2005astro.ph.10346T, 2016MNRAS.460.1270D}). DES is a photometric imaging survey that observed the southern hemisphere in five optical-NIR broadbands (\textit{grizY}) over six years (2013-2019). 
The processing of the raw images was performed by the DES Data Management (DESDM) team. We refer the reader to \cite{morganson2018dark, DES2018cosmology} for a detailed description of the image processing pipeline.
In particular, we use the fiducial DES Y3 weak lensing shape catalogue presented in \cite*{gatti2020dark}. The shear measurement pipeline used to create the catalogue is \Mcal{} \citep{HuffMcal2017, SheldonMcal2017}, which allows the self-calibration of the measured shapes against most of the shear and selection multiplicative biases by measuring the mean shear and selection response matrix of the sample. An additional multiplicative calibration (at the level of 2-3\%) to correct for detection and blending biases is provided based on image simulations \citep{maccran2020}. A number of null tests presented in \cite*{gatti2020dark} prove the catalogue to be robust against additive biases. The final sample comprises about a hundred million objects, for an effective number density of $n_{\rm eff} = 5.59$ galaxies/arcmin$^{2}$, spanning an effective area of 4143 square degrees. The galaxies of the DES Y3 shape catalogue are further divided into four tomographic bins and redshift estimates for each of the tomographic bins are provided by the SOMPZ method \citep*{Myles2020}. We present the normalised, tomographic redshift distributions in Figure~\ref{fig:nz}.

\subsection{N-Body simulation suite \darkgrid}
\label{sec:simulations}

\begin{figure}
\includegraphics[width=0.43\textwidth]{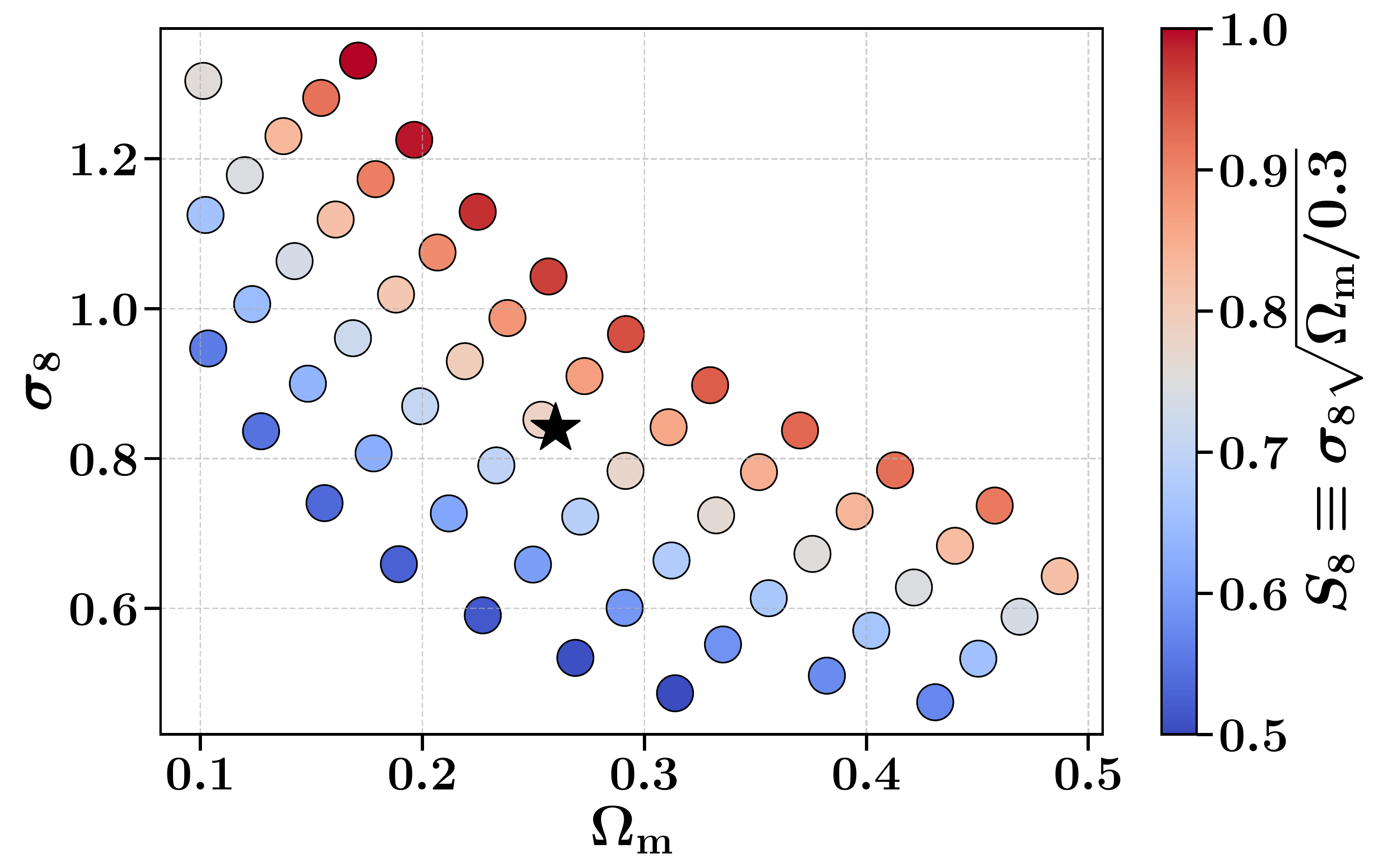}
\caption{Distribution of the \texttt{PKDGRAV3} simulations in the $\Omega_{\mathrm{m}}-\sigma_8$ plane.
The remaining cosmological parameters were fixed to the ($\Lambda$CDM,TT,TE,EE+lowE+lensing) results of Planck 2018 \citep{aghanim2020planck} in all simulations.
Fifty simulations were run at the fiducial cosmology ($\Omega_{\mathrm{m}} = 0.26, \sigma_8 = 0.84$); this cosmology is marked by the star. Five simulations were run at every other cosmology. The simulations are distributed around the fiducial cosmology along lines of approximately constant $S_8$. The colour of the points indicates the value of the $S_8$ parameter.}
\label{fig:sims}
\end{figure}

We rely on an emulator approach to predict the angular power spectra and peak counts at different cosmologies. 
The emulator is built on numerical predictions of the statistics from simulations.
Therefore, we require a suite of simulations spanning the studied cosmological parameter space, namely the $\Omega_{\mathrm{m}}-\sigma_8$ plane. 
We use the same simulation suite as was used by \citetalias{zurcher2021cosmological}, but extended with additional simulations; we dub this larger suite \darkgrid.
The simulations sample the $\Omega_{\mathrm{m}}-\sigma_8$ plane at 58 different cosmologies; their distribution is shown in Figure~\ref{fig:sims}. The simulation grid is centred at the fiducial cosmology inferred by the DES Y1 cosmic shear analysis \citep*{troxel2018dark} (marked by the star in Figure~\ref{fig:sims}) and the simulations are distributed along lines of approximately constant $S_8$.
Fifty independent full-sky simulations were run for the fiducial cosmology and five simulations for every other cosmology. 
The simulations at the fiducial cosmology are used to estimate the covariance matrix while the remaining simulations are used to train the emulator (see Section~\ref{sec:method}). \\

\noindent All simulations were produced using the publicly available code \pkdgrav\ \citep{potter2017pkdgrav3}. This dark-matter-only N-Body code features
a full-tree algorithm and a fast multipole expansion, yielding a run-time 
that increases linearly with the number of particles in the simulation. \pkdgrav\ runs on CPUs and GPUs simultaneously.
In each simulation $768^3$ particles and a unit box with a side-length of 900 Mpc/h were used. 
To cover the necessary redshift range up to $z = 3.0$ a large enough cosmological volume must be sampled; hence, the unit box is replicated up to 14 times along each dimension ($14^3$ replicas in total) using periodic boundary conditions. At the fiducial cosmology ten replicas along each dimension are sufficient to achieve the necessary volume. Such a replication scheme is known to under-predict the variance on very large scales (see \cite{Fluri2019kids} for example). 
We confirm that our simulations recover the angular power spectrum as predicted from the theory code \textsc{CLASS} \citep{lesgourgues2011cosmic} on all scales considered in this analysis and beyond (see Appendix~\ref{sec:acc_sim_conv}).
Nevertheless, we apply a scale cut of $\ell \geq 30$ for the angular power spectra to avoid the inclusion of scales strongly affected by mode-mixing with super-survey modes that might be underestimated in the simulations. \\

\noindent Apart from the varying $\Omega_{\mathrm{m}}$ and $\sigma_8$ parameters, all remaining cosmological parameters are fixed to the ($\Lambda$CDM,TT,TE,EE+lowE+lensing) results of Planck 2018 \citep{aghanim2020planck}. This corresponds to a baryon density $\Omega_{\mathrm{b}}=0.0493$, a scalar spectral index $n_{\mathrm{s}}=0.9649$, a dark energy equation-of-state parameter $w = -1$, and a dimensionless Hubble parameter of $h=0.6736$\footnote{We define the Hubble constant as $H_0 = 100 \times h$ km s$^{-1}$ Mpc$^{-1}$.}. The dark energy density $\Omega_{\Lambda}$ is adapted for each cosmology to achieve a flat geometry. 
All simulations include three massive neutrino species with a mass of $m_{\nu}=0.02$ eV per species. A degenerate mass hierarchy was adopted. The neutrinos are modelled as 
a relativistic fluid in the simulations following the treatment described in \cite{tram2019fully}. This results in a neutrino density of $\Omega_{\nu} \approx 0.0014$ today. \\

\noindent The particle positions are returned in $\sim 87$ particle shells distributed between $z=3.0$ and $z=0.0$, using the lightcone mode of \pkdgrav.
The exact number of shells varies slightly with cosmology as the shells are equally spaced in proper time.
The default precision settings of \pkdgrav\ are used and the initial conditions of all simulations are generated at $z = 99.0$. \\

\noindent To model the additional measurement uncertainty from the remaining, unconstrained $\Lambda$CDM parameters that are fixed in the main simulation suite ($\Omega_{\mathrm{b}}$, $n_{\mathrm{s}}$ and $h$) we require an additional set of simulations. Each of the simulations in this suite varies only one of the mentioned parameters, holding the others fixed to their fiducial values. Each parameter direction is explored using simulations at four different locations distributed around the fiducial parameter values. This results in simulations with
\begin{linenomath} 
\begin{align*}
    \Omega_{\mathrm{b}} &\in \{ 0.0453, 0.0473, 0.0513, 0.0533\}, \\
    n_{\mathrm{s}} &\in \{ 0.9249, 0.9449, 0.9849, 1.0249\}, \\
    h &\in \{ 0.6336, 0.6536, 0.6936, 0.7136\}.
\end{align*}
\end{linenomath}
As for the main simulation grid we run five individual \pkdgrav\ simulations per cosmology. We use the same five initial conditions in all these simulations. We refer the reader to Section~\ref{sec:marginal} for the details of the modelling of the influence of these parameters on the summary statistics.

\section{Systematic effects}
\label{sec:systematics}

The results of cosmic shear studies are influenced by a variety of systematic effects
that arise from unaccounted physical effects as well as inaccuracies in the shear estimation process; if not treated, they will bias our outcomes.
We therefore take into account the major systematic effects that are known to bias cosmic shear results:
shear calibration bias, uncertainties in the redshift distribution estimation of the source galaxies, galaxy intrinsic alignment,
and baryonic physics.
Additionally, source clustering is known to potentially bias the results inferred using peak counts (see Section~\ref{sec:source_clustering}). \\

\noindent We include in the inference process a treatment of the multiplicative shear bias, photometric redshift uncertainty, and galaxy intrinsic alignment. We incorporate the amplitude of the galaxy intrinsic alignment signal $A_{\mathrm{IA}}$ as an additional parameter in our analysis due to its strong influence on the outcomes of cosmic shear studies (see e.g. \citep{zurcher2021cosmological}). The values of the remaining systematics parameters considered in this work have been found to be largely unconstrained in weak lensing measurements (see e.g. \citep*{troxel2018dark}). Hence, we model their influence on the summary statistics in a cosmology independent way and marginalise them out over their priors (see Table~\ref{tab:priors}). \\

\noindent We adopt a different strategy to mitigate a second set of potential systematics, namely those due to baryonic physics, source clustering and additive shear biases: we apply appropriate scale cuts to the statistics to control the induced bias in parameters of interest. 
In the following, we give an overview of these systematic effects and we describe how they are treated in this study.

\subsection{Photometric redshift uncertainty}
\label{sec:photo_z}
Surveys aiming at performing cosmological measurements using cosmic shear require the imaging of a large number of galaxies. For such surveys, determining the redshifts of the galaxies using spectroscopy is currently not
feasible, and instead the redshifts are measured using photometry. 
Inaccuracies or catastrophic failures in the redshift measurement can propagate to the redshift 
distributions of the source galaxies. 
It was demonstrated that an incorrect redshift distribution can alter the cosmological constraints inferred in cosmic shear studies (see e.g. \citep{huterer2006systematic, choi2016cfhtlens, hildebrandt2020kids+}). 
We take the uncertainty of the redshift distributions into account by introducing the nuisance parameters $\Delta_{z, i}$ in the analysis. The index $i$ runs over the four tomographic bins. The parameters $\Delta_{z, i}$ incorporate the uncertainty of the redshift distributions by describing a shift of the distributions $n_i(z)$ according to
\begin{equation}
n_i'(z) = n_i(z - \Delta_{z, i}).
\end{equation}
This method of incorporating the uncertainty of the redshift distributions into the analysis was used in previous studies (see \citep*{troxel2018dark}, for example)
and was demonstrated to be adequate for the DES Y3 shape catalogue \citep{y3-hyperrank, des2021cosmic}.
We treat the influence of the nuisance parameters $\Delta_{z, i}$ as a second-order effect and
neglect its dependence on cosmology.
The relative variation (as a function of $\Delta_{z, i}$) of $d_j$ (the $j$-th element of the data vector belonging to tomographic bin $i$) is denoted $f_{i, j}$, so that
\begin{equation}
\label{eq:model2}
d_j(\Omega_{\mathrm{m}}, \sigma_8, \Delta_{z, i}) = d_j(\Omega_{\mathrm{m}}, \sigma_8, \Delta_{z, i}=0) (1 + f_{i, j}(\Delta_{z, i})).
\end{equation}
We model $f_{i, j}$ as a quadratic polynomial:
\begin{equation}
\label{eq:model1}
f_{i, j}(\Delta_{z, i}) = c^1_{i, j} \Delta_{z, i} + c^2_{i, j} \Delta_{z, i}^2.
\end{equation}
The coefficients $c^1_{i, j}$ and $c^2_{i, j}$ are fit individually for each element of the data vector,
based on a set of simulations at the fiducial cosmology in which the $n_i(z)$ has been shifted by $\Delta_{z, i}$. The simulations span the range $\Delta_{z, i} \in [-0.1, 0.1]$ in nine linearly spaced steps with 200 realisations each. \\

\noindent We test that the chosen model is sufficiently accurate (so that its imperfections should not alter the results of the study) using a `leave-one-out' cross-validation strategy (see Appendix~\ref{sec:emulator_tests}).
The priors used for the individual $\Delta_{z, i}$ parameters are listed in Table~\ref{tab:priors}.

\subsection{Shear bias}
The inference of the cosmic shear field from the observed galaxy shapes requires a detailed understanding of how the intrinsic shapes of the galaxies are altered due to gravitational lensing. The Earth's atmosphere, the telescope, and the detector itself further alter the observed shapes of the galaxies. Their influence is typically modelled using the point spread function (PSF) \citep{bernstein2002shapes}. Furthermore, noise rectification or misspecification of the noise model can introduce biases in the measurements \citep*{hirata2004galaxy, bernstein2010shape, refregier2012noise, melchior2012means}.
The measured shear $\vec{\gamma}_{\mathrm{obs}}$ of a single galaxy is commonly modelled as being composed of three terms:
\begin{equation}
    \vec{\gamma}_{\mathrm{obs}} = m \thinspace \vec{\gamma} + \vec{c} + \delta \vec{e}_{\mathrm{noise}},
\end{equation}
where $\vec{\gamma}$ denotes the actual cosmic shear \citep{heymans2006shear, mandelbaum2014third}. The noise component $\delta \vec{e}_{\mathrm{noise}}$ is modelled to have zero mean and its contribution to the averaged
shear signal $\langle \vec{\gamma}_{\mathrm{obs}} \rangle$ is expected to vanish ($\langle \delta \vec{e}_{\mathrm{noise}} \rangle = 0$), if the average is taken over a large enough number of galaxies.
The multiplicative and additive shear biases are denoted as $m$ and $\vec{c}$, respectively.
Both biases can arise from a variety of sources. An error in the estimation of the size of the PSF can introduce a multiplicative bias, whereas an error in the estimation of its ellipticity can contribute to the additive bias. Other sources of bias include selection and detection effects as well as calibration errors in the shear estimation process itself \citep{kaiser2000new, bernstein2002shapes,Kacprzak2012noisebias, bernstein2016accurate, hoekstra2017study, SheldonMcal2017, fenech2017calibration}. \\

\noindent The DES Y3 \Mcal\ shape catalogue used in this study underwent extensive testing for shear biases by \cite*{gatti2020dark}.
The \Mcal\ shear estimation method does not account for a potential shear dependence in the detections and blending of sources.
Therefore, \cite{maccran2020} used image simulations to detect and correct for remaining sources of additive biases.
Additionally, \cite*{gatti2020dark} performed a series of null tests on the catalogue level to investigate known sources of shear biases, such as stellar contamination.
A null test using the B-mode shear signal was performed and an investigation on the correlations between galaxy properties and the shear signal was carried out.
Lastly, a set of PSF diagnostics were used to assess the accuracy of the PSF estimation and to model additive biases arising from insufficient PSF modelling.
However, \cite*{gatti2020dark} point out that some small additive biases from PSF mismodelling 
still remain. While they do not correct the shape catalogue for these biases, they do provide
an estimate of their contribution to the galaxy shapes.

\noindent Hence, we confirm that 
the estimated additive shear bias does not change the outcomes of this study (see Appendix~\ref{sec:add_shear_bias_test}) and we refrain from including an additional treatment of the additive shear bias in the inference process.

\noindent On the other hand, we treat the tomographic multiplicative shear biases $m_i$ as nuisance parameters, as even small multiplicative biases are expected to alter the results of cosmic shear studies. The index $i$ denotes the tomographic bin.
We treat the multiplicative shear bias as a second-order effect independent of cosmology and use a treatment analogous to that used for the photometric redshift uncertainty (see Section~\ref{sec:photo_z}). The fitting is based on 
a set of simulations at the fiducial cosmology that span the range $m_i \in [-0.1, 0.1]$ in nine linearly spaced steps with 200 realisations each. 
The multiplicative shear bias is incorporated in the simulations by altering the cosmological convergence signal 
according to 
\begin{equation}
\kappa_{m_i} = (1 + m_i)\kappa_{m_i=0}.
\end{equation}
The accuracy of the model is tested in Appendix~\ref{sec:emulator_tests} and the priors used on the parameters $m_i$ are included in Table~\ref{tab:priors}.

\subsection{Galaxy intrinsic alignment}
\label{sec:intrinsic_alignment}
Weak gravitational lensing causes distortions of the shapes of galaxies of $\sim 1\%$ of the intrinsic ellipticities of the galaxies. Therefore, cosmic shear studies must gain in statistical power 
by averaging over multiple galaxies in a patch of the sky. Under the assumption that 
the intrinsic shapes of the galaxies are randomly distributed, the intrinsic shapes average out and the
cosmic shear signal becomes measurable.

\noindent This assumption does not hold true in reality as the intrinsic ellipticities of galaxies are correlated
with each other as well as with the large-scale structure.
This effect, which is typically not incorporated in dark-matter-only simulations, is referred to as galaxy intrinsic alignment (IA).
The IA signal can be broken down into two components: 1) intrinsic-intrinsic (II), describing the correlation between the ellipticities of the galaxies and the large-scale structure, and 2) gravitational-intrinsic (GI), describing the correlation between the sheared background galaxies and the ellipticities of the foreground galaxies \citep{heavens2000intrinsic}. \\

\noindent We include a treatment of IA in the inference process as IA is known to potentially bias cosmological parameter constraints inferred in cosmic shear studies if neglected \citep{heavens2000intrinsic}.
We use a map-level implementation of the non-linear intrinsic alignment model (NLA), which was first introduced in \cite{Fluri2019kids}, to generate IA signals from the dark matter simulations. 
We refer the reader to \citetalias{zurcher2021cosmological} for the details of the implementation.
The NLA model was developed by \citet*{hirata2004intrinsic, bridle2007dark, joachimi2011constraints}
and has three model parameters:
$A_{\mathrm{IA}}$, the IA amplitude, governs the overall strength of the signal,
while $\eta$ and $\beta$ allow a dependency of the IA signal on the galaxy's redshift and luminosity, respectively. The dependencies are modelled around arbitrary pivot parameters $z_0$ and $L_0$. 
We include the IA amplitude $A_{\mathrm{IA}}$ and $\eta$ as nuisance parameters in our inference pipeline but we neglect
the luminosity dependence of the IA signal by setting $\beta=0$.
This parameter choice for the modelling of the IA signal was used previously in \citet*{troxel2018dark}.
The redshift dependence is modelled around the median redshift $z_{\mathrm{med}} \approx 0.6$ of the global redshift distribution of the galaxies in the DES Y3 data. \\

\noindent The influence of $A_{\mathrm{IA}}$ on the summary statistics is modelled using simulations and incorporated in the emulator (see Section~\ref{sec:emulator}).
The value of $A_{\mathrm{IA}}$ is measured in the inference process.
On the other hand, we treat the redshift dependence of the IA signal as a second-order, cosmology-independent effect. 
The effect of $\eta$ on the data vector
level is modelled, analogously to the effect of photometric redshift uncertainty and multiplicative shear bias, using a
quadratic polynomial (see Section~\ref{sec:photo_z}). 
The coefficients are fit using a set of simulations at the central cosmology that span the range $\eta \in [-5, 5]$ in nine linearly spaced steps with 200 realisations each. 
The amplitude of the IA signal was fixed to $A_{\mathrm{IA}}=1$ in these simulations.
The accuracy of the model is tested in Appendix~\ref{sec:emulator_tests} and the priors used on $A_{\mathrm{IA}}$ and $\eta$ are listed in Table~\ref{tab:priors}.

\subsection{Baryons}
\label{sec:baryons}
The presence of baryons heavily affects the small scale fluctuations of the observable Universe. 
While radiative cooling can lead to a faster collapse of haloes and therefore steeper halo profiles \citep{yang2013baryon}, feedback effects arising from
active galactic nuclei or stellar winds and supernovae can counteract the collapse of structures \citep{osato2015impact}.
The individual contributions of the different baryonic effects depend strongly on the mass of the host halo and its redshift as well as the 
feedback model adopted \citep{mccarthy2016bahamas}. 
The complexity of the modelling of such baryonic physics often restricts the accessibility of small scales in cosmic shear studies.
While there exist models such as \texttt{HMCODE} \citep{mead2015accurate} that include baryonic effects in the theory predictions for the angular power spectra,
no such models are currently available for peak counts.
In a forward-modelling approach as followed in this work, the baryonic effects would have to be added to the dark-matter-only simulations. 
Recent, successful approaches to alter dark-matter-only simulations to 
mimic the effect of baryonic physics include the use of parametric models to change the positions of the simulated particles \citep{schneider2019quantifying} and the use of deep learning 
techniques to paint the baryon effects onto the lensing maps as inferred
from hydrodynamic simulations \citep{troster2019painting}.
The impact of baryons on the peak counts was studied in \cite{weiss2019effects}, who found that it becomes small for large smoothing scales. \\

\noindent Due to the large uncertainty in the feedback model the inclusion of baryonic effects in the analysis would require the introduction of several additional nuisance parameters, increasing the computational cost of the
analysis significantly.
Instead, we decided to quantify the impact of baryonic effects on the resulting cosmological parameter constraints using baryon-contaminated mock simulations. 
We then restrict ourselves to the use of scales that are not significantly biased by baryonic physics. 
The details of the test and the results are outlined in Appendix~\ref{sec:baryon_test}.
Based on the results of the test, we restrict ourselves to spherical harmonics with $\ell \leq 578$ for the angular power spectra and to peaks with a full-width-at-half-maximum ($\mathrm{FWHM}$) $\geq 7.9$ arcmin, avoiding the need for a full treatment of baryonic effects.

\subsection{Source clustering}
\label{sec:source_clustering}
The strength of the lensing signal depends on the arrangement of the source galaxies and the corresponding foreground lenses along the line-of-sight. The lensing signal reaches its maximum when the distance between the observer and the lens is 
equal to the distance between the lens and the source galaxy. 
Hence, the strength of the lensing signal of a certain lens depends on the redshift distribution of the 
corresponding source galaxies behind the lens. 
In order to accurately reproduce the lensing signal measured in the data, the redshift distribution of the source galaxies in the simulations at the position of a lens of a certain strength must match the corresponding distribution in the data. \\

\noindent While we match the global, tomographic redshift distributions of the source galaxies in our simulations with those in the data, we do not take into account the variation of the redshift distributions across the survey region.
Such variations can be caused by the varying depth of the survey across the sky, as well as by source clustering and blending.

\noindent Source clustering refers to the alteration of the local redshift distribution of the source galaxies at the location of a massive galaxy cluster due to a large number of galaxies residing at the redshift of the cluster.

\noindent While source clustering increases the number of galaxies at the redshift of the lens, blending counteracts this effect to some extent, as the increased number of galaxies in the same region of the sky inevitably causes the loss of some galaxies due to overlaps in their light profiles. \\

\noindent Neither blending nor source clustering are reproduced realistically in our simulations; the galaxies in the simulations
are located in the same positions as in the DES Y3 shape catalogue
and are not correlated with the dark matter density (see Section~\ref{sec:mass_mapping}).
In a non-tomographic analysis or in a case of highly-overlapping redshift bins, this can lead to a difference in the $\kappa$ measurement at positions of clusters, where the shear signal would be diluted by the cluster members, which carry no shear signal associated with the cluster.
The effect of source clustering can be accounted for by modifying the signal-to-noise-ratio (SNR) of the detected lensing peaks according 
to a `boost factor' correction \citep{mandelbaum2005systematic}. 
However, in the case of a sufficient separation of sources and lenses, this effect is negligible. \\

\noindent As we use a tomographic analysis in this work and past studies have found blending to be sub-dominant compared to source clustering, we expect blending to have marginal impact on our results (\citepalias{kacprzak2016cosmology}).
That is why we only investigate the influence of source clustering in our analysis.
To infer the boost factors we follow the approach outlined in Appendix C of \citetalias{kacprzak2016cosmology}.
Using the boost factor corrections, we measure the impact of source clustering on the cosmological constraints inferred using
peak counts. We find that source clustering does not constitute a significant source of bias in this survey setup and hence we 
neglect it in this analysis (see Appendix~\ref{sec:boost_factor_corr}).

\section{Method} 
\label{sec:method}

We infer cosmological constraints from the DES Y3 data using two map-based statistics: angular power spectra and peak counts.
To avoid needing analytical predictions for the peak counts we 
rely on a forward modelling approach. Based on a suite of numerical simulations, we forward-model DES Y3-like mass maps for different cosmologies, conserving the original survey properties (e.g. galaxy number density, shape noise properties, etc.).
By measuring the statistics on the simulated mass maps we obtain predictions for the selected cosmologies sampled with simulations.
An emulator is used to predict the summary statistics at other cosmologies. \\

\noindent Our analysis closely follows the approach outlined in \citetalias{zurcher2021cosmological}. 
In the following we introduce the mass map simulation procedure, the calculation of the summary statistics, and the inference process, including a description of the emulator, covariance, and likelihood.
We use an updated version of the \texttt{NGSF}\footnote{\url{https://cosmo-gitlab.phys.ethz.ch/cosmo_public/NGSF}}
(Non-Gaussian Statistics Framework) software 
as well as \texttt{estats}\footnote{\url{https://cosmo-gitlab.phys.ethz.ch/cosmo_public/estats}} (as introduced in \citetalias{zurcher2021cosmological}).
The complete codebase is publicly available to ensure the reproducibility of the presented results.

\subsection{Mass map simulations}
\label{sec:mass_mapping}
The \texttt{PKDGRAV3} simulation suite introduced in Section~\ref{sec:simulations} is used to predict DES Y3-like mass maps 
at 58 different cosmologies distributed in the $\Omega_{\mathrm{m}}-\sigma_8$ plane. 
We use the \texttt{UFalcon}\footnote{\url{https://cosmology.ethz.ch/research/software-lab/UFalcon.html}} software to convert the discrete particle density shells of the \pkdgrav\ simulations into mass maps in the same fashion as \citetalias{zurcher2021cosmological}.
We refer the reader to \cite{sgier2019fast} for a complete description of the \texttt{UFalcon} software. \\
\noindent Using the Born approximation, \texttt{UFalcon} avoids relying on a full ray-tracing treatment that would otherwise be necessary to produce the mass maps. 
The use of the Born approximation might introduce some deterioration of the accuracy of the simulated mass maps as compared
to a full ray tracing treatment.
Although \cite{petri2017validity} have demonstrated that the bias introduced by such inaccuracies is negligible in the presence of shape noise, even for a LSST-like survey, we test that the angular power spectra of the simulated mass maps agree with predictions
from a state-of-the-art theory code (see Appendix~\ref{sec:acc_sim_conv}). \\

\begin{figure*}
\begin{center}
\includegraphics[width=1\textwidth]{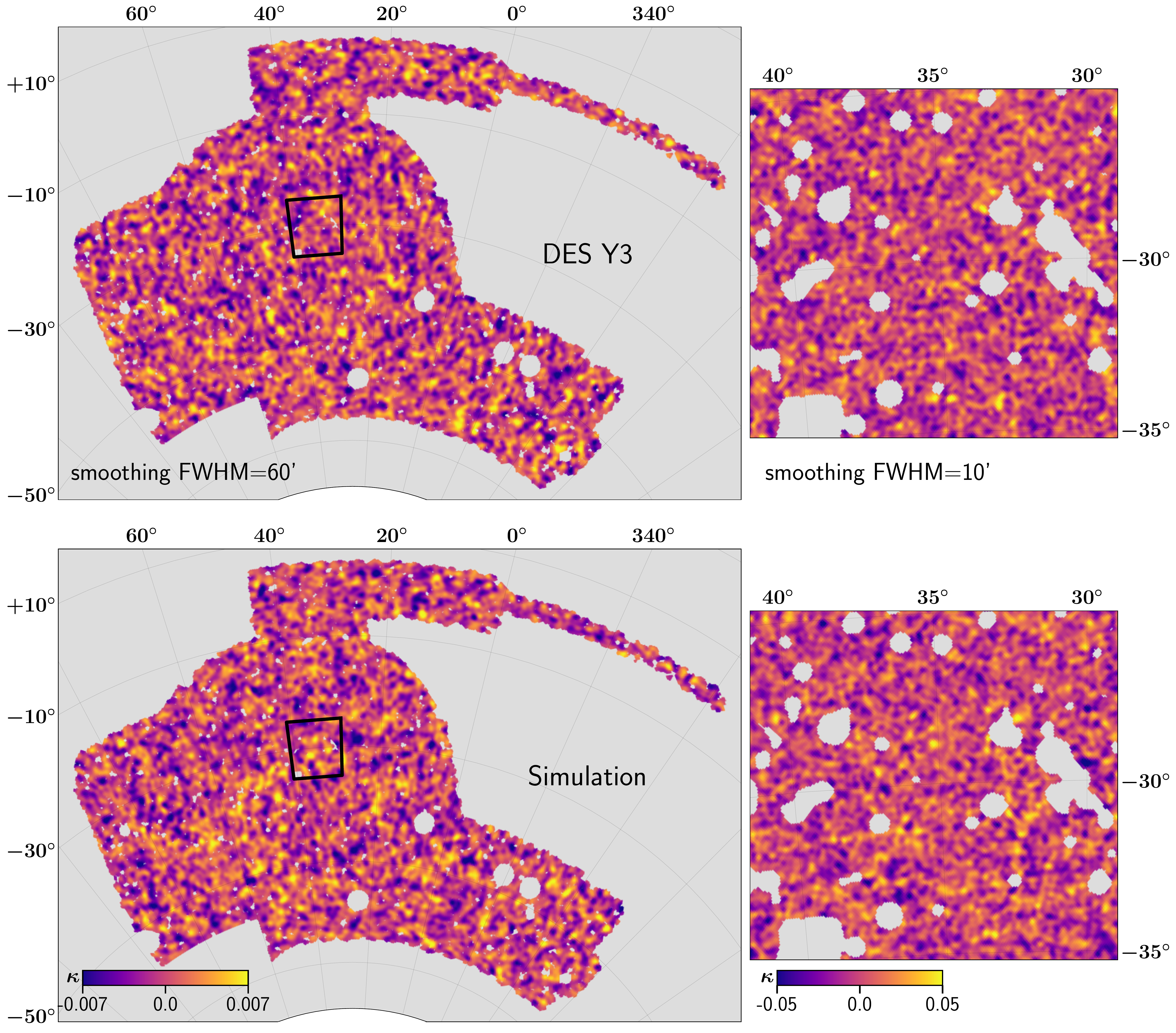}
\end{center}
\caption{Comparison of the non-tomographic DES Y3 mass map and a simulated mass map from \darkgrid, 
produced using \pkdgrav\ and following the procedure outlined in Section~\ref{sec:mass_mapping}. The maps were smoothed using a Gaussian smoothing kernel with a full-width-at-half-maximum (FWHM) of 60 and 10 arcmin for the full and the zoomed in maps, respectively.}
\label{fig:mass_maps}
\end{figure*}

\noindent In order to realistically reconstruct the properties of the original DES Y3 mass maps, 
a statistically equivalent shape noise component has to be added to the cosmological 
convergence signal. We perform the addition of shape noise and cosmological signal in 
shear space. Hence the cosmological convergence signal $\kappa$
must first be converted to a shear signal $\vec{\gamma}$; this is done using the spherical Kaiser-Squires (KS) mass mapping method \citep{kaiser1993mapping, wallis2017mapping}. The shape noise signal $\vec{e}_{\mathrm{noise}}$ is drawn from the original 
DES Y3 shape catalogue by rotating the galaxy ellipticities $\vec{e}_{j}$ in place by a randomly drawn phase $\phi_{j}$ for each galaxy. The addition of shape noise and the cosmological signal is performed at the pixel level to obtain the total simulated shear signal $\vec{e}_{\mathrm{sim}}$:
\begin{equation}
\vec{e}_{\mathrm{sim}} = \vec{e}_{\mathrm{noise}} + \vec{\gamma} = \frac{\sum_{{j}=1}^N w_{j} \vec{e}_{j}\exp(i\phi_{j})}{\sum_{{j}=1}^N w_{j}} + \vec{\gamma}.
\end{equation}
The individual galaxies are weighted according to their \Mcal\ weights $w_j$.
The final, forward modelled mass map $\kappa_{\mathrm{sim}}$ is then obtained by applying the spherical KS mass mapping method on the $\vec{e}_{\mathrm{sim}}$ map.
The pixelization of the sphere was performed using the \texttt{HEALPIX}\footnote{\url{http://healpix.sf.net}} software
\citep{gorski2005healpix} with a resolution of \texttt{NSIDE} = 1024.
The mass maps simulated by this procedure contain a statistically equivalent shape noise component and have the same survey mask as the original DES Y3 mass maps. A visual comparison of such a simulated mass map to the DES Y3 mass map is shown in Figure~\ref{fig:mass_maps}.

\noindent To optimally use the full-sky simulations we rotate the DES Y3 galaxy coordinates on the sky in order to produce four independent simulations of the survey from a single simulation in \darkgrid. As demonstrated in Appendix~\ref{sec:acc_sim_conv}, these
rotations leave the angular power spectra of the maps unaffected.

\subsection{Summary statistics}
Two-point summary statistics, such as the angular power spectra, are sufficient statistics in the case of a homogeneous, isotropic Gaussian random field with zero mean. This assumption is violated for mass maps due to the non-linear nature of gravitational collapse at late times leading to the maps becoming non-Gaussian. 
It has been found that additional statistics, such as peak counts, ought to be considered in order to fully describe the mass maps \citep{petri2014impact}. 
In this work, we extract information from the mass maps using two statistics: the angular power 
spectrum and peak counts. 
In the following, we introduce the theoretical background of the two statistics and describe how we measure them from the mass maps in this analysis.

\paragraph*{Angular power spectrum}
Two-point statistics have been well studied and 
have been used extensively and successfully to extract cosmological information from mass maps in weak lensing surveys (see \cite*{heymans2013cfhtlens, hildebrandt2017kids, troxel2018dark, hikage2019cosmology, Heymans2021kids, des2021cosmic, secco2021dark} for example). One of their main advantages is that they can be well modelled from the accurately predictable power spectrum using the Limber approximation \citep{limber1953analysis}. It was found that the Limber approximation is appropriate for current weak lensing surveys \citep{lemos2017effect}.
In this study we use the angular power spectrum (the Fourier analogue of the real-space angular two-point correlation function). \\

\noindent The angular power spectrum $C_{\ell}$ of the convergence field $\kappa$ for a multipole $\ell \geq 0$ can be estimated as
\begin{equation}
C_{\ell} = \frac{1}{2\ell + 1}\sum_{m=-\ell}^{\ell} |a_{lm}|^2, 
\end{equation}
given the decomposition of $\kappa$ into its spherical harmonic components $a_{lm}$.
As gravitational lensing only produces curl-free modes the convergence field $\kappa$
is commonly decomposed into a curl-free component $\kappa_{\mathrm{E}}$ and
a divergence-free component $\kappa_{\mathrm{B}}$. In the presence of a finite-size survey mask, mode mixing can lead to a small part of the cosmological signal leaking into the B-modes as well as the production of EB-modes. However, we only consider the E-modes ($C^{\mathrm{E}}_{\ell}$) for cosmological inference in this work. 

\noindent Gravitational lensing does not produce any B-modes ($C^{\mathrm{B}}_{\ell}$), but systematic effects, arising for example from imperfections in the shear-calibration process or selection biases, can do so.
Hence, B-modes are often used to test for unaccounted systematic effects in weak lensing studies (see for example \cite*{zuntz2018dark} or \cite*{gatti2020dark}). We perform a null test on the B-mode signal as part of our unblinding procedure in Section~\ref{sec:blinding}.
The angular power spectra are measured in 32 square-root-spaced bins between $\ell = 8$ and $\ell = 2048$.
However, not all bins are 
used in the inference procedure (see Section~\ref{sec:blinding}).
We measure the angular power spectra from the mass maps using the
\texttt{anafast} routine of the \texttt{healpy} software \citep{zonca2019healpy}.

\paragraph*{Peak counts}
Massive structures of the local Universe such as dark matter haloes get imprinted on mass maps as local maxima, known as \textit{peaks}. The study of peaks provides a way to
extract information from such highly non-linear structures that is largely complementary to the information captured by two-point statistics \citep{tyson1990detection, miralda1991correlation, kaiser1993mapping,Yang2011information}.
A straightforward way to use peaks for cosmological inference is to 
measure the number of peaks of a mass map.
The term \textit{peak function} has emerged in recent literature for such a record of the number of peaks as a function of either the signal-to-noise ratio ($\mathrm{SNR}=\kappa / \sigma_{\kappa}$) or the convergence $\kappa$ of the peaks. Here, $\sigma_{\kappa}$ denotes the standard deviation of the mass map.
We choose to bin the detected peaks as a function of $\kappa$ as this binning has been demonstrated to carry a stronger cosmological signal due to the self-similarity of 
the peak functions at different cosmologies in the SNR binning scheme (\citetalias{zurcher2021cosmological}).
The peak function is divided into 15 equally spaced bins. In order to suppress the contribution from shot noise and obtain an approximately Gaussian likelihood the ranges of the outermost bins were chosen such that 
at least 30 peaks are recorded in each bin, independent of cosmology. 
We restrict ourselves to peaks with SNR $\leq 4$, as it is known from past studies that peaks with an SNR $> 4$ are strongly affected by source clustering, leading to biased results (\citetalias{kacprzak2016cosmology}, \citet{harnois2020cosmic}). \\

\noindent There are different ways to detect peaks on weak lensing mass maps.
Some studies record peaks as local maxima on aperture-mass maps (see for example \citetalias{kacprzak2016cosmology}, \citet{martinet2020probing, harnois2020cosmic}).
However, such optimised aperture mass filters only work under a flat-sky approximation. Instead, we detect peaks as local maxima on spherical mass maps directly. We measure peaks of different angular size
by applying Gaussian filters of different scales to the mass maps prior to the detection of the peaks. We use 12 such filters with a full-width-at-half-maximum (FWHM) ranging from 2.6 to 31.6 arcmin resulting in 12 peak functions per tomographic bin. Similarly, as for the angular power spectra, not all of the scales were used in 
the inference procedure (see Section~\ref{sec:blinding}).
We regard a pixel of the smoothed mass maps as a peak if its convergence value is 
higher than the value of its nearest-neighbour pixels. \\

\noindent We detect peaks separately on each one of the tomographic convergence maps $\kappa_i(\theta, \phi)$, where the index $i$ indicates the tomographic bin number. These maps
can be written in the basis of the spin-0 spherical harmonics $Y_{\ell m}(\theta, \phi)$ as
\begin{equation}
    \kappa_i(\theta, \phi) = \sum_{\ell = 0}^{\ell_{\mathrm{max}}}\sum_{m = 0}^{\ell} \hat{\kappa}_{i, \ell m} Y_{\ell m}(\theta, \phi),
\end{equation}
where the upper limit $\ell_{\mathrm{max}}$ is dictated by the pixel resolution of the maps ($\ell_{\mathrm{max}} = 3072$ in our case) \footnote{We note that we take the sum over the parameter $m$ only over the semi-positive range $m \in [0, \ell]$ instead of $m \in [-\ell, \ell]$ as it is commonly done. As the convergence maps are real-valued and the spherical harmonics $Y_{lm}$ satisfy the symmetry relation $Y_{lm}^*=(-1)^m Y_{l-m}$ all information is contained in the coefficients $\{a_{lm} | m \in [0, \ell]\}$. The remaining coefficients $\{a_{lm} | m \in [-\ell, -1]\}$ can be reconstructed from $\{a_{lm} | m \in [0, \ell]\}$.}.
The inclusion of peaks detected from maps constructed 
using multiple tomographic bins (hereafter called `cross-peaks') was demonstrated to provide additional information beyond using solely `auto-peaks' by \cite{martinet2020probing, harnois2020cosmic}. 
In this work we explore the potential of cross-peaks, identified on spherical, convolved convergence maps $\kappa_{ij}(\theta, \phi)$ 
\begin{equation}
    \kappa_{ij}(\theta, \phi) = \sum_{\ell = 0}^{\ell_{\mathrm{max}}}\sum_{m = 0}^{\ell} \hat{\kappa}_{i, \ell m} \hat{\kappa}_{j, \ell m} Y_{\ell m}(\theta, \phi),
\end{equation}
where the indices $i$ and $j$ indicate the two different tomographic bins.
We apply the same set of Gaussian filters to the convolved convergence maps before peak detection, as we do for the auto-peaks.
We include cross-peaks in the fiducial setup of this analysis. \\

\noindent Further information beyond the peak function can be extracted by including other peak-based statistics that are not used in this study such as for example the density profiles around mass map peaks \citep{marian2013cosmological}.

\subsection{Emulator}
\label{sec:emulator}
We require a means to predict the angular power spectra and the peak counts for different cosmologies as well as different configurations of nuisance parameters. As galaxy intrinsic alignment is strongly degenerate with cosmology, we model the amplitude of the intrinsic alignment signal $A_{\mathrm{IA}}$ as being dependent on the cosmological parameters $\Omega_{\mathrm{m}}$ and $\sigma_8$.
The influence of the remaining nuisance parameters on the statistics are treated as second-order effects and are modelled as being independent of cosmology to first order (see Section~\ref{sec:systematics}).
We train a Gaussian Process Regression (GPR) emulator to predict the values of the statistics for different inputs of $\Omega_{\mathrm{m}}$, $\sigma_8$, and $A_{\mathrm{IA}}$ (see e.g. \cite{quinonero2005unifying}).

\noindent The training of the GPR emulator is based on simulations at 522 different locations in the $\Omega_{\mathrm{m}}-\sigma_8-A_{\mathrm{IA}}$ space, of which 58 are the original \texttt{PKDGRAV3} simulations spanning the $\Omega_{\mathrm{m}}-\sigma_8-(A_{\mathrm{IA}}=0)$ subspace. A map-based implementation of the 
non-linear intrinsic alignment model (NLA) is used to generate $A_{\mathrm{IA}}\neq 0$ simulations from the \darkgrid\ to sample the remaining parameter space (see Section~\ref{sec:intrinsic_alignment}).
As each point in the parameter space is sampled using five independent simulations and subsequently four survey rotations
are applied (see Section~\ref{sec:mass_mapping}), $522 \times 5 \times 4 = 10\,440$ truly independent simulations are generated.
Additionally, ten shape-noise realisations are produced per simulation.
We note that this leads to simulations that are only pseudo-independent. While the noise signal is different for each simulation, some simulations contain the same cosmological signal (we discuss the implications of this in Section~\ref{sec:compression}).
The resulting $104\,400$ simulations are used to predict the mean values of the statistics at the 522 locations of the parameter space. The GPR emulator is then trained using these predictions.

\noindent We use the GPR implementation \texttt{GaussianProcessRegressor} of \texttt{scikit-learn} \citep{scikit-learn} and a radial-based function (RBF) kernel. The length scale is optimized prior to training to minimize fitting errors.
A different GPR emulator is trained individually for each element of the data vectors. \\

\noindent We report on the accuracy of the emulator in Figure~\ref{fig:interpolator} in Appendix~\ref{sec:emulator_tests}. 
The accuracy of the emulator was tested using a `leave-one-out' cross-validation strategy. 
We conclude that the emulator is sufficiently accurate that its imperfections will not alter the results of this study significantly. \\

\begin{figure*}
\begin{center}
\includegraphics[width=1\textwidth]{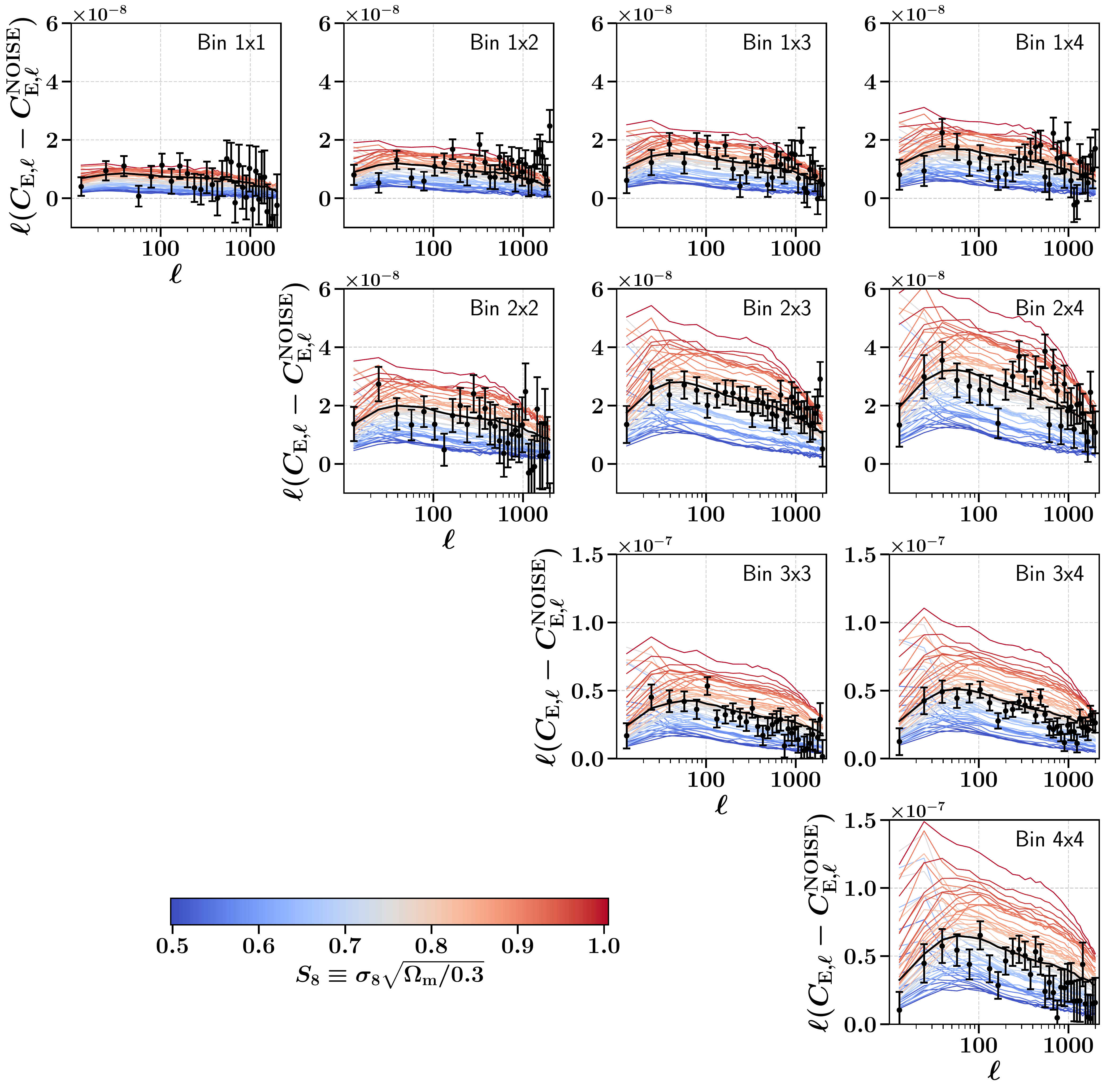}
\end{center}
\caption{Comparison between the predicted angular power spectra at different $S_8$ (coloured curves) and the power spectra measured from the DES Y3 mass maps (black data points). The error bars indicate the standard deviation estimated from the simulations at the fiducial cosmology. For ease of presentation the contribution from the shape noise signal has been subtracted from all spectra.
The shaded regions indicate the scales that were not used in the analysis due to the imposed scale cuts (see Section~\ref{sec:blinding}).
The best-fitting simulation to the data yields a global reduced $\chi^2 / \mathrm{dof}=1.49$ and a p-value of 
$p=17\%$ and is shown as a black curve in each panel.}
\label{fig:e_modes_CrossCLs}
\end{figure*}

\begin{figure*}
\begin{center}
\includegraphics[width=1\textwidth]{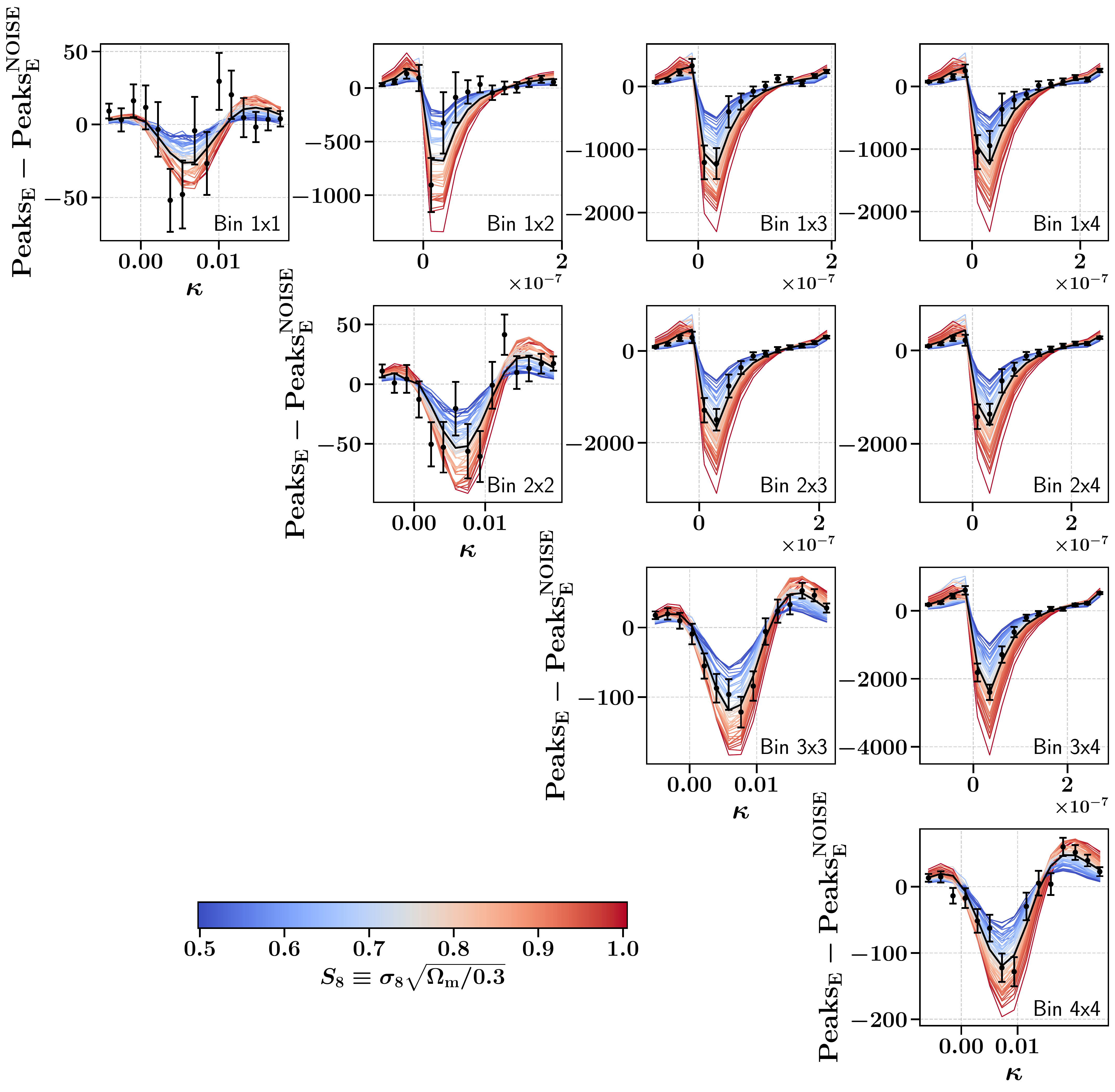}
\end{center}
\caption{Comparison between the predicted peak counts at different $S_8$ (coloured curves) and the peak counts measured from the DES Y3 mass maps (black data points). The error bars indicate the standard deviation estimated from the simulations at the fiducial cosmology. For the sake of presentation, the number of random peaks, estimated from noise-only simulations, were subtracted.
For brevity, we only present the results for a filter scale of $\mathrm{FWHM} = 21.1$ arcmin.
The best-fitting simulation to the data yields a global $\chi^2 / \mathrm{dof}=1.13$ and a p-value of $p=34\%$ and is shown as a black curve in each panel.}
\label{fig:e_modes_Peaks}
\end{figure*}

\noindent The predicted angular power spectra and peak counts for different values of $S_8$ are presented in Figure~\ref{fig:e_modes_CrossCLs} and Figure~\ref{fig:e_modes_Peaks}, respectively. The predictions are
compared to the angular power spectra and peak counts measured from the DES Y3 mass maps.

\subsection{Approximate marginalisation over remaining $\Lambda$CDM parameters}
\label{sec:marginal}
While past studies have found that weak lensing measurements
can put tight constraints on a combination of $\Omega_{\mathrm{m}}$ and $\sigma_8$ they also showed that the remaining $\Lambda$CDM parameters are largely unconstrained by weak lensing data (see \citet*{troxel2018dark} for example).
Hence, we refrain from measuring the remaining $\Lambda$CDM parameters ($\Omega_{\mathrm{b}}$, $n_{\mathrm{s}}$ and $h$) in this analysis. Instead, we model their influence on the summary statistics in a similar fashion as for the photometric redshift uncertainty, multiplicative shear bias, and redshift dependence of the galaxy intrinsic alignment signal. \\
\noindent We model the dependency of the summary statistics on the individual parameters at the \darkgrid\ fiducial cosmology.
Following the same strategy as for the emulation of the systematic effects (see Equation~\ref{eq:model2} and Equation~\ref{eq:model1}), for each parameter $\theta \in \{\Omega_{\mathrm{b}}, n_{\mathrm{s}}, h\}$ and each data vector element j, we define $f_{j}(\theta)$ to be the relative variation in $d_{j}$ as a function of $\theta$, so that
\begin{equation}
d_{j}(\Omega_{\mathrm{m}}, \sigma_8, \theta) = d_{j}(\Omega_{\mathrm{m}}, \sigma_8) (1 + f_{j}(\theta)).
\end{equation}
We then model $f_{j}(\theta)$ as a quadratic polynomial:
\begin{equation}
f_{j}(\theta) = c^1_{j} (\theta - \theta_{\mathrm{fiducial}}) + c^2_{j} (\theta - \theta_{\mathrm{fiducial}})^2,
\label{eq:marg_model}
\end{equation}
where $\theta_{\mathrm{fiducial}}$ indicates the Planck 2018 \citep{aghanim2020planck} value of the parameter in question.
Note that this treatment can be understood to be an expansion of the likelihood function in the parameters $\Omega_{\mathrm{b}}, n_{\mathrm{s}}$ and $h$ around their fiducial values.
Hence, the parameters $\Omega_{\mathrm{b}}$, $n_{\mathrm{s}}$ and $h$ are treated as nuisance parameters in the inference process and are marginalised out over their priors.
As the parameters $\Omega_{\mathrm{b}}$, $n_{\mathrm{s}}$ and $h$ are fixed to the values found in the Planck 2018 \citep{aghanim2020planck} study in all simulations used to train the GPR emulator, we choose normal priors centred at these values. The widths of the priors are chosen as ten times the width of the Planck 2018 \citep{aghanim2020planck} posteriors, primarily to account for the additional uncertainty on the $h$ parameter due to the tension between early and late universe measurements. The priors used are listed in Table~\ref{tab:priors}. We confirm that the chosen marginalisation scheme does not bias our results by running a set of tests in which the values of $\Omega_{\mathrm{b}}, n_{\mathrm{s}}$ and $h$ were fixed to their fiducial values (see Table~\ref{tab:parameters}). Furthermore, we test that the chosen marginalisation scheme does not exhibit a significant dependence on cosmology (see Appendix~\ref{sec:cos_dep_marg}). \\

\noindent We do not constrain parameters outside of the $\Lambda$CDM model in this study and we leave that to future work.  
However, we mention that \citet{harnois2020cosmic} found that the additional non-Gaussian information extracted using peak counts can help to constrain the dark matter equation of state parameter $w$. Especially, cross-peaks are expected to greatly improve such constraints \citep{martinet2020probing}. \\

\noindent Counts of peaks with a high SNR $\gtrsim 4$ have also been identified as a powerful means to constrain the sum of neutrino masses, \citep{li2019constraining, ajani2020constraining}.
While we do not infer the sum of neutrino masses in this study, we are potentially biased by a sum of neutrino masses that is different from $\sum m_{\nu} = 0.06$ eV as adopted in the simulations. \citet{fong2019impact} found that in a Stage-3-like weak lensing survey setup only peaks with an SNR > 3.5 are significantly affected by changes in the sum of neutrino masses and that effects from baryonic physics dominate otherwise. As we have already chosen scale cuts that mitigate the effects from baryonic physics and most of the cosmological constraining power is obtained from peaks with low or medium SNR we apply no additional cuts to accommodate for this effect.

\subsection{Covariance matrix}
\label{sec:cov_matrix}
\begin{figure}
\includegraphics[width=0.45\textwidth]{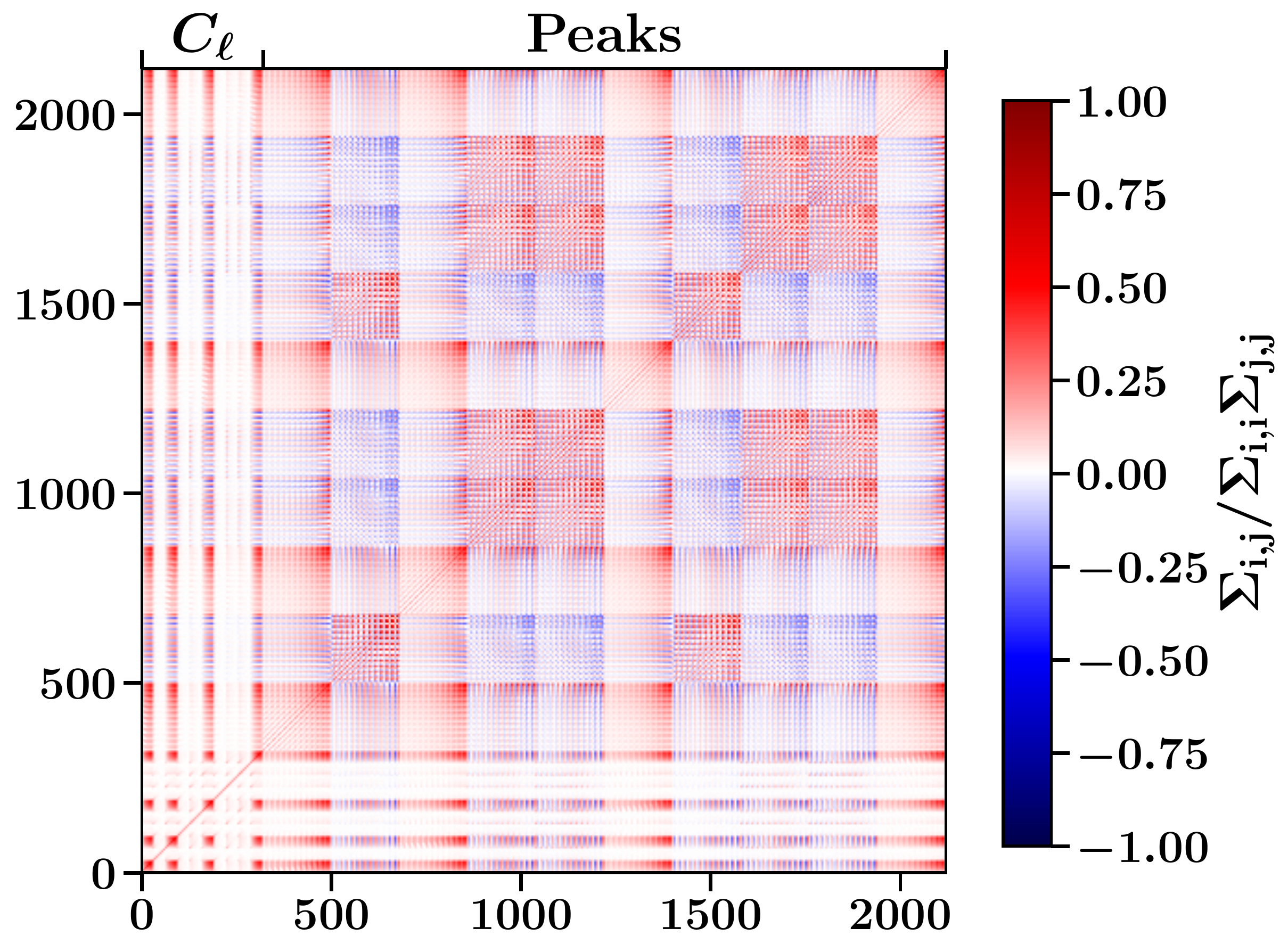}
\caption{The combined, tomographic correlation matrix $C_{{i, j}} \equiv \Sigma_{{i, j}}/\sqrt{\Sigma_{i, i} \Sigma_{j, j}}$
including all angular power spectra and peak counts. The ordering of the tomographic bin combinations for the angular power spectra and peak counts is from left to right 1x1, 1x2, 2x2, 1x3, 2x3, 3x3, 1x4, 2x4, 3x4 and 4x4. For each power spectrum 
all 32 bins ranging from $\ell = 8$ to $\ell = 2048$ are shown.
For each tomographic bin combination the peak counts are visualised for all scales
31.6, 29.0, 26.4, 23.7, 21.1, 18.5, 15.8, 13.2, 10.5, 7.9, 5.3 and 2.6 arcmin from left to right.
The correlation matrix is estimated based on $10\,000$ simulations at the central cosmology ($\Omega_{\mathrm{m}}=0.26, \sigma_8=0.84$).
We note that not all data vector bins shown here are used in the inference (see Section~\ref{sec:blinding}).}
\label{fig:corr_mat}
\end{figure}

A stable estimate of the covariance matrix is needed to accurately infer cosmological parameters.
We use a set of $N_{\mathrm{c}} = 10\,000$ pseudo-independent, simulated realisations $\vec{X}_{i}$ of the summary statistics at the central cosmology ($\Omega_{\mathrm{m}}=0.26, \sigma_8=0.84$)
to robustly estimate the covariance matrix $\Sigma$ according to
\begin{equation}
\hat{\Sigma} = \frac{1}{N_{\mathrm{c}} - 1} \sum_{{i}=1}^{N_{\mathrm{c}}} (\vec{X}_{i} - \hat{\vec{X}}_{\mathrm{M}}) (\vec{X}_{i} - \hat{\vec{X}}_{\mathrm{M}})^T,
\end{equation}
where $\hat{\vec{X}}_{\mathrm{M}}$ is the estimated mean data vector at the fiducial cosmology.
The $10\,000$ simulations are generated from 50 independent simulations. Four survey rotations as well
as 50 shape noise realisations are used per simulation, yielding the final $10\,000$ simulations that are used to estimate $\Sigma$.
Figure~\ref{fig:corr_mat} shows the combined, tomographic correlation matrix $C_{{i, j}} \equiv \Sigma_{{i, j}}/\sqrt{\Sigma_{i, i} \Sigma_{j, j}}$ including all angular power spectra and peak counts at all scales. \\
\noindent We find similarly positive correlations between peaks identified on different scales and using different redshift bins as in past studies (see \citetalias{zurcher2021cosmological} for example). 
Further, we observe that the cross-peaks are only mildly correlated with the auto-peaks indicating that they indeed probe a different kind of information. However, we also find some strong correlations between different cross-peaks similarly as in \cite{harnois2020cosmic}. \\

\noindent The covariance matrix has two major constituents; (1) A highly correlated shape-noise component originating from the uncertainty of the intrinsic shapes of the source galaxies as
well as measurement noise, and (2) a cosmic variance term accounting for the randomness in the cosmic matter density field itself. \\
\noindent We compare the angular power spectrum part of our estimated covariance matrix to the covariance matrix used
in \citet{doux2021harmonic}.
While we estimate our covariance matrix from simulations, \citet{doux2021harmonic} use an analytical covariance matrix. 
They calculate the Gaussian part of the covariance matrix with \texttt{NaMaster} \citep{alonso2019unified}. The non-Gaussian part is computed using the \texttt{CosmoLike} software \citep{eifler2014combining, krause2017cosmolike}, taking into account super-sample covariance and the contribution from the connected four-point covariance originating from the shear field trispectrum.
We find that the two covariance matrices agree very well, despite the differences in the two approaches (see \citet{doux2021harmonic} for details). 
The agreement further strengthens our belief in the accuracy of the simulations and the used methodology. Since the analytical covariance matrix includes an accurate estimate of the super-sample covariance the agreement especially alleviates prior concerns about the underestimation of such super-survey modes in the \texttt{PKDGRAV} simulations (see Section~\ref{sec:simulations}).

\subsection{Data compression}
\label{sec:compression}
As we estimate the covariance matrix from a finite number of data vector realisations the resulting covariance matrix is noisy. 
To reduce the noise we apply data compression in order to shorten the length of the data vectors prior to the computation of the covariance matrix.
We compress the data vectors using the MOPED data compression algorithm \citep{tegmark1997karhunen, heavens2000massive, gualdi2018maximal}. 
We implement the MOPED compression following the algorithm described in \citet*{Gatti2020moments}.
The algorithm yields a compression matrix that reduces the length of the original data vectors to the number of model parameters that are inferred in the analysis (15 in our case). The calculation of the MOPED compression matrix follows a Karhunen–Lo\`eve algorithm as in \cite{heavens2000massive} and \cite*{Gatti2020moments}. 
The $i$-th component of the compressed data vector $\vec{X}^{\mathrm{comp}}$ is calculated from the uncompressed data vector $\vec{X}$ as
\begin{equation}
    \vec{X}_{i}^{\mathrm{comp}} = \frac{\partial \vec{X}^{\mathrm{T}}}{\partial p_{i}}\Sigma^{-1} \vec{X},
\end{equation}
where $p_{i}$ denotes the $i$-th model parameter. The partial derivatives $\frac{\partial \vec{X}}{\partial p_{i}}$ are calculated analytically using the GPR emulator.

\noindent The MOPED compression algorithm requires an estimate of the precision matrix $\Sigma^{-1}$ in the original data space. We use the 10\,000 pseudo-independent simulations at the fiducial cosmology to estimate this precision matrix. Out of these 10\,000 simulations only 200 of them are truly independent (see Section~\ref{sec:cov_matrix}).
This might raise the concern that the contribution from cosmic variance might be underestimated since only 200 different cosmological signals are used in the estimate. However, said precision matrix is only used to calculate the MOPED compression matrix. A potential bias caused by an imperfect compression due to an underestimation of the cosmic variance is alleviated by a simultaneous broadening of the parameter constraints.
For the estimation of the precision matrix in the MOPED space only the 200 truly independent realisations at the fiducial cosmology are used.



\subsection{Likelihood}
\label{sec:likelihood}
We model the likelihood as Gaussian assuming a Gaussian noise model. 
This assumption is approximately justified by the central limit theorem and by choosing the bins of the peak functions such that for each cosmology at least 30 peaks are recorded in each bin.
However, as we do not predict the covariance matrix analytically but rather estimate it from simulations, the intrinsic uncertainty of the covariance matrix itself needs to be incorporated in the likelihood function as well.
According to \cite{sellentin2015parameter} the resulting likelihood is no longer Gaussian but instead is a modified version of a multivariate t-distribution. 
Furthermore, the additional uncertainty due to the estimation of the statistics 
at different cosmologies must be taken into account, leading to an additional correction of the likelihood \citep{jeffrey2019parameter}.
Given a location $\pi$ in parameter space, the probability of observing the data vector $\vec{X}$ reads
\begin{equation}
L(\vec{X} | \pi) \propto \left( 1 + \frac{N_{\pi}}{(N_{\pi} + 1)(N_{\mathrm{c}} - 1)} Q \right)^{-N_{\mathrm{c}}/2},
\end{equation}
with 
\begin{equation*}
    Q = (\vec{X} - \hat{\vec{X}}(\pi))^T\hat{\Sigma}^{-1}(\vec{X} - \hat{\vec{X}}(\pi)).
\end{equation*}
Here, $\hat{\vec{X}}(\pi)$ denotes the prediction of the data vector at location $\pi$. The prediction of $\hat{\vec{X}}(\pi)$ is based on at least 200 simulations depending on $\pi$ ($N_{\pi} \geq 200$). $N_c$ denotes the number of simulations used to estimate the covariance matrix in the MOPED space ($N_c = 200$). \\

\noindent We efficiently sample the parameter space using the Markov Chain Monte Carlo (MCMC) sampler \texttt{emcee} \citep{daniel2013emcee}. All constraints presented in this work were generated using 60 walkers with an individual chain length of $200\,000$ samples. 
We follow the suggestions in the \texttt{emcee} documentation and use the integrated auto-correlation times of the individual chains as a measure of convergence of the chains. Following the \texttt{emcee} documentation we confirm that the chains are long enough to obtain a stable estimate of the integrated auto-correlation times.
We regard all chains used in this study as converged.
The priors used throughout the analysis are listed in Table~\ref{tab:priors}.

\begin{table}
\centering
\caption{Priors used throughout this work.
$\mathcal{U}$ denotes a flat prior with the indicated lower and upper bounds, while $\mathcal{N}$ denotes a normal prior with the indicated mean and scale.}
\begin{adjustbox}{width=0.25\textwidth}
\begin{tabular}{l c}
 \hline
{Parameter} & {Prior} \\
 \hline
 $\Omega_{\mathrm{m}}$ & $\mathcal{U}(0.1, 0.5)$ \\
 $\sigma_8$ & $\mathcal{U}(0.3, 1.4)$ \\
 $\Omega_{\mathrm{b} } \times 10^2$ & $\mathcal{N}(4.39, 0.33)$ \\
 $n_{\mathrm{s}}$ & $\mathcal{N}(0.9649, 0.042)$ \\
 $h$ & $\mathcal{N}(0.6736, 0.054)$ \\
 \hline
 $A_{\mathrm{IA}}$ & $\mathcal{U}(-6, 6)$ \\
 $\eta$ & $\mathcal{U}(-5, 5)$ \\
 $m_1 \times 10^2$ & $\mathcal{N}(-0.63, 0.91)$ \\
 $m_2 \times 10^2$ & $\mathcal{N}(-1.98, 0.78)$ \\
 $m_3 \times 10^2$ & $\mathcal{N}(-2.41, 0.76)$ \\
 $m_4 \times 10^2$ & $\mathcal{N}(-3.69, 0.76)$ \\
 $\Delta_{z,1} \times 10^2$ & $\mathcal{N}(0.0, 1.8)$ \\
 $\Delta_{z,2} \times 10^2$ & $\mathcal{N}(0.0, 1.5)$ \\
 $\Delta_{z,3} \times 10^2$ & $\mathcal{N}(0.0, 1.1)$ \\
 $\Delta_{z,4} \times 10^2$ & $\mathcal{N}(0.0, 1.7)$ \\
 \hline
\label{tab:priors}
\end{tabular}
\end{adjustbox}
\end{table}

\section{Blinding} 
\label{sec:blinding}

In order to avoid intentional or unintentional confirmation biases as well as fine tuning of the data by the experimenters, we followed a strict two-stage blinding scheme throughout the analysis.

\paragraph*{Stage 1} We built the analysis framework using a blinded shape catalogue. The catalogue was blinded according to the scheme used in \citet{zuntz2018dark} and \citet*{gatti2020dark}.
The ellipticities $e$ of the galaxies in the catalogue were transformed as $|\eta| \equiv 2\, \mathrm{arctanh}|e| \rightarrow f |\eta|$, using a fixed but unknown $0.9 < f < 1.1$. This transformation preserves the confinement of the ellipticities to the unit disc, while rescaling all inferred shear values. Hence, the transformation leaves the performance of systematic tests intact, while changing the results of the analysis \citep*{zuntz2018dark}. 
Using the blinded catalogue the analysis pipeline was developed and tested. In order to continue to stage 2 the following criteria had to be met:
\begin{itemize}
    \item The angular power spectra calculated from the simulated full sky mass maps have to agree with the predictions from theory. Similarly, the
    angular power spectra measured from the partial-sky, fully forward-modelled mass maps have to agree with the theory predictions. We use a pseudo-$C_{\ell}$ approach to incorporate the masking effects into the theory predictions \citep{hikage2011shear}. The tests are described in Appendix~\ref{sec:acc_sim_conv}.
    \item The accuracy of the GPR emulator as well as the polynomial models used to emulate the second-order systematic effects has to be sufficient. 
    We follow a `leave-one-out' cross-validation strategy in the tests, which are outlined in more detail in Appendix~\ref{sec:emulator_tests}.
    \item Using synthetic data vector realisations the analysis pipeline has to be able to recover the input cosmology within the uncertainty bounds.
\end{itemize}

\paragraph*{Stage 2} After the criteria defined in stage 1 were met, the shape catalogue
was unblinded to perform a series of systematic tests, without unblinding the E-mode data vectors used for cosmological inference nor the cosmological constraints themselves.
It was investigated how the cosmological constraints react to contamination by systematic effects that are not included in the analysis. The tested effects include:
\begin{itemize}
    \item Additive shear biases (see Appendix~\ref{sec:add_shear_bias_test})
    \item Baryonic physics (see Appendix~\ref{sec:baryon_test})
    \item Source clustering (see Appendix~\ref{sec:boost_factor_corr}).
\end{itemize}
The details of the tests appear in the corresponding appendices.
Contamination due to source clustering or an additive shear bias component did not lead to a significant shift of the cosmological constraints. 
No additional scale cuts were imposed based on these tests.
However, neglecting the impact of baryonic physics caused a significant shift of the $\Omega_{\mathrm{m}} - S_8$ constraints.
Hence, as we do not model the impact of baryons in this study, we restrict ourselves to scales that are only mildly affected by baryonic physics. 
This led to the decision to exclude multipoles $\ell > 578$ in the angular power spectra and peaks with a $\mathrm{FWHM}< 7.9$ arcmin. After these scale cuts, we estimate the shift of the cosmological constraints due to the un-modelled baryonic physics to be smaller than $0.3 \sigma$ (see Appendix~\ref{sec:baryon_test}). The imposed threshold of $0.3 \thinspace \sigma$ is in accordance with the unblinding criteria defined in Appendix D of \citet{des20213x2pt}. \\

\noindent Additionally, a B-mode null test was performed. We compare the measured DES Y3 B-mode signal to a pure shape noise signal that we estimate from simulations without the addition of a cosmological signal. 
As gravitational lensing does not produce any B-modes, the signals are expected to match. 
A disagreement would point towards a remaining contamination of the data by some unaccounted systematic effect.
We present the noise-subtracted B-mode signals of the angular power spectra and the peak counts in Appendix~\ref{sec:b-modes}. Following the unblinding criteria defined in Appendix D of \citet{des20213x2pt} we find the B-mode null test to be passed. \\

\noindent The DES Y3 shape catalogue used in this study underwent extensive testing by \cite*{gatti2020dark}.
We conducted further systematic tests (baryonic physics, source clustering and additive shear bias), as well as map-level null B-mode tests using angular power spectra and peak counts.
However, unknown or un-modelled systematic effects can further bias
our results. We test for such unaccounted systematic effects by performing a set of blinded robustness tests leaving out a different subset of the data vector each time. More specifically, we test for a scale dependent systematic effect by splitting the data vectors into 
a small-scale and a large-scale sample. For the angular power spectra and the peak counts the small-scale samples contain $\ell \in [258, 578]$ and $\mathrm{FWHM} \in [7.9, 18.5]$ arcmin, respectively, while the large-scale samples contain $\ell \in [30, 257]$ and $\mathrm{FWHM} \in [21.1, 31.6]$ arcmin, respectively.
Further, we test for an unaccounted redshift dependent systematic effect by leaving out all data vector elements involving a certain tomographic bin at a time.
None of the alterations led to a significant shift of the constraints. \\

\noindent Lastly, before unblinding the E-mode data vectors and constraints, we performed a `goodness-of-fit test' to check if the complexity of our model is able to reproduce the data sufficiently well.
The best-fitting model was chosen as the prediction of the emulator that yields the maximum posterior given the data.
The best-fit models are included in Figure~\ref{fig:e_modes_CrossCLs} and Figure~\ref{fig:e_modes_Peaks} alongside the predictions from simulations for different cosmologies.
For the angular power spectra we find $\chi^2 / \mathrm{dof}  = 1.49$ and $p = 17\%$ while for the peak counts we find  $\chi^2 / \mathrm{dof}  = 1.13$ and $p = 34\%$, passing the imposed requirement of $p > 1\%$, that was chosen in accordance with the unblinding criteria defined in Appendix D of \citet{des20213x2pt}.

\section{Cosmological Constraints} 
\label{sec:results}

In this work we use angular power spectra and peak counts to constrain the the matter density $\Omega_{\mathrm{m}}$ and the amplitude of density fluctuations $\sigma_8$ of the Universe, as well as the structure growth parameter $S_8 \equiv \sigma_8\sqrt{\Omega_{\mathrm{m}}/0.3}$. 
As the remaining $\Lambda$CDM parameters ($\Omega_{\mathrm{b}}$, $n_{\mathrm{s}}$ and $h$) are mostly unconstrained by weak lensing data, we take the additional uncertainty into account by treating them as nuisance parameters and marginalising them out over their priors (see Section~\ref{sec:marginal}).
Further, we take into account the major systematic effects known to bias weak lensing studies: photometric redshift uncertainty, multiplicative shear bias and galaxy intrinsic alignment (see Section~\ref{sec:systematics}). \\

\noindent In this section, we present our fiducial cosmological parameter constraints and our findings regarding galaxy intrinsic alignment. Furthermore, we discuss the outcomes of a range of internal consistency tests that we performed and we compare our findings to the results found in other studies that make use of the DES Y3 data as well as external data sets. \\

\noindent In the following, the numerical 1D-posteriors of the parameters are presented as the median parameter values as well as their projected 68\% confidence region. A complete record of the numerical results of this study can be found in Table~\ref{tab:parameters} and a visual comparison of the constraints on $S_8$ is presented in Figure~\ref{fig:S8s}.

\subsection{Fiducial results}
\label{sec:fiducials}
We selected the fiducial data vector configurations for the angular power spectra and peak counts based on the results of previous studies, as well as the tests described in Section~\ref{sec:blinding}.
While we measured the angular power spectra in 32 square-root-spaced bins between $\ell=8$ and $\ell=2048$ we do not use all scales in the analysis; instead, we decided to only use the scales $\ell \in [30, 578]$. 
The lower scale cut $\ell \geq 30$ is driven primarily by the decision to exclude scales that might be affected by mixing with super-survey modes (see Section~\ref{sec:simulations}). 
The upper scale cut $\ell \leq 578$ is imposed to stay unbiased to baryonic physics (see Section~\ref{sec:baryon_test}).
\noindent In order to be sensitive to peaks of different spatial extent, the mass maps are smoothed with a set of Gaussian kernels
prior to identifying the peaks. Initially 12 such kernels were used, with $\mathrm{FWHM} = [31.6, 29.0, 26.4, 23.7, 21.1, 18.5, 15.8, 13.2, 10.5, 7.9, 5.3, 2.6]$ arcmin. Again, a scale cut is applied in order to avoid biases from baryonic physics. Hence, we restrict ourselves to using peaks with a $\mathrm{FWHM} \geq 7.9$ arcmin (see Section~\ref{sec:baryon_test}). \\

\noindent We present our fiducial constraints on the cosmological parameters $\Omega_{\mathrm{m}}$, $\sigma_8$ and $S_8$ in the left-hand plot in Figure~\ref{fig:triangle}. For completeness, we include the corresponding constraints of the full cosmological parameter space in Appendix~\ref{sec:full_contours}.
Using solely angular power spectra we find
\begin{linenomath} 
\begin{align*}
    \Omega_{\mathrm{m}} &= 0.278^{+0.080}_{-0.11} \\
    \sigma_8 &= 0.85^{+0.11}_{-0.18} \\
    S_8 &= 0.783^{+0.026}_{-0.019}\thinspace (\mathrm{precision} \ 2.9\%), \\
\end{align*}
\end{linenomath}
while the peaks analysis yields
\begin{linenomath} 
\begin{align*}
    \Omega_{\mathrm{m}} &= 0.252^{+0.030}_{-0.066} \\
    \sigma_8 &= 0.867^{+0.10}_{-0.068} \\
    S_8 &= 0.780\pm 0.016 \thinspace (\mathrm{precision} \ 2.1\%). \\
\end{align*}
\end{linenomath}
We note that the results obtained using the two different summary statistics are well in agreement with each other, without tensions in any of the parameters. Furthermore, we note that the peaks analysis yields tighter constraints on all studied parameters as expected from past studies (see e.g \citetalias{zurcher2021cosmological}, \cite{harnois2020cosmic}), with the $S_8$ constraint tightening up by ~29\% over the angular power spectra analysis. This is different from past studies like \citetalias{kacprzak2016cosmology} where peaks and 2-point statistics led to similar constraints. The peaks gain more in constraining power as the survey area increases compared to the 2-point statistics. \\ 
\noindent We also observe a mild breaking of the $\Omega_{\mathrm{m}}-\sigma_8$ degeneracy due to the non-Gaussian information extracted using the peak counts. The degeneracy breaking is not as pronounced as in past studies due to the scale cut of $\mathrm{FWHM} \geq 7.9$ arcmin. Another way to help to break the $\Omega_{\mathrm{m}}-\sigma_8$ degeneracy is the addition of small-scale shear ratios \citep{sanchez2021} as demonstrated in \citet{des2021cosmic, secco2021dark} and \citet{gatti2021moments}. However, we do not include shear ratios in this study. \\
\noindent The combination of angular power spectra and peak counts yields
\begin{linenomath} 
\begin{align*}
    \Omega_{\mathrm{m}} &= 0.276^{+0.034}_{-0.086} \\
    \sigma_8 &= 0.850^{+0.13}_{-0.068} \\
    S_8 &= 0.797^{+0.015}_{-0.013} \thinspace (\mathrm{precision} \ 1.8\%), \\
\end{align*}
\end{linenomath}
leading to a further improvement of the $S_8$ constraint by ~13\% over the peaks-only analysis.
We find a shift of ~$1 \thinspace \sigma$ towards larger $S_8$ values. We attribute this to a strong break of the $S_8-A_{\mathrm{IA}}$ degeneracy by the combination of the two summary statistics. This effect is discussed in more detail in Section~\ref{sec:IA_contours}.

\subsection{Constraints on galaxy intrinsic alignment}
\label{sec:IA_contours}
\begin{figure*}
\begin{center}
\includegraphics[width=0.45\textwidth]{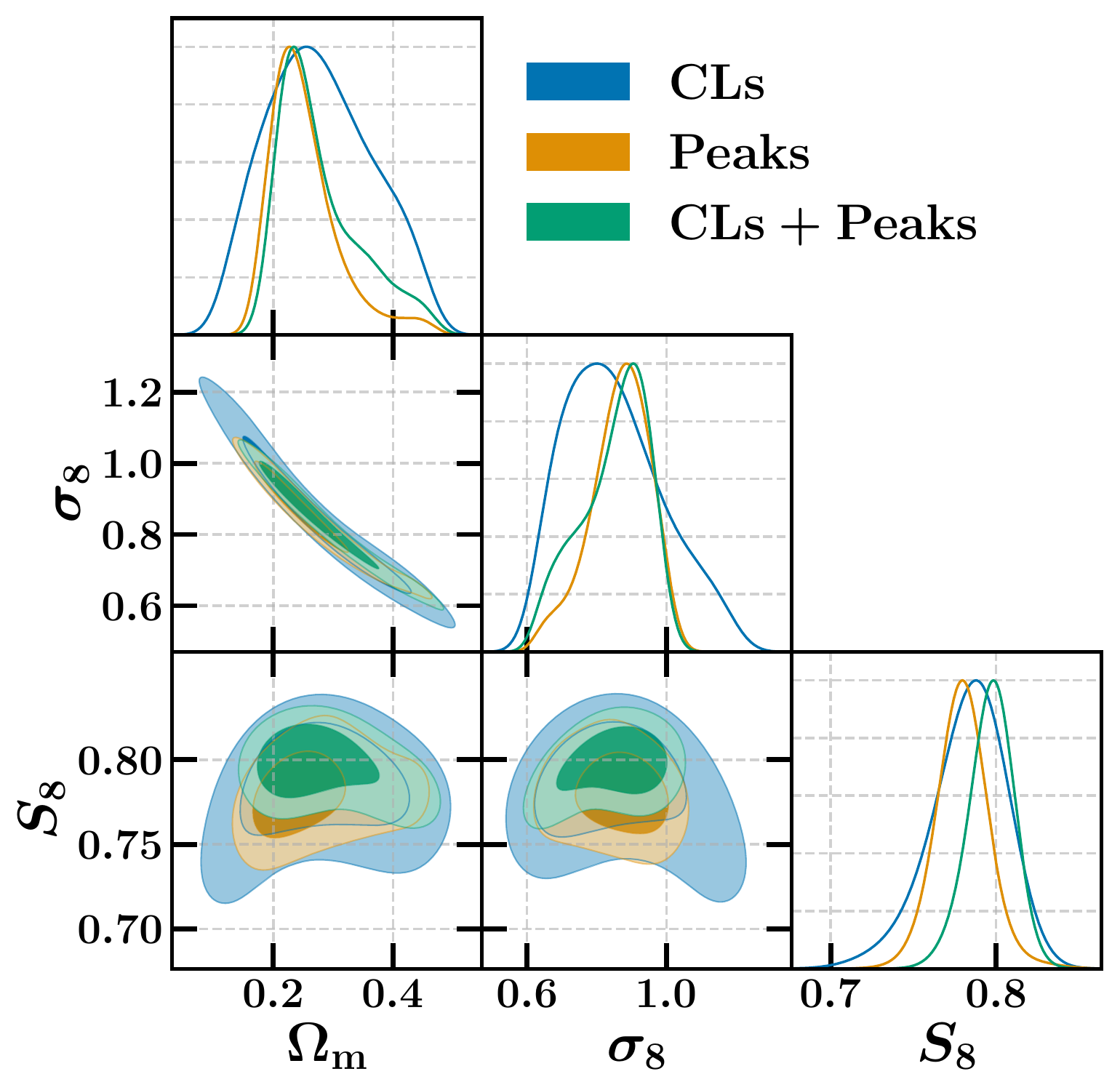}
\includegraphics[width=0.45\textwidth]{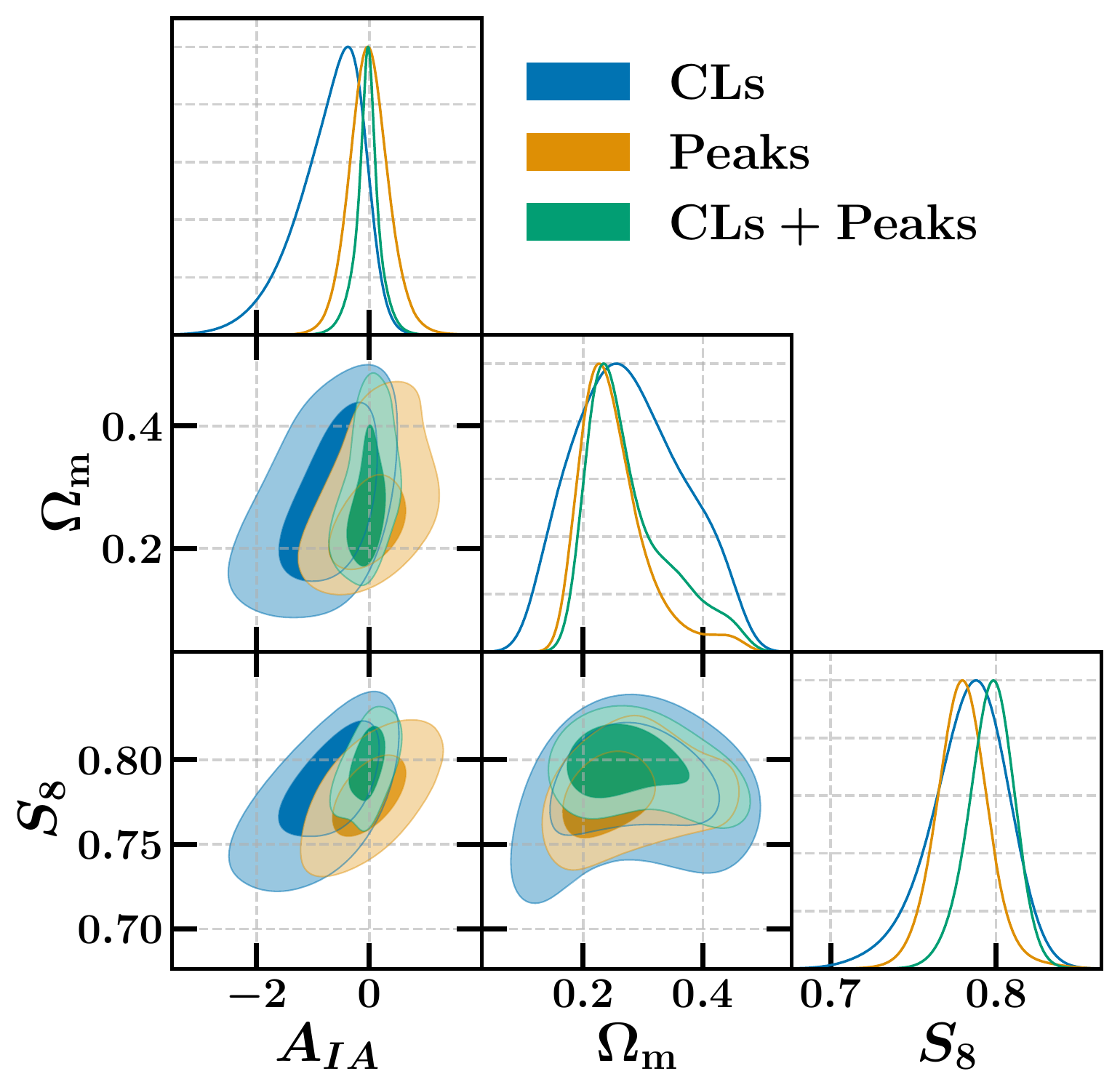}
\end{center}
\caption{\textbf{Left:} Fiducial constraints on the matter density $\Omega_{\mathrm{m}}$, the amplitude of density fluctuations $\sigma_8$, and the structure growth parameter $S_8 \equiv \sigma_8\sqrt{\Omega_{\mathrm{m}}/0.3}$ inferred using the angular power spectra (CLs), peak counts (Peaks), and both summary statistics (CLs + Peaks). \textbf{Right:} Constraints on the galaxy intrinsic alignment amplitude $A_{\mathrm{IA}}$ as well as the degeneracy parameter $S_8$ and $\Omega_{\mathrm{m}}$ as inferred using angular power spectra (CLs), peak counts (Peaks), and both (CLs + Peaks). The contour levels in both plots indicate the 68\% and 95\% confidence regions of the constraints.}
\label{fig:triangle}
\end{figure*}

Galaxy intrinsic alignment is one of the major systematic effects driving the uncertainty of the cosmological parameter constraints in cosmic shear analyses and is known to potentially bias the cosmological parameter constraints \citep{heavens2000intrinsic}. We present, in the right-hand plot in Figure~\ref{fig:triangle}, the constraints found on the amplitude of the intrinsic alignment signal $A_{\mathrm{IA}}$ and its correlation with the inferred cosmological parameters.
We find a strong correlation between $S_8$ and $A_{\mathrm{IA}}$ for both summary statistics with lower $A_{\mathrm{IA}}$ values leading to lower $S_8$ values. While both summary statistics find an $A_{\mathrm{IA}}$ constraint consistent with zero, the angular power spectra analysis prefers values lower than the peaks analysis ($A_{\mathrm{IA}}=-0.72^{+0.72}_{-0.39}$ for angular power spectra and $A_{\mathrm{IA}}= 0.11^{+0.22}_{-0.49}$ for peaks).
We find a tight constraint of $A_{\mathrm{IA}}=-0.03\pm 0.23$ and we observe the hoped-for breaking of the $S_8-A_{\mathrm{IA}}$ degeneracy when the two summary statistics are combined. As can be seen from the right-hand plot in Figure~\ref{fig:triangle} this also leads to a shift of ~$1 \thinspace \sigma$ of the $S_8$ constraint towards larger values when compared to the peaks-only case. \\

\noindent Furthermore, we find that the additional tomographic information obtained with the cross-tomographic peaks strongly contributes to constraining $A_{\mathrm{IA}}$, tightening the $A_{\mathrm{IA}}$ constraints by ~43\% over the auto-peaks only case.
Nevertheless, a similar break of the $S_8-A_{\mathrm{IA}}$ degeneracy can also be observed when the angular power spectra and auto-peaks are combined, but with the angular power spectra dominating the $A_{\mathrm{IA}}$ constraints. In this case a lower value of $A_{\mathrm{IA}}=-0.28^{+0.32}_{-0.18}$ is obtained due to the dominance of the angular power spectra. \\

\noindent While the observed breaking of the $S_8-A_{\mathrm{IA}}$ degeneracy provides a promising way for future weak lensing analyses to improve the robustness of cosmological parameter constraints with respect to galaxy intrinsic alignment, we note that we used a rather simple galaxy intrinsic alignment model in this analysis.
It is left to future studies to investigate if this holds true when more realistic galaxy intrinsic alignment models, such as the Tidal Alignment and Tidal Torquing model (TATT \citep{blazek2019beyond}), are used.

\subsection{Internal consistency checks}
\label{sec:robustness}
A suite of blinded consistency tests using alternative data vector configurations were performed to test for unaccounted systematic effects as discussed in Section~\ref{sec:blinding}.
We repeat these tests after unblinding to assess the internal consistency of the data. The resulting parameter constraints are included in Table~\ref{tab:parameters}.

\begin{figure*}
\begin{center}
\includegraphics[height=0.45\textwidth]{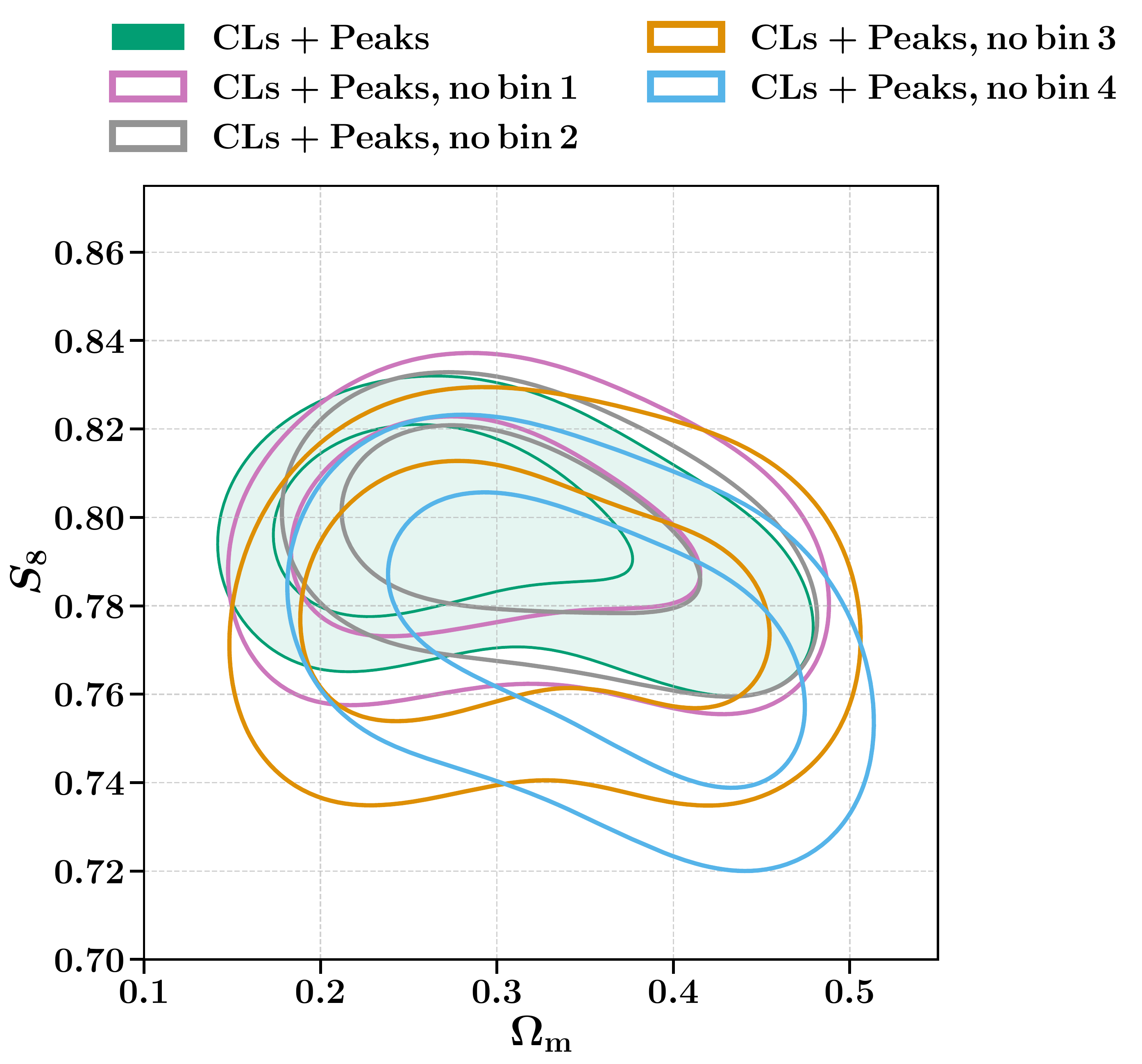}
\includegraphics[height=0.45\textwidth]{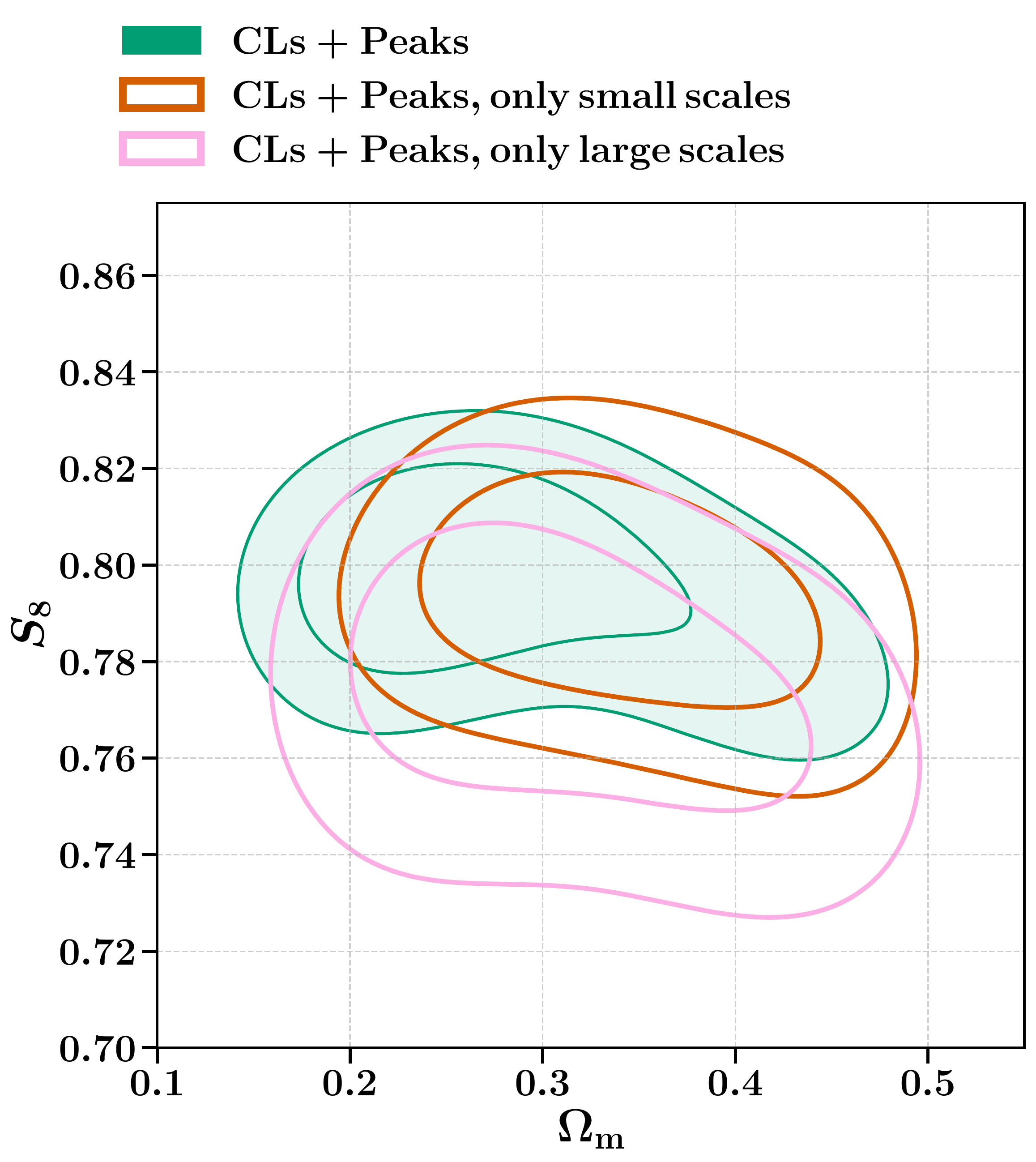} 
\end{center}
\caption{\textbf{Left:} Robustness of the $\Omega_{\mathrm{m}} - S_8$ constraints inferred using the combination of angular power spectra and peak counts (CLs + Peaks) to the removal of individual redshift bins. \textbf{Right:} Analogous robustness test of the $\Omega_{\mathrm{m}} - S_8$ constraints inferred using the combination of angular power spectra and peak counts (CLs + Peaks) to removal of either all large ($\ell \in [30, 257]$ and $\mathrm{FWHM} \in [21.1, 31.6]$ arcmin) or small ($\ell \in [258, 578]$ and $\mathrm{FWHM} \in [7.9, 18.5]$ arcmin) scales. The contour levels in both plots indicate the 68\% and 95\% confidence regions of the constraints. }
\label{fig:robustness}
\end{figure*}

\paragraph*{Tomographic bins} We obtain cosmological constraints leaving out subsets of the data vector that are associated with a certain tomographic bin. The results in the $\Omega_{\mathrm{m}}-S_8$ plane are presented in the left-hand plot of Figure~\ref{fig:robustness}. With tomographic bins 1 and 2 contributing little to the overall constraining power, their removal from the data vector has little impact on the results.  
On the other hand, we find that removing either the contributions from bin 3 or bin 4 leads to a shift towards smaller $S_8$ values by over $1 \thinspace \sigma$, while increasing the uncertainty by $\sim50\%$. The shift in $S_8$ can be explained by the significantly increased uncertainty on $\Omega_{\mathrm{m}}$. As most of the constraining power on $\Omega_{\mathrm{m}}$ is gained from bins 3 and 4, their removal from the data vector allows for larger $\Omega_{\mathrm{m}}$ values to become acceptable, which leads to the observed shift in $S_8$. This effect is illustrated in Figure~\ref{fig:robustness}. While our findings are similar to the trends observed in \citet{des2021cosmic} the shift in $S_8$ is larger in our case if either bin 3 or bin 4 are removed. We attribute this to the additional small scale information from shear ratios that is included by \citet{des2021cosmic}, which disfavours large $\Omega_{\mathrm{m}}$ values.

\paragraph*{Angular scales} We split the data vector into a small and a large scale sample as described in Section~\ref{sec:blinding}. The results in the $\Omega_{\mathrm{m}}-S_8$ plane for the combination of angular power spectra and peak counts are presented in the right-hand plot of Figure~\ref{fig:robustness}, while the individual comparisons are presented in Appendix~\ref{sec:separate_scales}.
We find a similar trend as \citet{des2021cosmic, secco2021dark}, with the removal of the large scales leaving $S_8$ intact, while the removal of the small scales leads to a shift in $S_8$ towards smaller values. We also note that the the uncertainty increases significantly more if the small scales are removed, indicating once more that most information is gained from the smaller scales.

\subsection{Comparison to other studies using DES Y3 data}
\label{sec:internal}

We compare the parameter constraints found in this study using the combination of angular power spectra and peak counts to the results of other analyses that make use of the DES Y3 data. 
We find that all results agree well between the different studies, indicating a high level of internal consistency within the DES Y3 data.
The parameter constraints found in the considered studies were added to Table~\ref{tab:parameters} and a visual comparison of the cosmological constraints is shown in the left-hand plot in Figure~\ref{fig:contours}.

\paragraph*{DES Y3 $\xi_{\pm}$} \label{sec:des_y3_cosmic_shear_comp} We compare our results with the fiducial $\Lambda$CDM constraints found by the DES Y3 cosmic shear analysis \citep{des2021cosmic} that uses angular two-point shear correlation functions as well as small-scale shear ratios. We note that a direct comparison between our results and the DES Y3 $\xi_{\pm}$ results is not meaningful for several reasons: 1) we use normal priors on $\Omega_{\mathrm{b}}$, $n_{\mathrm{s}}$ and $h$, while the DES Y3 $\xi_{\pm}$ analysis uses flat priors, 2) we adopt a fixed sum of the neutrino masses (see Section~\ref{sec:simulations}), while the DES Y3 $\xi_{\pm}$ analysis keeps the sum of the neutrino masses as a free parameter, 3) we model the galaxy intrinsic alignment signal in our simulations using the NLA model, while the DES Y3 $\xi_{\pm}$ analysis uses the more complex TATT model, 4) we impose scale cuts in harmonic space, while the DES Y3 $\xi_{\pm}$ analysis imposes scale cuts in real space, further complicating the comparison. \\
\noindent Therefore, we additionally compare to a modified version of the DES Y3 $\xi_{\pm}$ analysis (which we will refer to as `DES Y3 $\xi_{\pm}$ (comp)') in which the sum of the neutrino masses is fixed to the same value adopted in our analysis and the galaxy intrinsic alignment model is changed to NLA. These modifications account for the primary sources of potential shifts in the parameter constraints that might arise due to the differences in the analysis choices making the comparison between the median parameter constraints more meaningful. However, due to the different scale cuts and prior choices a direct comparison between the found confidence regions is not straightforward. \\
\noindent Including these modifications we find that our results are in good agreement with the DES Y3 $\xi_{\pm}$ (comp) constraints, yielding no differences beyond $1 \thinspace \sigma$ in any of the constrained parameters. Both studies find an intrinsic alignment signal that is consistent with each other and with a null signal. We also note that \citep{des2021cosmic} obtains even tighter $A_{\mathrm{IA}}$ constraints thanks to the inclusion of small scale shear ratios in the analysis.

\paragraph*{DES Y3 3x2pt} \label{sec:des_y3_3x2pt_comp} The DES Y3 3x2pt analysis uses information from galaxy-clustering and galaxy-galaxy lensing in addition to cosmic shear to constrain cosmology \citep{des20213x2pt}. The same caveats as for the comparison between this study and the DES Y3 $\xi_{\pm}$ analysis also apply for the comparison with the DES Y3 3x2pt results.
Therefore, we again compare our results not only to the fiducial DES Y3 3x2pt analysis but also to a modified version called `DES Y3 3x2pt (comp)'. The same modifications as for DES Y3 $\xi_{\pm}$ (comp) were made. The change from the TATT to the NLA model results in an increase in the constraining power. As we did not adapt the scale cuts to reflect this it has to be noted that the DES Y3 3x2pt (comp) analysis does not pass the criteria defined in Appendix D of \citet{des20213x2pt} and the results might be biased. As a consequence we decided to center the constraints at the fiducial cosmology adapted in the simulations ($\Omega_{\mathrm{m}}=0.26, \sigma_8=0.84$). \\



\paragraph*{DES Y3 moments} We use peak counts to extract non-Gaussian information from the convergence field, but other summary statistics have also emerged as powerful tools to capture non-Gaussian information of the convergence field. \citet{gatti2021moments} use the second and third moments of the convergence field to capture Gaussian as well as non-Gaussian information from the DES Y3 shear data. Additionally, \citet{gatti2021moments} use small-scale shear ratios to further tighten their cosmological constraints.
As \citet{gatti2021moments} use the NLA galaxy intrinsic alignment model and keep the sum of the neutrino masses fixed in their analysis we directly compare to their fiducial results without any modifications. 
We again report no tension in any of the inferred parameters beyond $1 \thinspace \sigma$ and both studies measure a galaxy intrinsic alignment signal that is consistent with zero.

\subsection{Comparison to external studies}
\label{sec:comparison}
\begin{figure*}
\begin{center}
\includegraphics[height=0.45\textwidth]{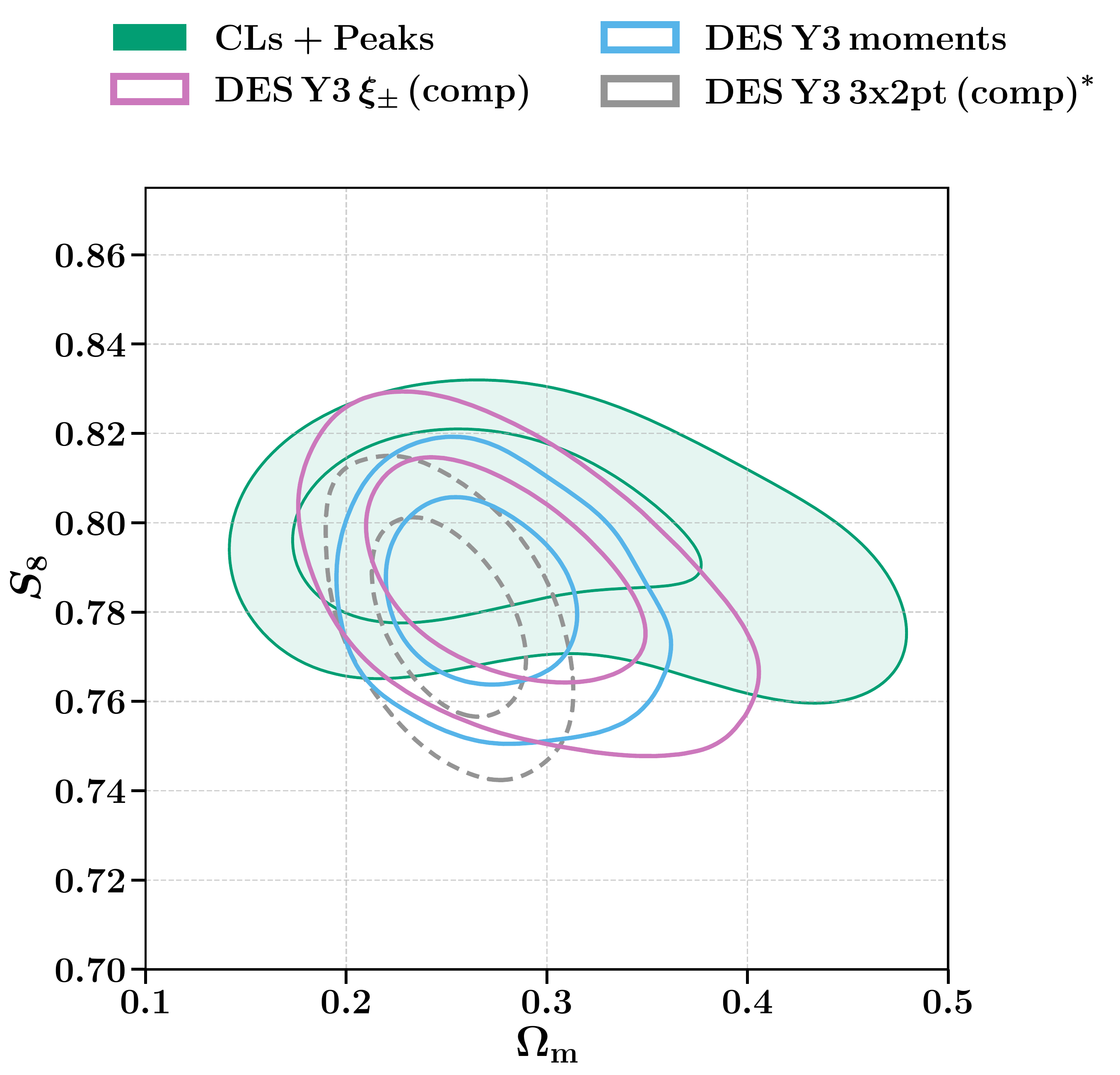}
\includegraphics[height=0.45\textwidth]{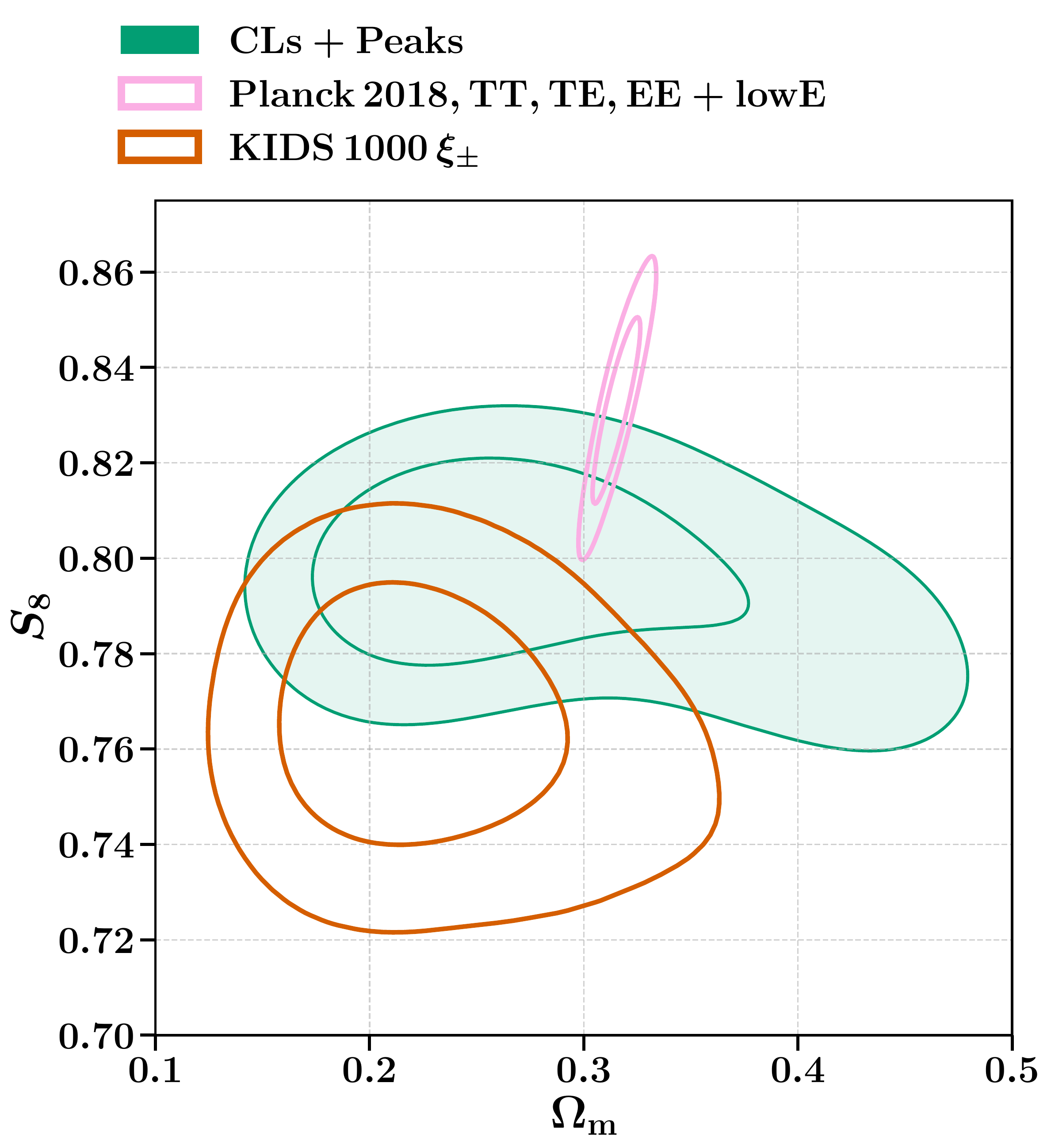} 
\end{center}
\caption{\textbf{Left:} A comparison between the fiducial $\Omega_{\mathrm{m}} - S_8$ constraints inferred in this study using the combination of angular power spectra and peak counts (CLs + Peaks) and the results from other studies using DES Y3 data. The constraints labelled with `(comp)' have been altered to enable a better comparison with our analysis (see Section~\ref{sec:des_y3_cosmic_shear_comp}). The DES Y3 3x2pt (comp) analysis does not pass the criteria defined in \citet{des20213x2pt} and has been centered at the fiducial cosmology adapted in the simulations ($\Omega_{\mathrm{m}}=0.26, \sigma_8=0.84$) (see Section~\ref{sec:des_y3_3x2pt_comp}).
\textbf{Right:} Another comparison between the fiducial $\Omega_{\mathrm{m}} - S_8$ constraints inferred in this study using the combination of angular power spectra and peak counts (CLs + Peaks) and the results from external studies using data other than the DES Y3 data. The contour levels in both plots indicate the 68\% and 95\% confidence regions of the constraints. }
\label{fig:contours}
\end{figure*}

\begin{figure*}
\begin{center}
\includegraphics[width=0.8\textwidth]{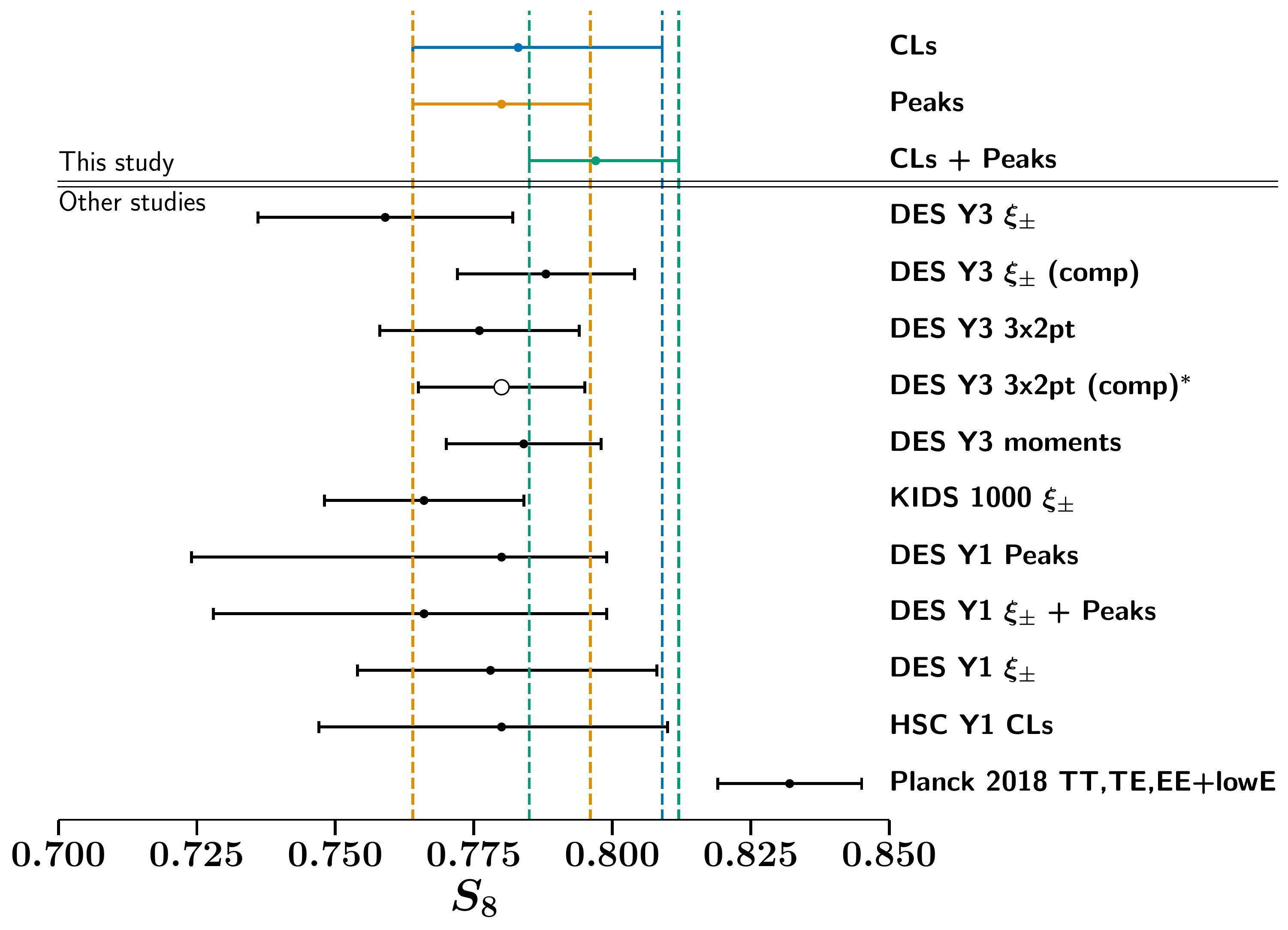}
\end{center}
\caption{A comparison between the constraints on the structure growth parameter $S_8 \equiv \sigma_8\sqrt{\Omega_{\mathrm{m}}/0.3}$ inferred in this study using angular power spectra (CLs), peak counts (Peaks), and both (CLs + Peaks) and the results from other studies.
The constraints labelled with `(comp)' have been altered to enable a better comparison with our analysis (see Section~\ref{sec:des_y3_cosmic_shear_comp}).
The DES Y3 3x2pt (comp) analysis does not pass the criteria defined in \citet{des20213x2pt} and has been centered at the fiducial cosmology adapted in the simulations ($S_8=0.78$) (see Section~\ref{sec:des_y3_3x2pt_comp}).
The results are reported as the median value of the $S_8$ posteriors with the error bars indicating the 68\% confidence limits of the constraints.} 
\label{fig:S8s}
\end{figure*}

The increased sensitivity of current LSS surveys brought to light some moderate tensions between the measurements of the structure growth parameter $S_8$ in the different surveys. Hence, we compare our findings to the recent results of the KIDS 1000 survey \citep{Asgari2021kids}. With the precision of LSS surveys on the $S_8$ parameter approaching the precision of CMB experiments mild tensions between LSS and CMB studies are arising as well. Therefore, we also compare our results to the Planck 2018 study \citep{aghanim2020planck}.
We estimate the tensions between our results and the external studies using the \texttt{tensiometer} software, allowing for a reliable, multi-dimensional estimate of the tensions taking into account non-Gaussianities in the posterior distributions \citep{raveri2019concordance, raveri2020quantifying, raveri2021non}.
A visual comparison between the results of this study and the findings of the KIDS 1000 and Planck 2018 surveys is presented in the right-hand plot of Figure~\ref{fig:contours}.
We restrict the discussion to the comparisons with KIDS 1000 and Planck 2018, but we also include the results from other surveys in Table~\ref{tab:parameters} and Figure~\ref{fig:S8s} that might be of interest to the reader.

\paragraph*{KIDS 1000 $\xi_{\pm}$} The KIDS 1000 cosmic shear study uses three different summary statistics to constrain cosmology from the cosmic shear field: COSEBIs, band powers and the shear two-point correlation functions \citep{Asgari2021kids}. We compare our results to the most constraining KIDS 1000 result obtained using the shear two-point correlation functions. The KIDS 1000 survey uses a similar inference setup as our study keeping the sum of neutrino masses fixed and using the NLA galaxy intrinsic alignment model. Hence, we do not require any modifications for a meaningful comparison. 
We find our results to be in agreement with the KIDS 1000 results at the $0.7 \thinspace \sigma$ level.
However, we stress again that a comparison is not straightforward due to the non-trivial differences in the scale cuts, as well as systematic effects (such as galaxy intrinsic alignments) being treated differently in the KIDS analysis.
Furthermore, the KIDS and DES surveys should not be treated as being fully independent, which further complicates the comparison.

\paragraph*{Planck 2018} We compare our constraints to the ($\Lambda$CDM,TT,TE,EE+lowE+lensing) results of Planck 2018 \citep{aghanim2020planck}, finding a mild tension of 1.5$\sigma$. The measured tension is slightly lower than that recorded in other weak lensing studies. This can be attributed to the near zero value of $A_{\mathrm{IA}}$ and the breaking of the $S_8-A_{\mathrm{IA}}$ degeneracy achieved by the combination of angular power spectra and peak counts (see right-hand plot in Figure~\ref{fig:triangle}).

\begin{table*} 
    \centering
    \caption{Constraints on the cosmological parameters inferred in this study ($\Omega_{\mathrm{m}}$, $\sigma_8$, $S_8$) as well as the galaxy intrinsic alignment amplitude ($A_{\mathrm{IA}}$). The posteriors on the remaining parameters are prior dominated and not presented here. For comparison, the results from other surveys are listed as well.
    The constraints labelled with `(comp)' have been altered to enable a better comparison with our analysis (see Section~\ref{sec:des_y3_cosmic_shear_comp}). Since the DES Y3 3x2pt (comp) analysis does not pass the criteria defined in \citet{des20213x2pt} and might be biased we only quote the found uncertainties on the parameters but not the central values (see Section~\ref{sec:des_y3_3x2pt_comp}).
    Each result is given as the median parameter value along with its 68\% confidence limits.}
\begin{adjustbox}{width=0.75\textwidth}
    \label{tab:parameters}
    \begin{tabular}{c  l  c  c  c  c }
    \hline
    & & $\Omega_{\mathrm{m}}$ & $\sigma_8$ & $S_8$ & $A_{\mathrm{IA}}$ \\ 
\hline
\rule{{0pt}}{{4ex}} \multirow{5}{*}{\rotatebox{90}{\textbf{Fiducials}}} & \textcolor{CLs}{CLs, fiducial} & \textcolor{CLs}{$ 0.278^{+0.080}_{-0.11}$} & \textcolor{CLs}{$ 0.85^{+0.11}_{-0.18}$} & \textcolor{CLs}{$ 0.783^{+0.026}_{-0.019}$} & \textcolor{CLs}{$ -0.72^{+0.72}_{-0.39}$} \\ 
\rule{{0pt}}{{4ex}} &  \textcolor{Peaks}{Peaks, fiducial} & \textcolor{Peaks}{$ 0.252^{+0.030}_{-0.066}$} & \textcolor{Peaks}{$ 0.867^{+0.10}_{-0.068}$} &  \textcolor{Peaks}{$ 0.780\pm 0.016$} & \textcolor{Peaks}{$ 0.11^{+0.22}_{-0.49}$} \\ 
\rule{{0pt}}{{4ex}}  & \textcolor{CLsPeaks}{CLs + Peaks, fiducial} &  \textcolor{CLsPeaks}{$ 0.276^{+0.034}_{-0.086}$} & \textcolor{CLsPeaks}{$ 0.850^{+0.13}_{-0.068}$} & \textcolor{CLsPeaks}{$ 0.797^{+0.015}_{-0.013}$} & \textcolor{CLsPeaks}{$ -0.03\pm 0.23$} \\ 
\hline
\rule{{0pt}}{{4ex}}   \multirow{17}{*}{\rotatebox{90}{\textbf{Variations}}} & CLs + Peaks, only small scales & $ 0.341^{+0.058}_{-0.079}$ & $ 0.756^{+0.070}_{-0.094}$ & $ 0.794\pm 0.017$ & $ 0.17^{+0.25}_{-0.38}$ \\ 
\rule{{0pt}}{{4ex}}  &   CLs + Peaks, only large scales & $ 0.314^{+0.057}_{-0.096}$ & $ 0.775^{+0.097}_{-0.11}$ & $ 0.776^{+0.021}_{-0.019}$ & $ -0.14^{+0.30}_{-0.20}$ \\ 
\rule{{0pt}}{{4ex}}   &  CLs + Peaks, no bin 1 & $ 0.294^{+0.044}_{-0.094}$ & $ 0.821^{+0.12}_{-0.085}$ & $ 0.795\pm 0.017$ & $ 0.08\pm 0.49$ \\ 
\rule{{0pt}}{{4ex}}   &  CLs + Peaks, no bin 2 & $ 0.312^{+0.045}_{-0.082}$ & $ 0.795^{+0.099}_{-0.085}$ & $ 0.797^{+0.015}_{-0.014}$ & $ 0.10^{+0.20}_{-0.25}$ \\ 
\rule{{0pt}}{{4ex}}  &   CLs + Peaks, no bin 3 & $ 0.31^{+0.12}_{-0.11}$ & $ 0.78^{+0.11}_{-0.13}$ & $ 0.781\pm 0.020$ & $ -0.21^{+0.33}_{-0.27}$ \\ 
\rule{{0pt}}{{4ex}}  &   CLs + Peaks, no bin 4 & $ 0.358^{+0.095}_{-0.063}$ & $ 0.722^{+0.062}_{-0.12}$ & $ 0.773\pm 0.021$ & $ -0.09^{+0.24}_{-0.16}$ \\ 
\rule{{0pt}}{{4ex}}   &  CLs + Peaks, fixed $\Omega_{\mathrm{b}}, n_{\mathrm{s}}, h$ & $ 0.258^{+0.033}_{-0.071}$ & $ 0.877^{+0.11}_{-0.073}$ & $ 0.799\pm 0.015$ & $ 0.05\pm 0.28$ \\ 
\rule{{0pt}}{{4ex}}   &  CLs + Peaks, only auto-peaks & $ 0.374^{+0.090}_{-0.047}$ & $ 0.708^{+0.046}_{-0.10}$ & $ 0.777\pm 0.018$ & $ -0.28^{+0.32}_{-0.18}$ \\ 
\rule{{0pt}}{{4ex}}  &   Peaks, only auto-peaks & $ 0.256^{+0.044}_{-0.095}$ & $ 0.85^{+0.14}_{-0.10}$ & $ 0.761\pm 0.022$ & $ 0.59^{+0.37}_{-0.87}$ \\ 
\hline
\rule{{0pt}}{{4ex}}   \multirow{24}{*}[-0.4ex]{\rotatebox{90}{\textbf{Other studies}}} & DES Y3 $\xi_{\pm}$ \citep{{des2021cosmic}} & $ 0.290^{+0.041}_{-0.061}$ & $ 0.783^{+0.074}_{-0.091}$ & $ 0.759\pm 0.023$ &  - \\ 
\rule{{0pt}}{{4ex}}  &   DES Y3 $\xi_{\pm}$ (comp) & $ 0.279^{+0.037}_{-0.053}$ & $ 0.825\pm 0.078$ & $ 0.788\pm 0.016$ & $ -0.08^{+0.19}_{-0.13}$ \\ 
\rule{{0pt}}{{4ex}}   &  DES Y3 3x2pt \citep{{des20213x2pt}} & $ 0.339\pm 0.031$ & $ 0.733^{+0.039}_{-0.050}$ & $ 0.776\pm 0.018$ &  - \\ 
\rule{{0pt}}{{4ex}}    & DES Y3 3x2pt (comp)$^*$ & $ \pm 0.025$ & $^{+0.035}_{-0.042}$ & $ \pm 0.015$ & $ ^{+0.069}_{-0.078}$ \\ 
\rule{{0pt}}{{4ex}}  &   DES Y3 moments \citep{{des2021cosmic}} & $ 0.269^{+0.026}_{-0.036}$ & $ 0.832\pm 0.055$ & $ 0.784\pm 0.014$ & $ -0.09^{+0.21}_{-0.13}$ \\ 
\rule{{0pt}}{{4ex}}   &  KIDS 1000 $\xi_{\pm}$ \citep{{Heymans2021kids}} & $ 0.227^{+0.033}_{-0.053}$ & $ 0.894\pm 0.095$ & $ 0.766\pm 0.018$ & $ 0.34^{+0.38}_{-0.31}$ \\ 
\rule{{0pt}}{{4ex}}   &   DES Y1 Peaks  & - & - & $0.78^{+0.06}_{-0.02}$ & - \\ 
& \citep{{harnois2020cosmic}} &  & & & \\ 

\rule{{0pt}}{{4ex}}  &   DES Y1 $\xi_{\pm}$ + Peaks  & - & - & $0.77^{+0.04}_{-0.03}$ & - \\ 
& \citep{{harnois2020cosmic}} &  & & & \\ 

\rule{{0pt}}{{4ex}}  &   DES Y1 $\xi_{\pm}$ \citep*{{troxel2018dark}} & $ 0.290^{+0.039}_{-0.065}$ & $ 0.802\pm 0.080$ & $ 0.778^{+0.030}_{-0.023}$ & $ 0.79^{+0.72}_{-0.48}$ \\ 
\rule{0pt}{4ex}     & HSC Y1 CLs \citep{hikage2019cosmology} & $0.16^{+0.04}_{-0.09}$ & $1.06^{+0.15}_{-0.28}$ & $0.78^{+0.03}_{-0.03}$ & - \\ 
\rule{{0pt}}{{4ex}} &    Planck 2018 TT,TE,EE + lowE + lensing & $ 0.3153\pm 0.0073$ & $ 0.8111\pm 0.0060$ & $ 0.832\pm 0.013$ &  - \\ 
& \citep{{aghanim2020planck}} &  & & & \\ 
    \hline
    \end{tabular}
    \end{adjustbox}
\end{table*}

\section{Summary} 
\label{sec:summary}
Lensing peaks are sensitive to the highly non-linear features of the mass maps that get imprinted through massive objects in the large scale structure (LSS) of the Universe.
As such, lensing peaks have been found to extract additional non-Gaussian information of the mass maps that is missed by the more commonly used 2-point statistics. We combine the constraining power of peak counts with that of angular power spectra, which primarily target the complementary, Gaussian part of the information. We also include additional redshift information from cross-tomographic peak counts.
We constrain the matter density $\Omega_{\mathrm{m}}$ as well as the amplitude of density fluctuations $\sigma_8$ of the Universe and the structure growth parameter $S_8$ (defined as $S_8 \equiv \sigma_8\sqrt{\Omega_{\mathrm{m}}/0.3}$ in this work) within the $\Lambda$CDM model.
It should be noted that peak counts experience a different $\Omega_{\mathrm{m}}-\sigma_8$ degeneracy than two-point statistics and the chosen definition of $S_8$ is not optimal for peak counts but most customary in weak lensing analyses.
We infer cosmological constraints using the first three years (Y3) of cosmic shear data recorded by the Dark Energy Survey (DES) containing about a hundred million galaxy shapes and spanning 4143 deg$^2$ of the southern sky. 
The uncertainty from the remaining $\Lambda$CDM parameters ($\Omega_{\mathrm{b}}$, $n_{\mathrm{s}}$ and $h$), that are largely unconstrained in weak lensing studies, is taken into account by marginalising them out over their priors. \\

\noindent Our method is based on a forward-modelling approach with numerical simulation. We simulate a large suite of DES Y3-like mass maps, dubbed \darkgrid, to predict the peak counts as well as the angular power spectra. 
We model these functions using a Gaussian Process Regression emulator.
We follow the methodology outlined in \citetalias{zurcher2021cosmological} to generate the simulated mass maps from a suite of \pkdgrav\ \citep{potter2017pkdgrav3} dark-matter-only N-Body simulations sampling the $\Omega_{\mathrm{m}} - \sigma_8$ space.
We confirm that our simulations accurately recover the angular power spectra predicted by the theory code \texttt{CLASS} \citep{lesgourgues2011cosmic}. Further, we consider the major weak lensing systematics in our analysis: multiplicative shear bias, photometric redshift uncertainty, and galaxy intrinsic alignment. While we infer the amplitude of the galaxy intrinsic alignment signal $A_{\mathrm{IA}}$, we treat the redshift dependence of the signal as a nuisance parameter and marginalise over it. we test that our results are robust and not biased by either additive shear biases or source clustering. As we do not include a treatment of baryonic effects in this work, we estimate the impact of the presence of baryons on our results and impose stringent scale cuts to avoid potential biases.

\noindent Throughout the analysis we follow a strict two-stage blinding scheme. Before unblinding we performed a range of robustness checks, a null B-mode test, and a `goodness-of-fit test'. \\

\noindent This work together with \citet{gatti2021moments} demonstrates the potential of using additional summary statistics beyond the commonly used 2-point functions to constrain cosmology from cosmic shear data. The main conclusions of this work include:
\begin{itemize}
    \item In this study we successfully apply the methodology developed in \citetalias{zurcher2021cosmological} to the DES Y3 data. We perform a range of tests that validate the accuracy and applicability of the simulation used and the inference pipeline for this kind of data. We confirm that we recover the angular power spectra predicted from \texttt{CLASS} \citep{lesgourgues2011cosmic} using the simulated DES Y3 like mass maps. Further, we assess that the accuracy of the emulator used is adequate for this sort of analysis.
    \item We investigate several potential sources of systematic errors. While we conclude that our analysis is unlikely to be biased by source clustering effects or additive shear bias we apply stringent scale cuts to our data vectors to avoid biases from unmodelled baryonic physics.
    \item Furthermore, we run a set of robustness tests to test for remaining redshift or scale dependent systematic effects. We perform a map-level null B-mode test finding the noise-subtracted B-mode angular power spectra and peak counts of the DES Y3 data to be consistent with a null signal.
    \item Using a combination of convergence angular power spectra and peak counts, we measure the structure growth parameter $S_8 \equiv \sigma_8\sqrt{\Omega_{\mathrm{m}}/0.3} = 0.797^{+0.015}_{-0.013}$ at 68\% confidence within the $\Lambda$CDM model. This corresponds to a 1.8\% precision constraint and constitutes a ~38\% and ~13\% improvement over the angular power spectra and peaks only cases, respectively.
    \item We compare our results to the findings of other studies that constrain cosmology using DES Y3 data. We find agreement between our results and the other studies at the $1 \thinspace \sigma$ level in all cosmological parameters constrained in this work.
    \item Considering all constrained parameters in this study we find our constraints to be statistically consistent with the results from the KIDS 1000 \citep{Heymans2021kids} survey at the $0.7 \thinspace \sigma$ level. Furthermore, we record a mild tension of  $1.5 \thinspace \sigma$ between our findings and the results from Planck 2018 \citep{aghanim2020planck}. 
    \item We find that the combination of angular power spectra and peak counts tightly constrains the amplitude of the galaxy intrinsic alignment signal to $A_{\mathrm{IA}}=-0.03\pm 0.23$ and breaks the typically observed $S_8-A_{\mathrm{IA}}$ degeneracy, greatly improving cosmological constraints. We further notice that the addition of cross-tomographic peaks significantly contributes to the constraining power of weak lensing peaks on $A_{\mathrm{IA}}$.
\end{itemize}
Having witnessed the potential of using peak counts to constrain cosmology from cosmic shear data, we look ahead to the application of the developed methodology to future data such as the DES Year 6 release. Such future surveys will be able to resolve the small scale structure of the LSS even better than in current data. We plan to include a treatment of baryonic physics that allows us to incorporate and forward-model small-scale baryonic effects in our simulation pipeline and that will allow us to include more small-scale information, considerably improving the cosmological constraining power. \\ \noindent Another way to further increase the amount of cosmological information available is given by including other summary statistics such as Minkowski functionals or minima counts (\citetalias{zurcher2021cosmological}).\\
\noindent A compromise between computational cost and the number of cosmological parameters that can be constrained had to be made. This led to the decision to only measure the parameters $\Omega_{\mathrm{m}}$ and $\sigma_8$ that are constrained most strongly by weak lensing data. We hope to extend our analysis to the full $w$CDM parameter space in a future project. \\
\noindent We observed that the combination of angular power spectra and peak counts puts tight constraints on $A_{\mathrm{IA}}$ and breaks the degeneracy between $S_8$ and $A_{\mathrm{IA}}$. However, we assume the rather simple non-linear intrinsic alignment (NLA) model to incorporate galaxy intrinsic alignment in this work. It is left to future studies to check if this remains true in the context of more complex alignment models such as the Tidal Alignment and Tidal Torquing (TATT) model \citep{blazek2019beyond}.

\section*{Acknowledgements}

The ETH Zurich Cosmology group acknowledges support by grants 200021\_192243 and 200021\_169130 of the Swiss National Science Foundation. \\

\noindent Some of the results in this paper have been derived using the \texttt{healpy} and \texttt{HEALPix} packages. \\

\noindent In this study, we made use of the functionalities provided by
\texttt{numpy} \citep{walt2011numpy}, \texttt{scipy} \citep{virtanen2020scipy}, \texttt{matplotlib} \citep{hunter2007matplotlib}, and \texttt{scikit-learn} \citep{scikit-learn}. \\

\noindent We thank Antony Lewis for the distribution of \texttt{GetDist},
on which we relied to produce some of the plots presented in this work \citep{lewis2019getdist}. \\

\noindent We would also like to thank Uwe Schmitt from ETH Z\"urich for his support
with the GitLab server and CI engine. \\

\noindent Funding for the DES Projects has been provided by the U.S. Department of Energy, the U.S. National Science Foundation, the Ministry of Science and Education of Spain, the Science and Technology Facilities Council of the United Kingdom, the Higher Education Funding Council for England, the National Center for Supercomputing Applications at the University of Illinois at Urbana-Champaign, the Kavli Institute of Cosmological Physics at the University of Chicago, the Center for Cosmology and Astro-Particle Physic at the Ohio State University, the Mitchell Institute for Fundamental Physics and Astronomy at Texas A\&M University, Financiadora de Estudos e Projetos, Fundacao Carlos Chagas Filho de Amparo a Pesquisa do Estado do Rio de Janeiro, Conselho Nacional de Desenvolvimento Científico e Tecnologico and the Ministerio da Ciencia, Tecnologia e Inovacao, the Deutsche Forschungsgemeinschaft and the Collaborating Institutions in the Dark Energy Survey. The Collaborating Institutions are Argonne National Laboratory, the University of California at Santa Cruz, the University of Cambridge, Centro de Investigaciones Energéticas, Medioambientales y Tecnologicas-Madrid, the University of Chicago, University College London, the DES-Brazil Consortium, the University of Edinburgh, the Eidgenoessische Technische Hochschule (ETH) Zurich, Fermi National Accelerator Laboratory, the University of Illinois at Urbana-Champaign, the Institut de Ciencies de l’Espai (IEEC/CSIC), the Institut de Fisica d’Altes Energies, Lawrence Berkeley National Laboratory, the Ludwig-Maximilians Universität Muenchen and the associated Excellence Cluster Universe, the University of Michigan, NFS’s NOIRLab, the University of Nottingham, the Ohio State University, the University of Pennsylvania, the University of Portsmouth, SLAC National Accelerator Laboratory, Stanford University, the University of Sussex, Texas A\&M University, and the OzDES Membership Consortium. Based in part on observations at Cerro Tololo Inter-American Observatory at NSF’s NOIRLab (NOIRLab Prop.ID 2012B-0001; PI: J. Frieman), which is managed by the Association of Universities for Research in Astronomy (AURA) under a cooperative agreement with the National Science Foundation. \\

\noindent Based on observations obtained with Planck\footnote{\url{http://www.esa.int/Planck}}, an ESA science mission with instruments and contributions directly funded by ESA Member States, NASA, and Canada.

\section*{Data availability}
The simulated data used in this work has been generated using the public code \textsc{PKDGRAV} \citep{potter2017pkdgrav3}. The full \mcal\ catalogue will be made publicly available following publication, at the URL \url{https://des.ncsa.illinois.edu/releases}. 
The code used in this article is publicly available at the URL \url{https://cosmo-gitlab.phys.ethz.ch/cosmo_public}.


\appendix
\section{Full cosmological parameter space constraints}
\label{sec:full_contours}
\begin{figure*}
\begin{center}
\includegraphics[width=0.7\textwidth]{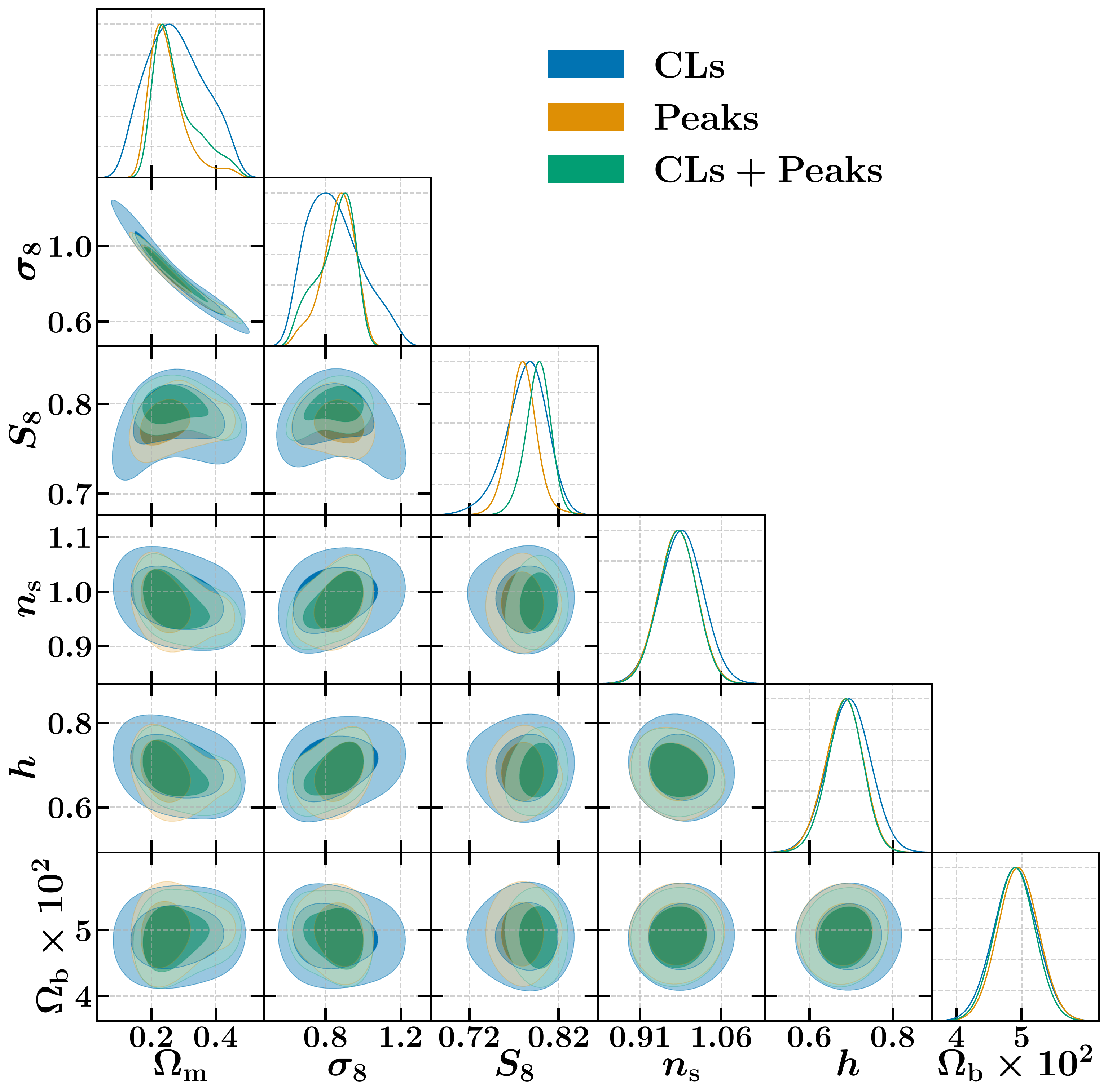}
\end{center}
\caption{The fiducial constraints on all cosmological parameters
obtained using angular power spectra (CLs), peak counts (Peaks), and their combination (CLs + Peaks). The contours indicate the 68\% and 95\% confidence regions. The constraints on $\Omega_{\mathrm{b}}$, $n_{\mathrm{s}}$ and $h$ are prior dominated (see Section~\ref{sec:marginal} for details).}
\label{fig:full_contours}
\end{figure*}

We present the constraints on the full cosmological parameter space for the fiducial data vector setups using the angular power spectra (CLs), peak counts (Peaks) and their combination (CLs + Peaks). The constraints are shown in Figure~\ref{fig:full_contours}.

\section{Emulator accuracy} 
\label{sec:emulator_tests}

We check that the precision of the GPR emulator (used to predict the summary statistics 
for parameter configurations that are not sampled by numerical simulations directly) is sufficient.
We do so following a `leave-one-out' cross-validation strategy. 
The accuracy of the emulator is assessed by building the emulator using all simulations 
except the simulations at one specific parameter configuration 
in the $\sigma_8 - A_{\mathrm{IA}} - S_8$ space. 
The error of the emulator is then calculated as the mean relative difference between 
the predicted data vector and the data vector obtained from the missing simulation as
\begin{equation}
    \epsilon_{\mathrm{emu}} = \mathrm{mean}\left(\frac{|d_{\mathrm{emu}} - d_{\mathrm{sim}}|}{|d_{\mathrm{sim}}|}\right),
    \label{eq:med_error}
\end{equation}
where the mean is taken over all elements of the data vector.

\begin{figure}
\includegraphics[width=0.47\textwidth]{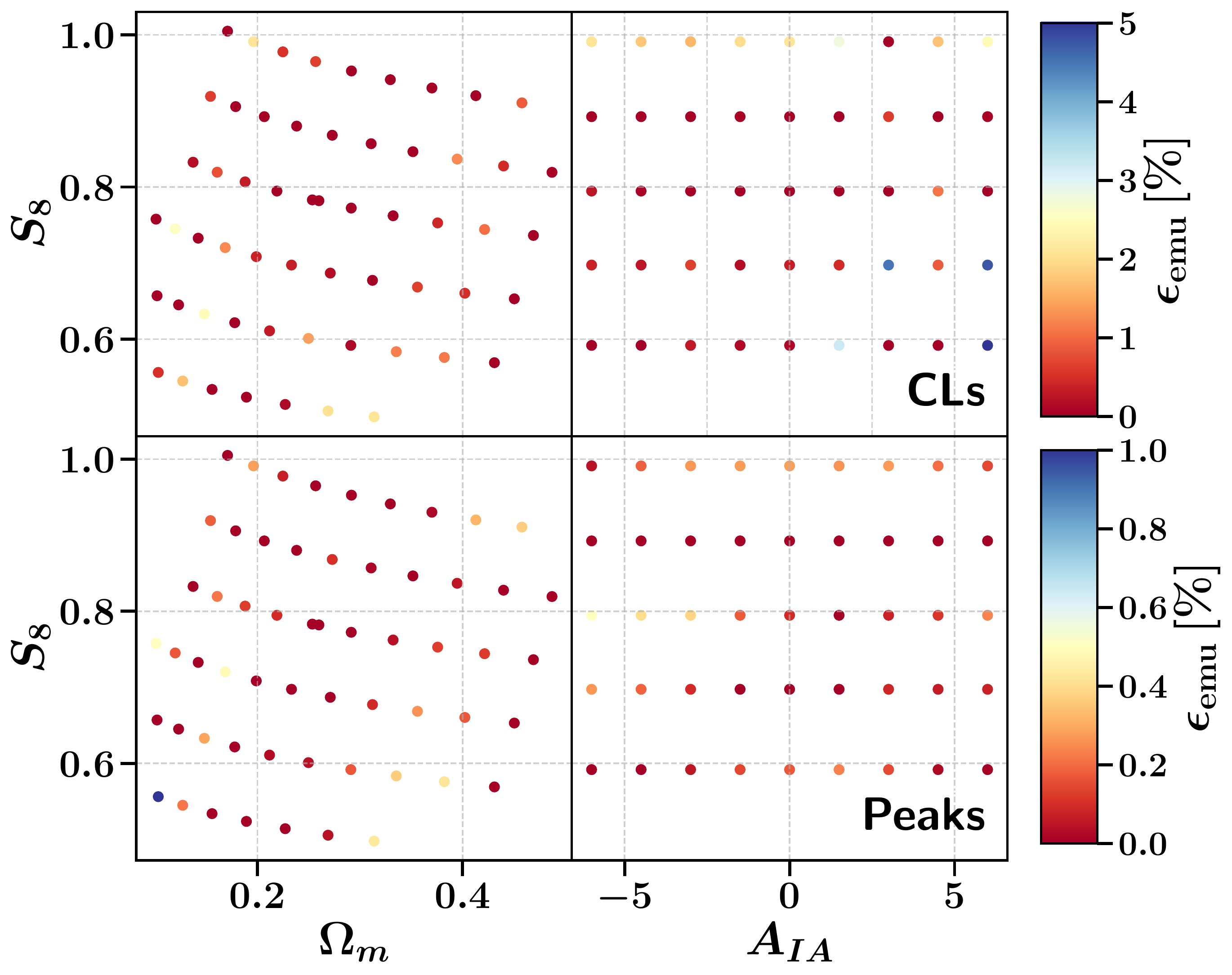}
\caption{Visualisation of the results of the accuracy tests for the GPR emulator in 
the $\Omega_{\mathrm{m}} - \sigma_8 - A_{\mathrm{IA}}$ space. 
The top row shows the results for the angular power spectra and the bottom row shows the results for the peak counts.
The left and right columns display the errors in the $\Omega_{\mathrm{m}} - S_8$ and $A_{\mathrm{IA}} - S_8$ sub-spaces, respectively. 
The colour of the data-points indicates the mean relative error $\epsilon_{\mathrm{emu}}$ of the GPR emulator,
calculated according to Equation~\ref{eq:med_error}.}
\label{fig:interpolator}
\end{figure}

\noindent We perform this test individually for the angular power spectra 
(top row of Figure~\ref{fig:interpolator}) and the peak counts 
(bottom row of Figure~\ref{fig:interpolator}). The maximum error found in both statistics is at the sub-percent level in most of the parameter space except for some points close to the borders of the investigated parameter space.
As the error of the GPR emulator prediction is subdominant compared to the precision of the DES Y3 measurement (being $\sim$ 30\% for the angular power spectra and $\sim$ 5\% for the peak counts) we 
conclude that our results are not significantly biased by the inaccuracy of the GPR emulator. \\

\begin{figure*}
\includegraphics[width=0.8\textwidth]{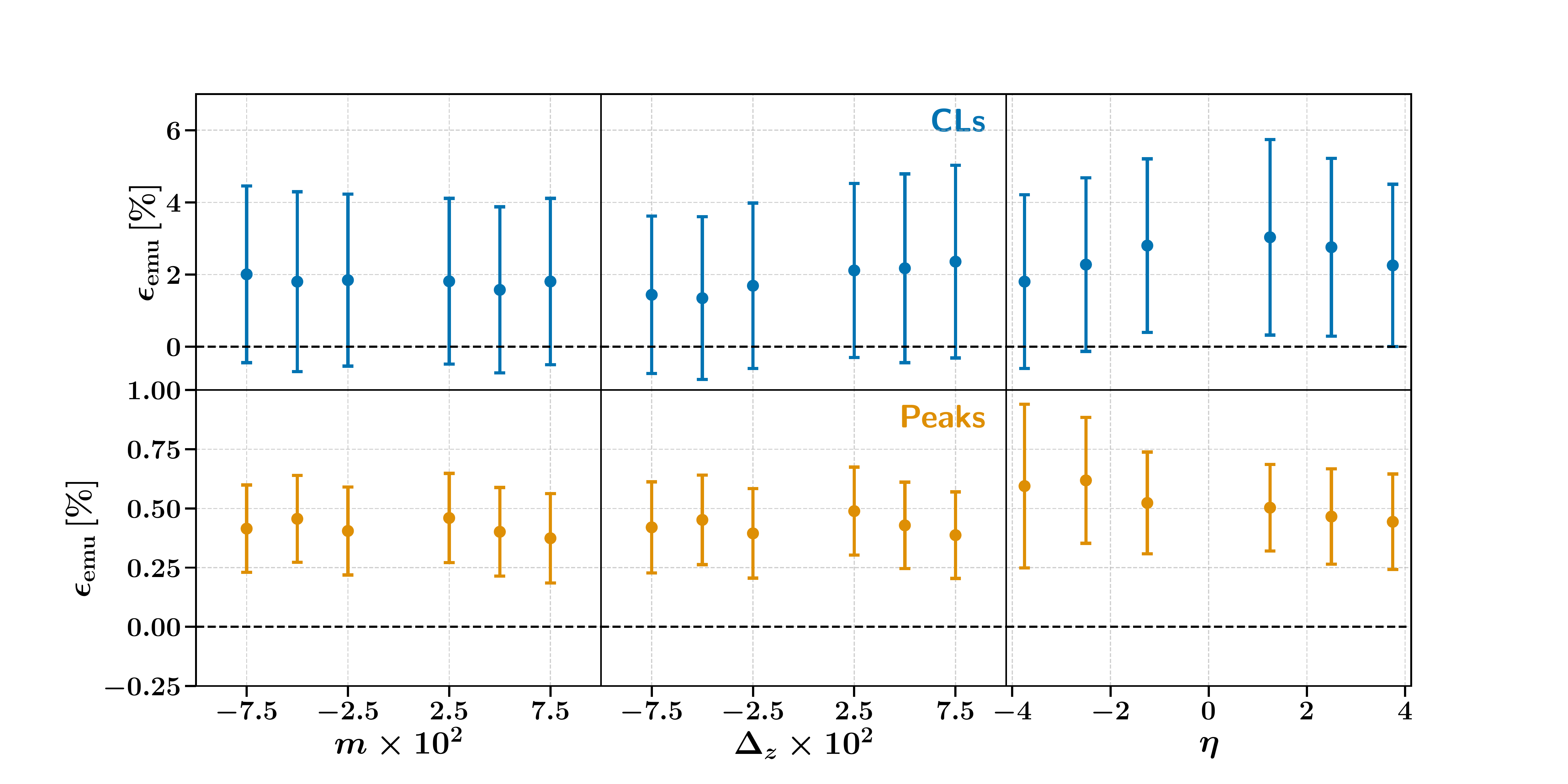}
\caption{The results of the accuracy tests for the polynomial models used for the emulation of the second-order systematic effects, namely: multiplicative shear bias ($m$), 
photometric redshift error ($\Delta_{\mathrm{z}}$), and the redshift dependence of the 
galaxy intrinsic alignment signal ($\eta$). The top and bottom rows display the results for the angular power spectra and the peak counts, respectively.}
\label{fig:fiducial_emu}
\end{figure*}

\noindent We apply the same testing strategy to investigate the accuracy of the polynomial models 
used for the emulation of the second-order systematic effects, namely: multiplicative shear bias ($m$), 
photometric redshift error ($\Delta_z$), and the redshift dependence of the 
galaxy intrinsic alignment signal ($\eta$). 
The results for the angular power spectra and the peak counts are presented in
the top and bottom rows of Figure~\ref{fig:fiducial_emu}, respectively.
Again, we conclude that the errors originating from the emulation are subdominant compared to the precision of the DES Y3 measurement. 
Hence, we conclude that the complexity of the polynomial models is adequate for our analysis.

\section{Additive shear bias}
\label{sec:add_shear_bias_test}

Mismodelling of the PSF may lead to additive biases in the inferred shapes of source galaxies. 
\citet*{gatti2020dark} model the additive bias $\delta \vec{e}_{\mathrm{PSF}}$ caused by PSF mismodelling in the DES Y3 shape catalogue as
\begin{equation}
\delta \vec{e}_{\mathrm{PSF}} = \alpha\vec{e}_{\mathrm{model}} + \beta (\vec{e}_* - \vec{e}_{\mathrm{model}}) + \eta \left( \vec{e}_* \frac{T_* - T_{\mathrm{model}}}{T_*} \right),
\label{eq:add_shear_bias}
\end{equation}
following \cite{paulin2008point} and \cite{jarvis2016science}.
The modelled ellipticity and size of the PSF are denoted as $\vec{e}_{\mathrm{model}}$ and $T_{\mathrm{model}}$, respectively. Correspondingly, $\vec{e}_*$ and $T_*$ denote the PSF quantities measured from stars directly. The variables $\alpha$, $\beta$, and $\eta$ are model parameters. Their values were inferred in \citet*{gatti2020dark} using a sample of reserved stars.
We list the tomographic values of $\alpha$, $\beta$, and $\eta$ in Table~\ref{tab:add_shear_params}. \\

\noindent We estimate the additive shear bias $\delta \vec{e}_{\mathrm{PSF}}$ of the DES Y3 shape catalogue according to Equation~\ref{eq:add_shear_bias} individually for each tomographic bin using the values of the model parameters reported in Table~\ref{tab:add_shear_params}. 
We check that the cosmological constraints do not shift significantly when the additive shear bias signal is added to the galaxy shapes according to
\begin{equation}
    \vec{e}_{\mathrm{mod}} = \vec{e} + \delta \vec{e}_{\mathrm{PSF}}.
\end{equation}
We use all multipoles $\ell \in [8, 2048]$ for the angular power spectra and all scales $\mathrm{FWHM} \in [2.6, 31.6]$ 
arcmin for the peak counts in this test.
We record a shift of the contours in the $\Omega_\mathrm{m} - S_8$ plane by less than 0.01$\sigma$ for both the angular power spectra and the peak counts.
We deem the estimated biases to be insignificant for our analysis and neglect additive shear bias in the inference process.

\begin{table}
\centering
\caption{Tomographic values of the additive PSF bias model parameters $\alpha$, $\beta$, and $\eta$ as reported by \citet*{gatti2020dark} for the DES Y3 shape catalogue.}
\begin{adjustbox}{width=0.35\textwidth}
\begin{tabular}{|c|r|r|r|}
 \hline
{Tomographic bin} & {$\alpha$} & {$\beta$} & {$\eta$} \\
 \hline
 1 & 0.009 & 0.57 & -4.52 \\
 2 & -0.0013 & 1.43 & -4.45 \\
 3 & -0.0029 & 2.4 & 3.03\\
 4 & 0.013 & 1.26 & 4.19\\
 \hline
\label{tab:add_shear_params}
\end{tabular}
\end{adjustbox}
\end{table}
\section{Boost factor corrections} 
\label{sec:boost_factor_corr}

\begin{figure*}
\begin{center}
\includegraphics[width=0.9\textwidth]{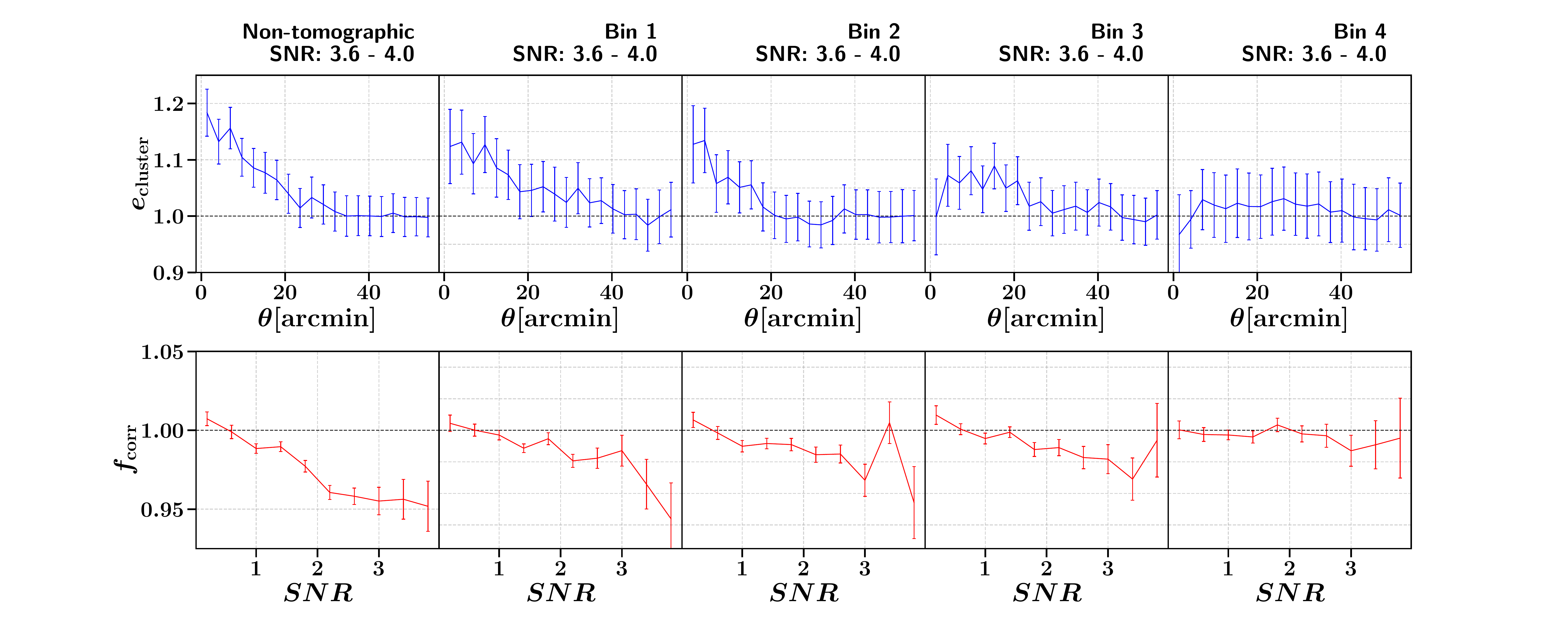}
\end{center}
\caption{\textbf{Top row:} The fractional excess of cluster galaxies around lensing peaks in the DES Y3 data relative to the number of galaxies in the field (according to Equation~\ref{eq:excess}). The signal is measured around lensing peaks with an SNR of 3.6 - 4.0
as a function of radial distance $\theta$ from the centre of the peaks. 
For brevity we only present the results for peaks detected using a filter scale of 21.1 arcmin. 
The clustering signal is strongest in the non-tomographic sample, while being 
less pronounced in the tomographic case.
\textbf{Bottom row:} The inferred boost factor corrections for the different tomographic bins as well as the non-tomographic sample as they apply to peaks with a scale of 21.1 arcmin. While we infer that a correction of up to $\sim 5 \%$ is necessary for the non-tomographic sample, the corrections become smaller in the tomographic case.}
\label{fig:boost_factor}
\end{figure*}

We estimate the boost factor corrections following the methodology outlined in Appendix C of \citetalias{kacprzak2016cosmology} but neglecting the contribution from blending.
We start by estimating the fractional excess of cluster galaxies $e_{\mathrm{cluster}}(\theta)$ around lensing peaks in the DES Y3 data relative to the number of galaxies in the field as
\begin{equation}
e_{\mathrm{cluster}}(\theta) = n^{\mathrm{DES}}(\theta) / n^{\mathrm{SIM}}(\theta).
\label{eq:excess}
\end{equation}
The fractional excess $e_{\mathrm{cluster}}(\theta)$ is a function of angular distance $\theta$
from the cluster centre. The number count of galaxies in the simulations $n^{\mathrm{SIM}}$ serves as a measure for the average number of galaxies around lensing peaks in the field, as the galaxy positions in our simulations are uncorrelated with the distribution of the dark matter.
We measure $e_{\mathrm{cluster}}(\theta)$ as a function of scale and SNR, since the excess may vary depending on the SNR as well as the angular size of the lensing peaks.
The galaxy excess functions $e_{\mathrm{cluster}}(\theta)$ inferred for an exemplary scale of 
21.1 arcmin and the highest recorded SNR bin are shown in the top row of Figure~\ref{fig:boost_factor}.
While we observe a clear clustering signal in the non-tomographic sample, the signal is less pronounced in the tomographic sample and gets weaker with increasing redshift.
As measured by \citetalias{kacprzak2016cosmology}, we observe that the clustering signal becomes stronger with increasing SNR of the lensing peaks. \\

\noindent Following the methodology outlined in greater detail in \citetalias{kacprzak2016cosmology}, we estimate the boost factor corrections 
$f_{\mathrm{corr}} = f_1 /f_2$ neglecting blending ($f_1 = 1$), where
\begin{align}
    f_2 = \frac{\sum_{\theta} n^{\mathrm{SIM}}(\theta) \thinspace \mathcal{N}(\theta ; (0, \mathrm{FWHM}))^2 \thinspace f_{\mathrm{cluster}}(\theta)}{\sum_{\theta} n^{\mathrm{SIM}}(\theta) \thinspace \mathcal{N}(\theta ; (0, \mathrm{FWHM}))^2}.
\end{align}
The quantity $f_{\mathrm{cluster}}(\theta)$ corresponds to the relative difference in clustering between the simulations and the data
\begin{equation}
    f_{\mathrm{cluster}}(\theta) = e_{\mathrm{cluster}}(\theta) - 1.
\end{equation}

\noindent The estimated boost factor corrections $f_{\mathrm{corr}}$ inferred for the exemplary filter scale of 21.1 arcmin are shown in the bottom
row of Figure~\ref{fig:boost_factor}. In the non-tomographic case we find a boost factor correction of up to $\sim 5\%$ for the highest SNR peaks, similar to the findings of \citetalias{kacprzak2016cosmology}. In the tomographic case the correction is smaller for each tomographic bin and further decreases with increasing redshift. \\

\begin{figure}
\includegraphics[width=0.47\textwidth]{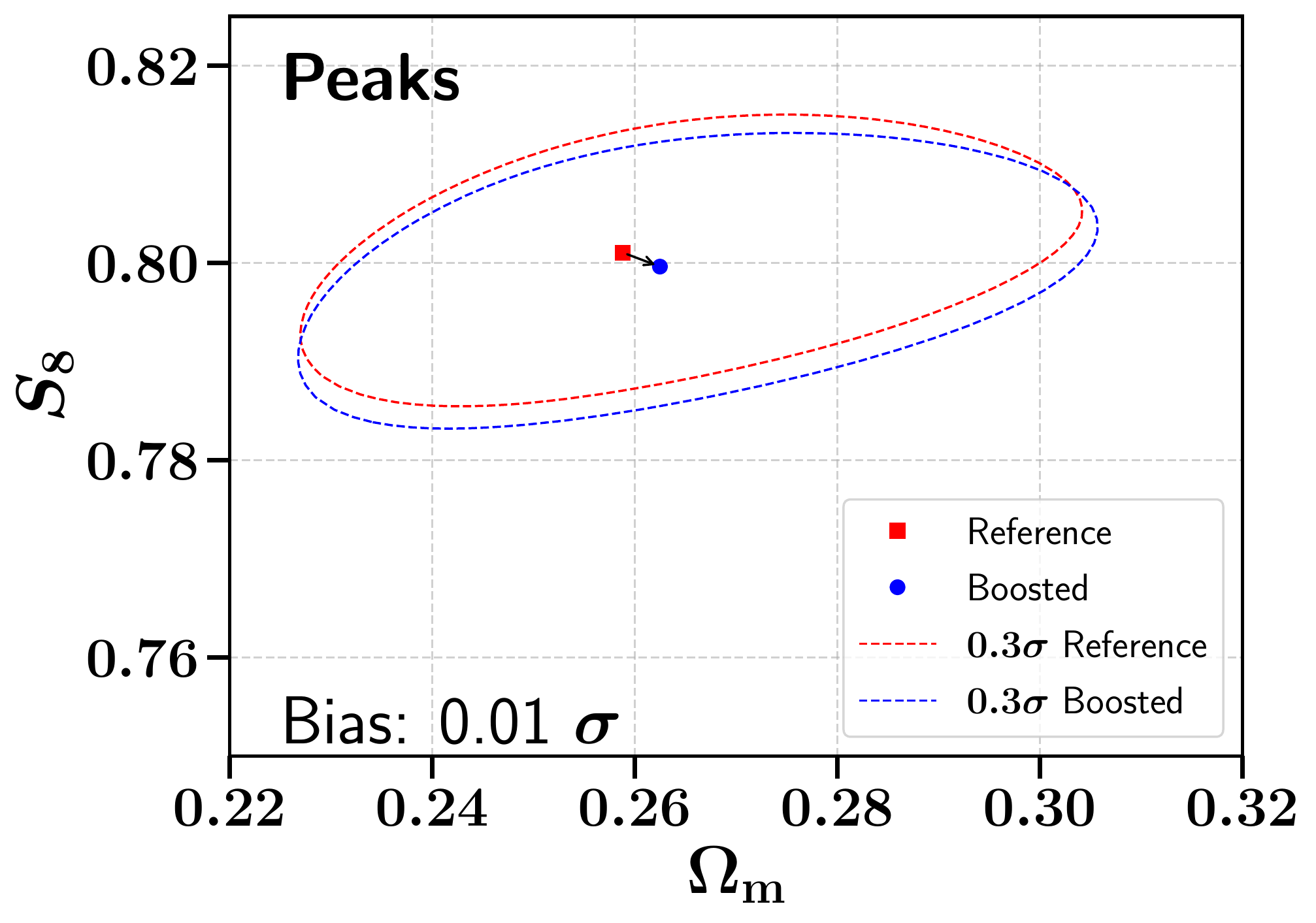}
\caption{Comparison of the cosmological constraints in the $\Omega_{\mathrm{m}} - S_8$ plane between a reference peak count measurement and the corresponding boost factor corrected version. The depicted contour levels indicate the two-dimensional 0.3 $\sigma$ distance from the best-fit parameter values. 
We record a shift of $\sim 0.01 \sigma$ when applying the boost factor corrections.}
\label{fig:clustering_bias}
\end{figure}

\noindent To assess the impact of source clustering on the inferred cosmological constraints, we compare the constraints obtained from a synthetic peak count data vector with and without applying the boost factor correction. The boost factor corrections are applied to the peak functions by adding $\Delta N$ peaks to each bin.
The correction $\Delta N$ is obtained analogously to \citetalias{kacprzak2016cosmology}, but as a function of $\kappa$ instead
of SNR
\begin{equation}
\Delta N(\kappa) = (1 - f_{\mathrm{corr}}) \frac{\mathrm{d} N}{\mathrm{d} \kappa} N(\kappa),    
\end{equation}
where we approximate $\frac{\mathrm{d} N}{\mathrm{d} \kappa} \approx \frac{\mathrm{d} N}{\mathrm{d} m}$, with $m$ being the multiplicative shear bias parameter. 
In performing this test, we consider all scales $\mathrm{FWHM} \in [2.6, 31.6]$ arcmin.
The comparison between the reference and the boost factor corrected constraints in the $\Omega_{\mathrm{m}} - S_8$ plane is shown in Figure~\ref{fig:clustering_bias}.
We find a shift of the contours by $\sim 0.01 \thinspace \sigma$. 
We therefore conclude that our analysis is not significantly biased by source clustering and we do not apply the boost factor correction in the fiducial analysis. 

\section{Accuracy of simulated mass maps} 
\label{sec:acc_sim_conv}
A unit box with a side-length of 900 Mpc/h and $768^3$ particles was used in the \pkdgrav\ N-Body simulations. In order to cover a large enough cosmological volume
the unit box has to be replicated up to 14 times per side for some cosmologies.  
It is expected that such a replication leads to the suppression of very large super-box modes. 
Further, the finite number of particles in the simulations leads to a shot-noise contribution. \\
\noindent The assembly of the particle shells into a projected convergence map is performed using \texttt{UFalcon} \citep{sgier2019fast}. \texttt{UFalcon} avoids a full ray-tracing treatment by relying on the Born approximation, which can further contribute to a deterioration of the accuracy of the simulated convergence maps. \\

\begin{figure*}
\begin{center}
\includegraphics[width=0.9\textwidth]{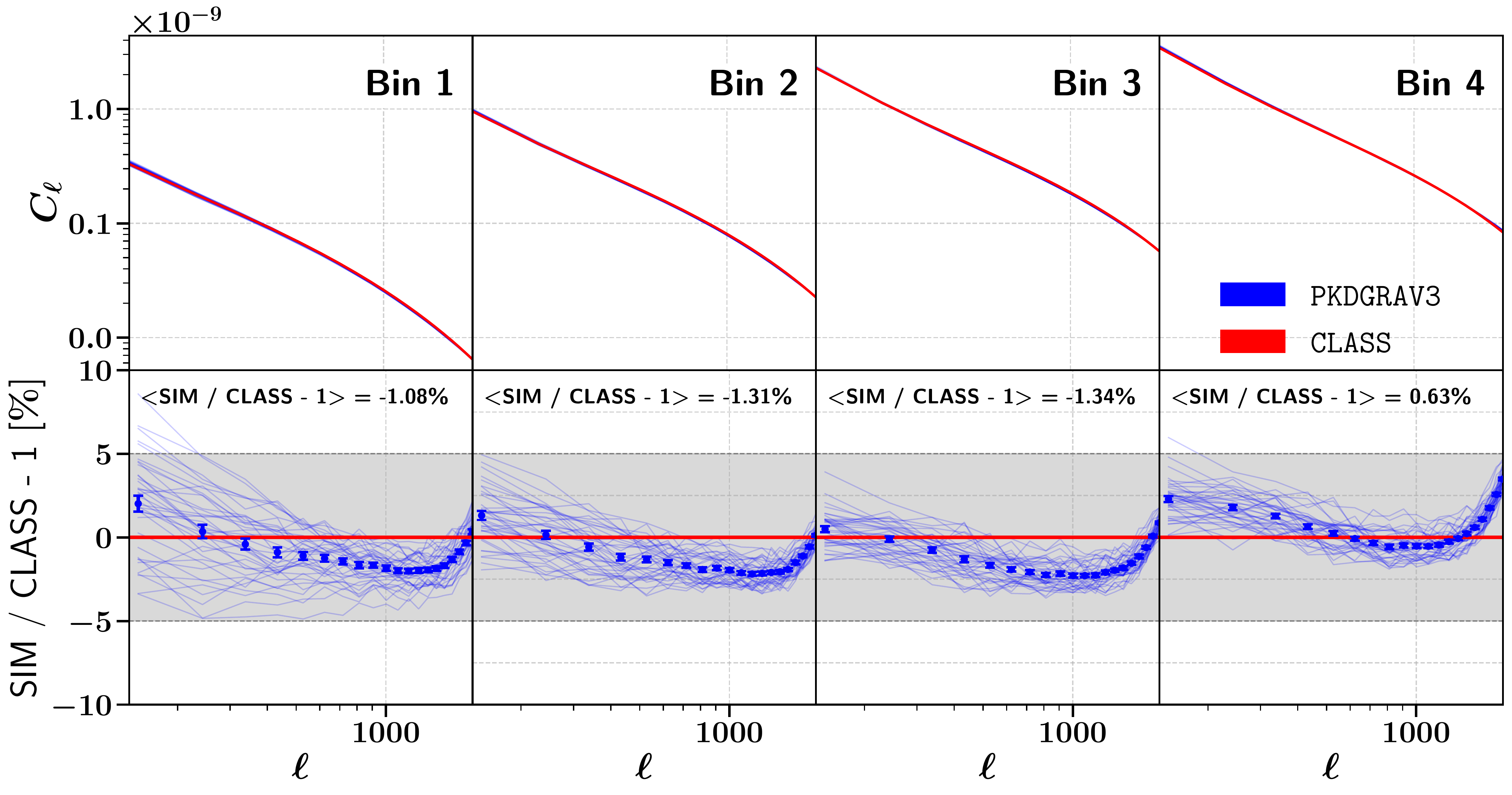}
\end{center}
\caption{Absolute and relative comparison of the power spectra measured from the simulated full-sky mass maps at the fiducial cosmology (blue curves) and the predicted power spectra calculated using \textsc{CLASS} (red curves). Each panel shows the comparison for a different tomographic bin.}
\label{fig:full_sky_cl}
\end{figure*}

\noindent We test the accuracy of the simulations by comparing the full-sky angular power spectra of the 50 simulations 
at the fiducial cosmology to the prediction of an independent theory code. The comparison is displayed in Figure~\ref{fig:full_sky_cl}. 
The theory prediction is obtained using the state-of-the-art theory code \textsc{CLASS} \citep{lesgourgues2011cosmic}. The \texttt{Halofit} code was used to predict the non-linear part of the power spectrum at small scales \citep{takahashi2012revising}.
We find that our simulations differ from the theory prediction by no more than $\approx 1.5\%$ for $\ell \in [8, 2048]$ on average.
The theory prediction for the matter power spectrum 
was found to vary by a few percent between different modelling implementations for the small scale signal \citep{martinelli2020euclid}.
As the difference in the angular power spectrum is expected to be of the same order of magnitude we conclude that the disagreement between our simulations and the theory prediction is comparable to the disagreement between different theory predictions themselves. 
Hence, we deem the accuracy of the full-sky mass maps acceptable for this analysis. \\

\begin{figure*}
\begin{center}
\includegraphics[width=0.9\textwidth]{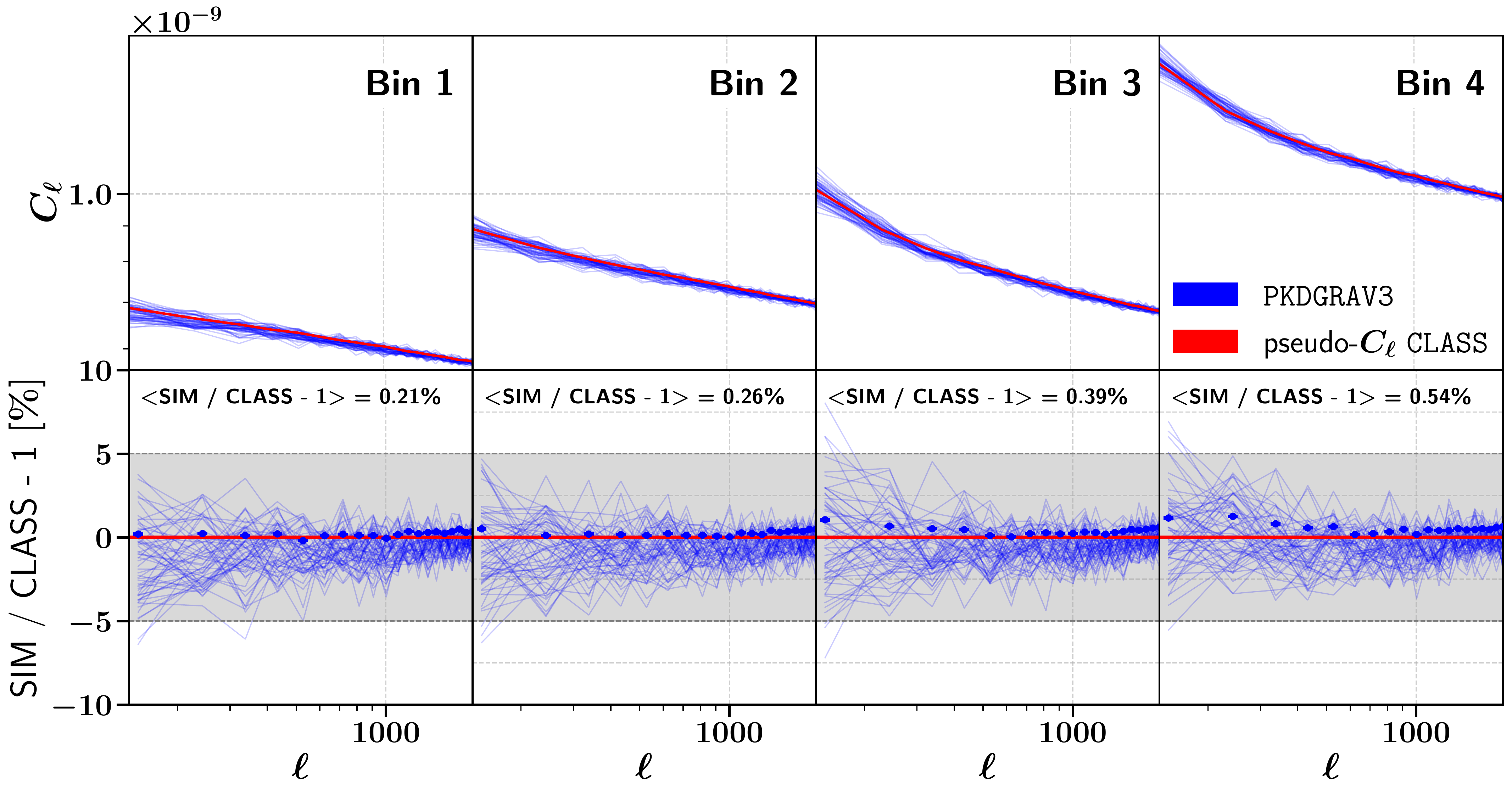}
\end{center}
\caption{Absolute and relative comparison of the power spectra measured from the masked, noisy mass maps at the fiducial cosmology (blue curves) and the predicted power spectra obtained using \textsc{CLASS} (red curves). The effects of the mask were propagated to the theory prediction using the pseudo-$C_{\ell}$ method \citep{hikage2011shear} implemented in \citet{sgier2020fast} and the average shape noise signal measured from simulations was added. Each panel shows the comparison for a different tomographic bin.}
\label{fig:masked_sky_cl}
\end{figure*}

\noindent We further test the accuracy of the fully forward-modelled mass maps.
We do so by comparing the angular power spectra measured from the 10,000 simulated mass maps at the fiducial cosmology to a theory prediction calculated using \textsc{CLASS}.
The effects arising from masking and shape noise are propagated to the theory prediction following the pseudo-$C_\ell$ method 
described in \citet{sgier2020fast}, based on \citet{brown2008cmb} and \citet{kogut2003first}. The comparison is shown in Figure~\ref{fig:masked_sky_cl}.
We find that the simulations agree with the theory prediction within $\approx 0.5\%$ for all considered scales. \\
We again conclude that the accuracy of the simulated mass maps is sufficient for this analysis, following the same argument as for the full-sky power spectra.

\section{Baryons} 
\label{sec:baryon_test}

The \pkdgrav\ simulations used to predict the angular 
power spectra and peak counts at different cosmologies are dark-matter-only simulations that do not include effects arising from the presence of baryonic particles. 
As described in Section~\ref{sec:baryons}, significant biases caused by the missing baryonic physics in the simulations may arise. 
\noindent We investigate the shift of the constraints in the $\Omega_{\mathrm{m}} - S_8$ plane caused by replacing a synthetic data vector with another synthetic, but baryon-contaminated, data vector at the same cosmology. Our methodology is similar to the approach used in \cite{des2021cosmic} where synthetic, baryon-contaminated shear data vectors were used to infer the necessary scale cuts.

\noindent The baryon-contaminated data vector was obtained from a simulation including 
baryonic corrections at the map level \citep{schneider2015new, schneider2019quantifying}. The baryon correction model 
was used to emulate these effects. This model allows us to 
mimic baryonic feedback effects by displacing particles around massive halos in an N-Body simulation. The altered halo profiles resemble realistic profiles
including effects from star formation and active galactic nucleus feedback.
The locations of the haloes in the N-body simulation were identified using the \texttt{AMIGA} halo finder algorithm \citep{knollmann2009ahf, klement2010halo}.
The baryonic correction model achieves good agreement with full hydrodynamical simulations as shown by \cite{schneider2019quantifying}.
One of the major parameters driving the strength of the baryonic corrections in the model is the gas fraction of the halos. \citet{schneider2019quantifying}
used X-Ray observations to constrain the gas fraction. However, the inferred gas fractions are prone to the uncertain hydrostatic mass bias. 
\citet{schneider2019quantifying} propose three different models with three different hydrostatic mass biases. 
We used the best-guess model in this analysis (called model B in \citet{schneider2019quantifying}). \\

\begin{figure*}
\begin{center}
\includegraphics[width=0.4\textwidth]{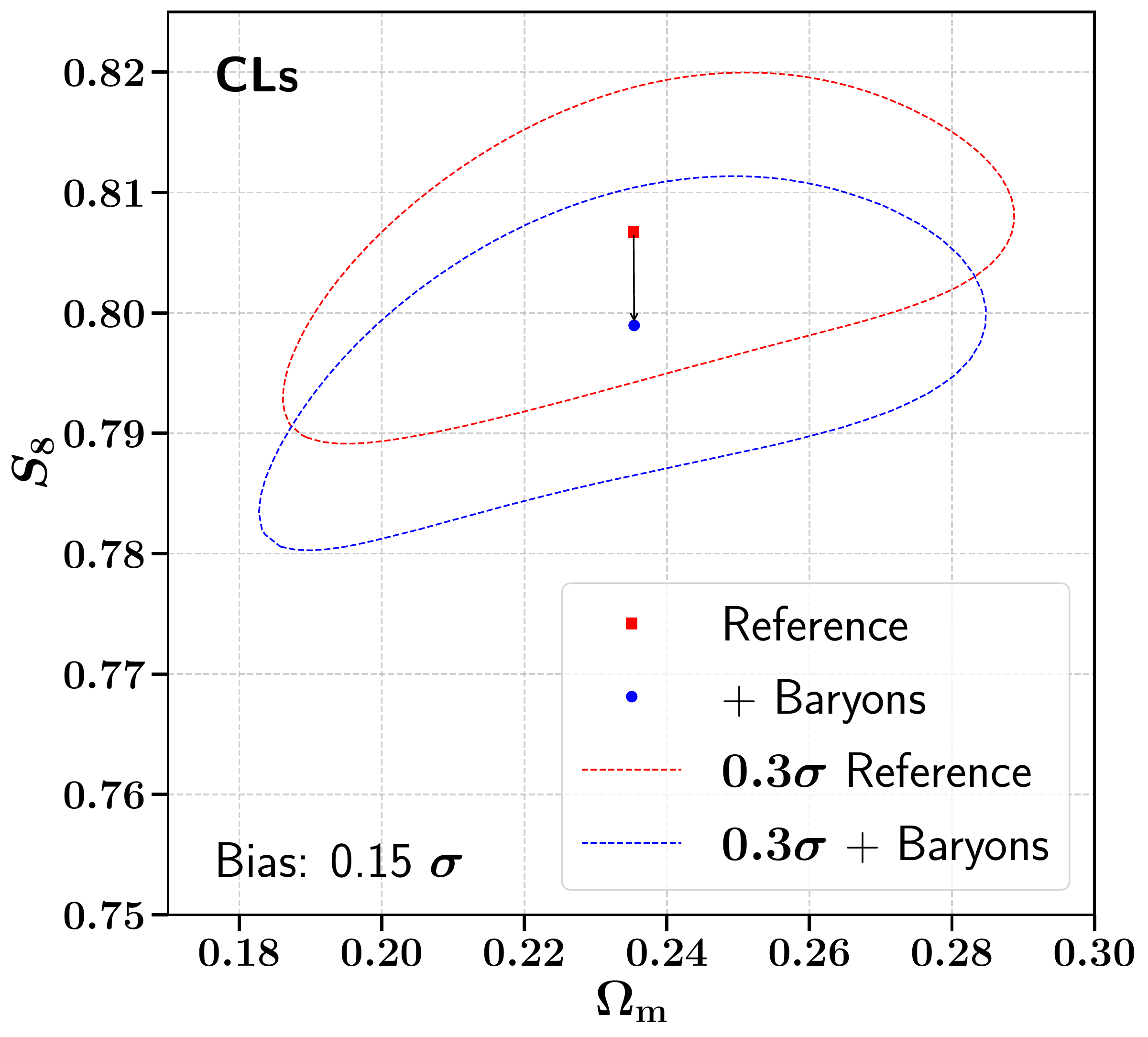}
\includegraphics[width=0.4\textwidth]{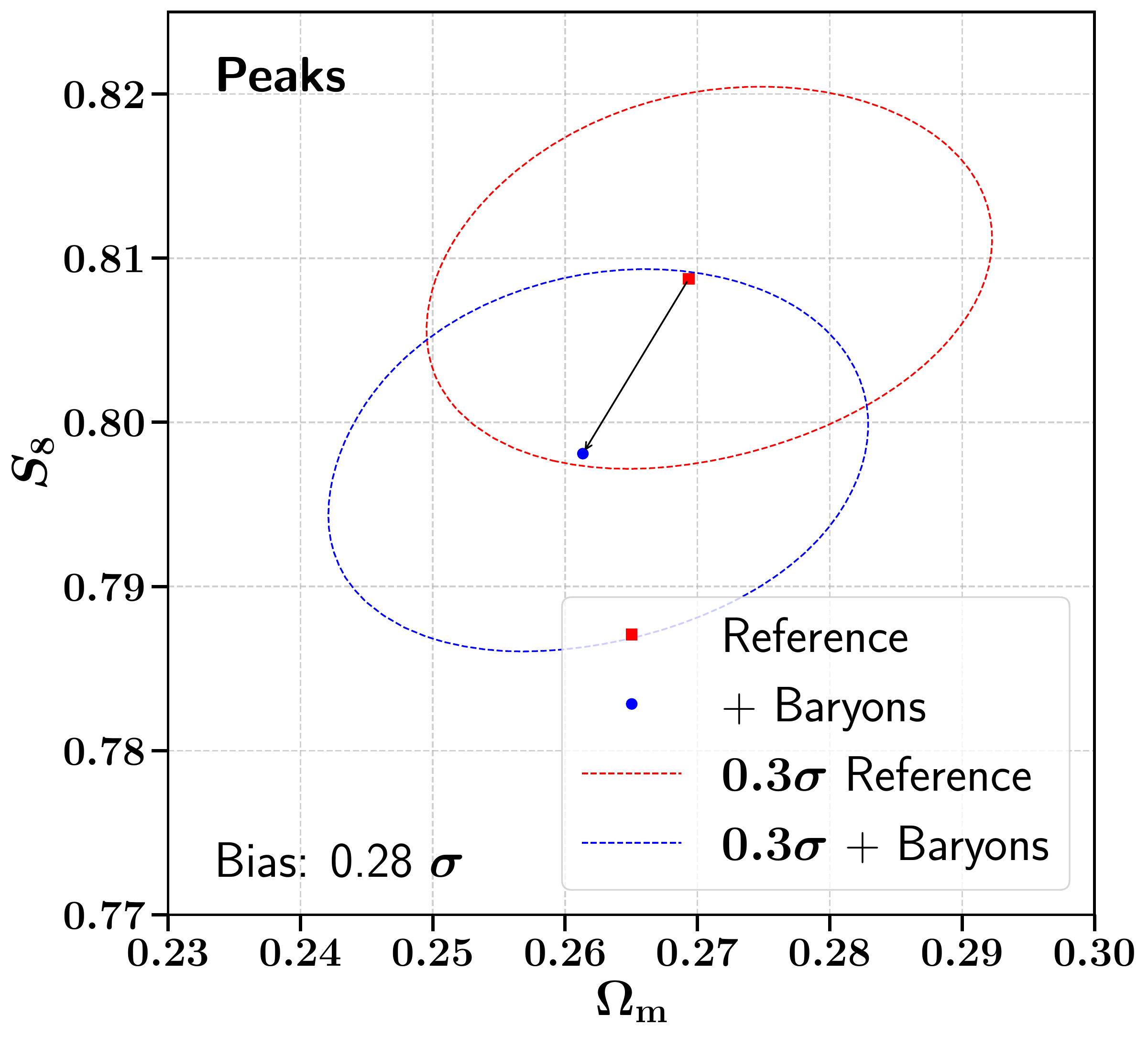}
\end{center}
\caption{Comparison between the obtained $\Omega_{\mathrm{m}} - S_8$ constraints when using a baryon-contaminated synthetic data vector and 
a corresponding uncontaminated reference data vector. The left and right images display the comparison of the constraints for the angular power spectra and the peak counts, respectively. The shift of the constraints is noted in the bottom left corner of the images. The scale cuts used in the production of these comparisons are $\ell \in [30, 578]$ for the angular power spectra and $\mathrm{FWHM} \in [7.9, 31.6]$ arcmin for the peak counts.}
\label{fig:baryon_tests}
\end{figure*}

\noindent The estimated shifts of the contours in the $\Omega_{\mathrm{m}} - S_8$ plane caused by the inclusion of baryonic effects for the angular power spectra and peak counts are illustrated in the left and right panels of Figure~\ref{fig:baryon_tests}, respectively.
The scale cuts used in the production of the constraints presented in Figure~\ref{fig:baryon_tests} were chosen such that
the resulting shifts of the contours for both statistics is $\leq 0.3\sigma$, which we deem acceptable for this analysis. The imposed threshold of $0.3 \sigma$ is in accordance with the unblinding criteria defined in Appendix D of \citet{des20213x2pt}
This corresponds to $\ell \in [30, 578]$ for the angular power spectra and $\mathrm{FWHM} \in [7.9, 31.6]$ arcmin for the peak counts. Our findings for the peak counts are in agreement with \citet{weiss2019effects}, who observed a non-negligible effect of baryonic physics on peak counts on scales below $\sim 8$ arcmin.
For the combination of angular power spectra and peak counts we record a shift of $0.29 \sigma$, which fulfills the requirement of $\leq 0.3 \sigma$.

\section{B-modes}
\label{sec:b-modes}
\begin{figure*}
\begin{center}
\includegraphics[width=0.9\textwidth]{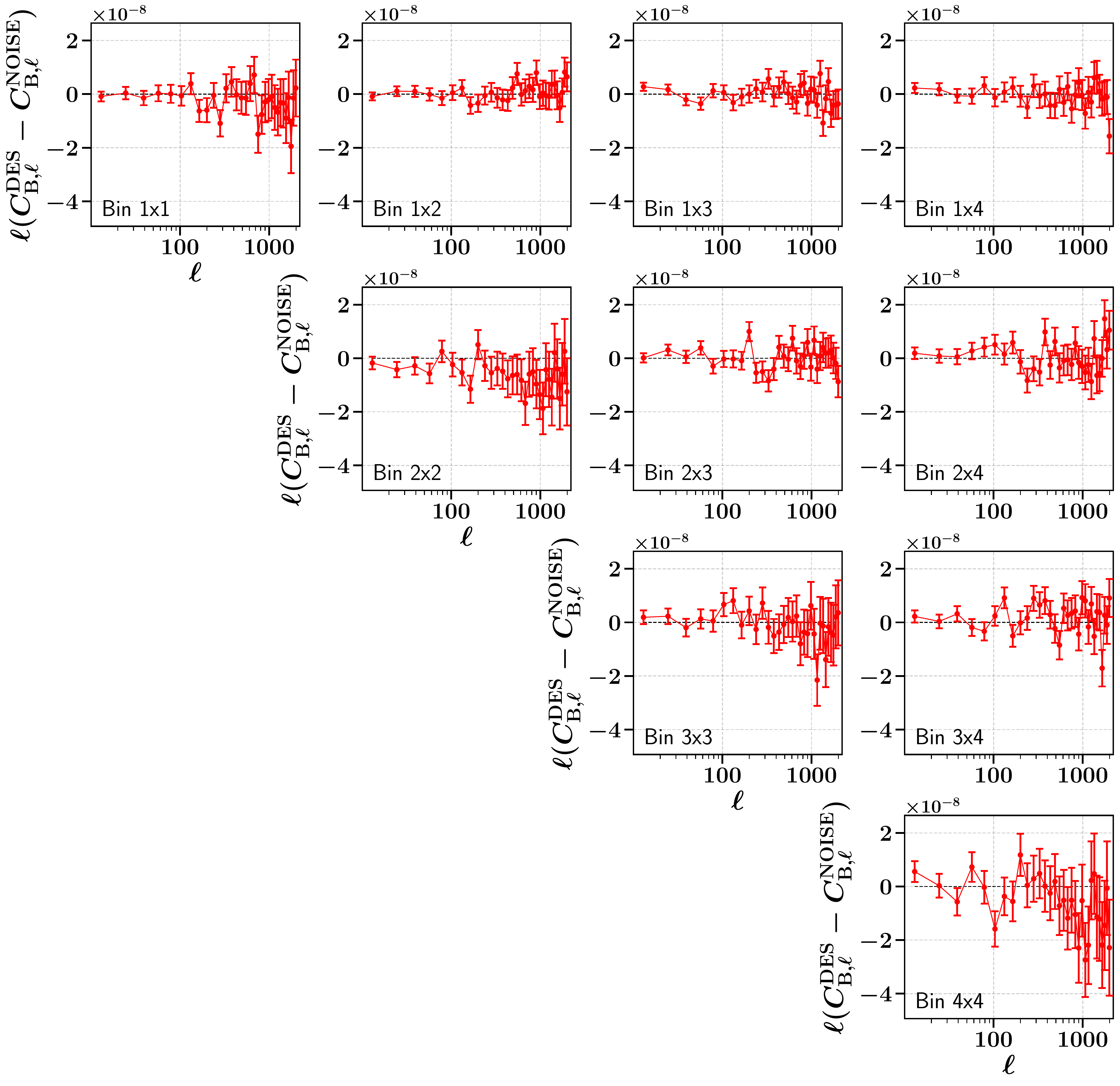}
\end{center}
\caption{The noise-subtracted angular power spectra DES Y3 B-modes.
The noise signal was inferred using simulations with only a shape noise component but no cosmological signal.
The shaded regions of the plots indicate where the scale cuts were imposed.
Considering all angular power spectra with $\ell \in [30, 578]$ we find a global $\chi^2 / \mathrm{dof} = 1.26$ and $p = 23\%$.}
\label{fig:b_modes_cls}
\end{figure*}

\begin{figure*}
\begin{center}
\includegraphics[width=1\textwidth]{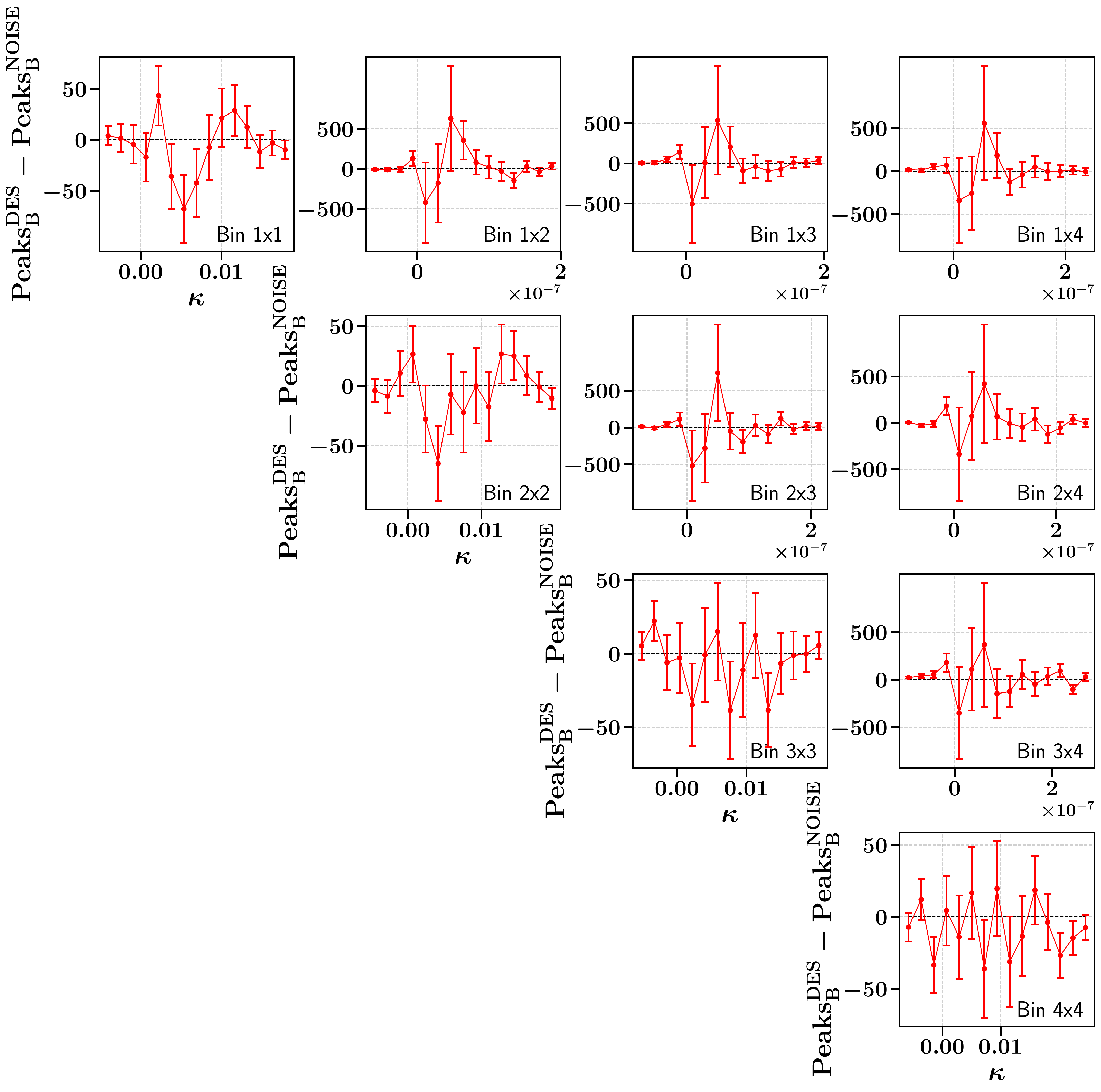}
\end{center}
\caption{The noise-subtracted DES Y3 peak count B-modes for a selected scale of $\mathrm{FWHM} = 21.1$ arcmin.
The noise signal was inferred using simulations with only a shape noise component but no cosmological signal.
Using the scales $\mathrm{FWHM} \in [7.9, 31.6]$ arcmin we find $\chi^2 / \mathrm{dof} = 1.14$ and $p = 31\%$.}
\label{fig:b_modes_peaks}
\end{figure*}

We present the noise-subtracted B-mode signals of the angular power spectra and the peak counts in Figure~\ref{fig:b_modes_cls} and Figure~\ref{fig:b_modes_peaks}, respectively.
We impose a requirement of $p > 1\%$ for the null test to pass. This is in accordance with the unblinding criteria defined in Appendix D of \citet{des20213x2pt}.
For the angular power spectra we find $\chi^2 / \mathrm{dof}  = 1.26$ and $p = 23\%$ while for the peak counts we find  $\chi^2 / \mathrm{dof}  = 1.14$ and $p = 31\%$.
Hence, we declare the B-mode null test as passed. \\

\section{Cosmology dependence of approximate marginalisation scheme} 
\label{sec:cos_dep_marg}

\begin{figure}
\includegraphics[width=0.47\textwidth]{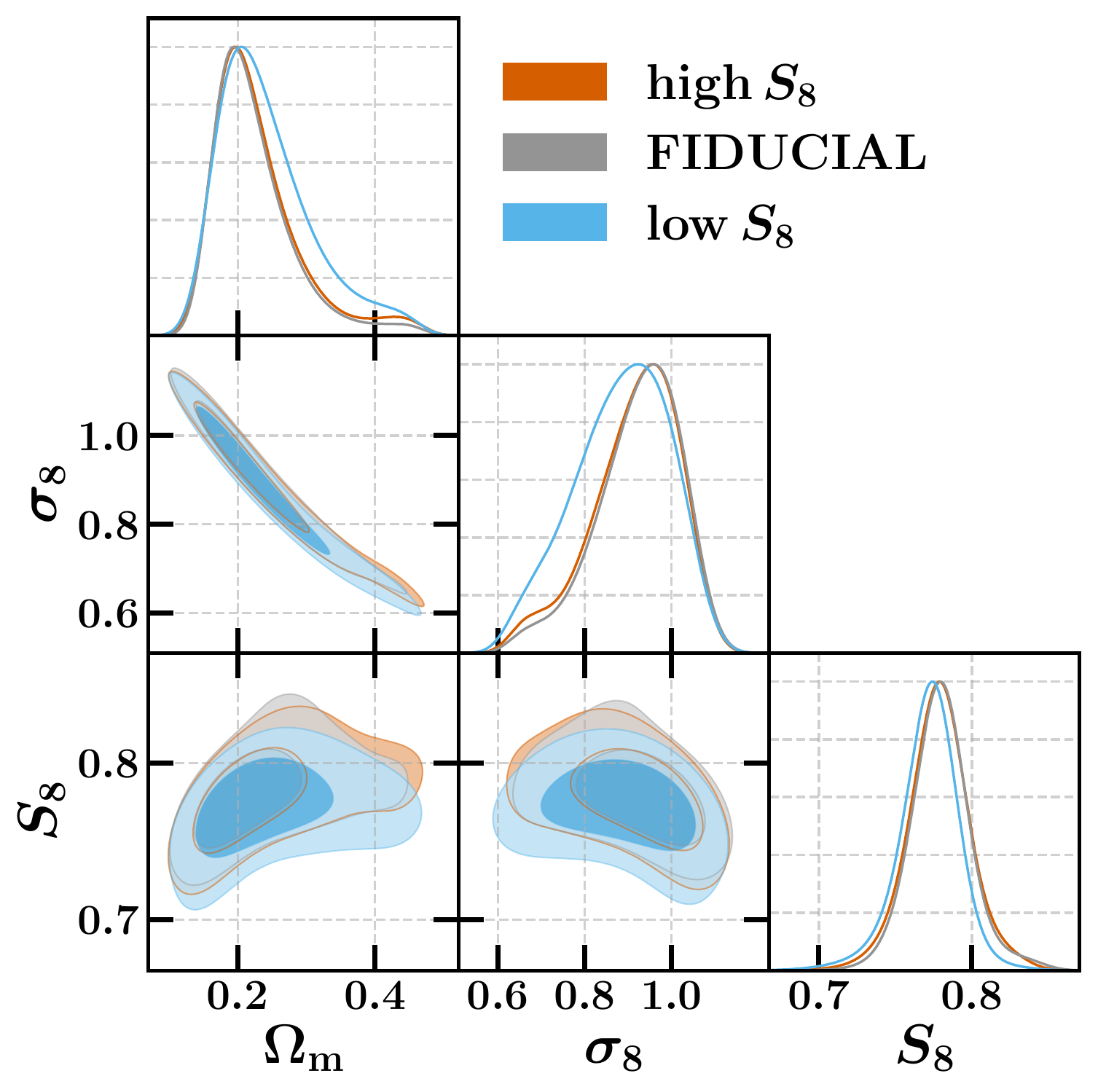}
\caption{Comparison of the cosmological constraints obtained using peak counts. To illustrate the cosmology dependence of the marginalisation the polynomials in Equation~\ref{eq:marg_model} were fit using simulations at three different central cosmologies (low $S_8 \sim 0.769$, fiducial $S_8 \sim 0.782$ and high $S_8 \sim 0.815$). The contour levels in both plots indicate the 68\% and 95\% confidence regions of the constraints.}
\label{fig:cos_marg}
\end{figure}

We confirm that the cosmology dependence of the used approximate marginalisation scheme does not bias the results found in this study. To do so, we fit the polynomials $f_j(\theta)$ given in Equation~\ref{eq:marg_model} at three different central cosmologies ($(\Omega_{\mathrm{m}}, \sigma_8) \in \{(0.26, 0.826), (0.26, 0.84), (0.26, 0.875) \}$). The index $j$ runs over the different elements of the data vector and $\theta \in \{\Omega_{\mathrm{b}}, n_{\mathrm{s}}, h\}$.
We then compare the cosmological constraints inferred using the three different fits (see Figure~\ref{fig:cos_marg} for the results for peaks for example).
We record a maximal shift of $0.1\sigma$ of the constraints in the $\Omega_{\mathrm{m}} - S_8$ plane, which we deem acceptable for this analysis in accordance with the criteria defined in Appendix D of \citet{des20213x2pt}

\section{Individual scale tests} 
\label{sec:separate_scales}
\begin{figure*}
\begin{center}
\includegraphics[height=0.45\textwidth]{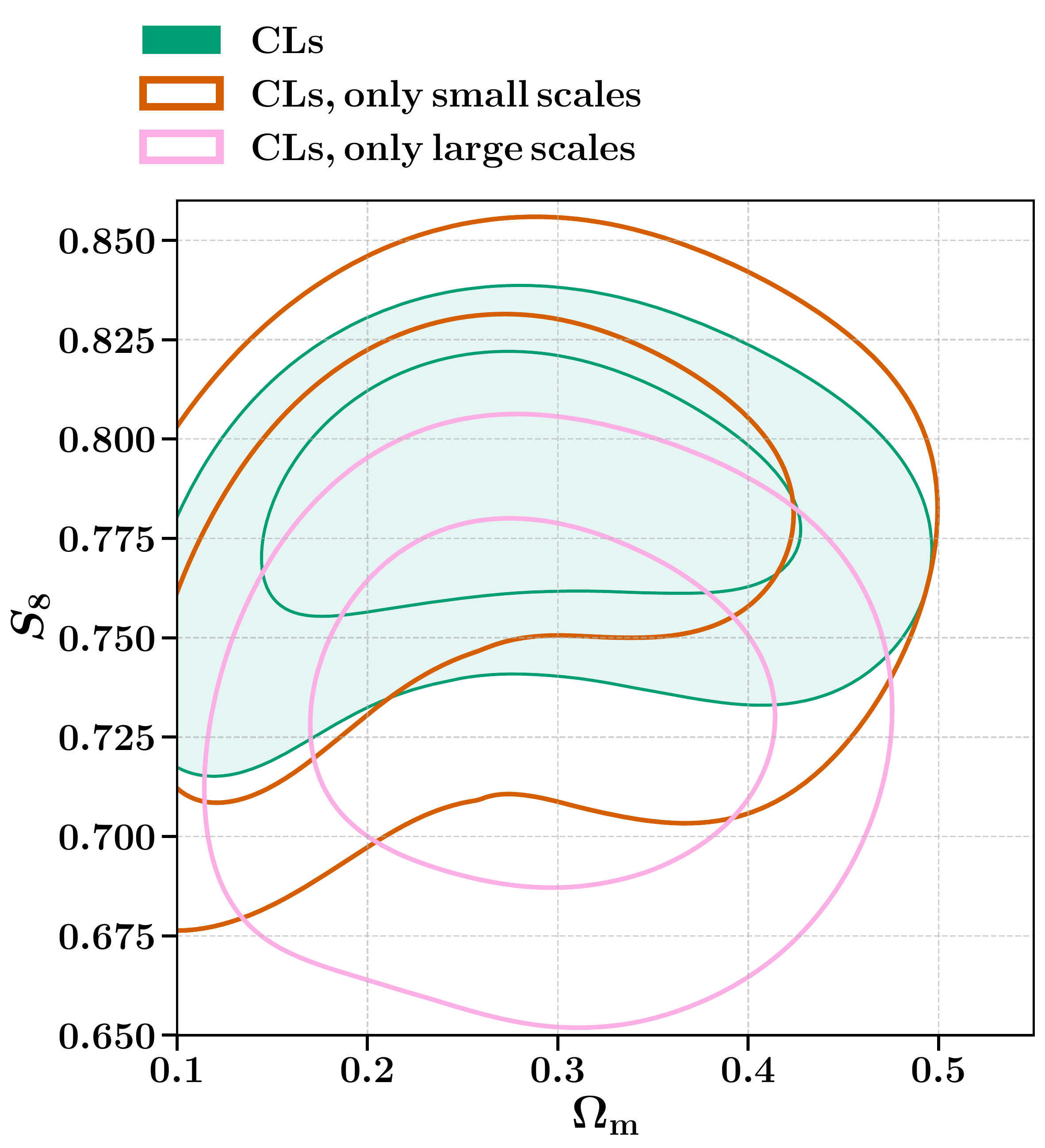}
\includegraphics[height=0.45\textwidth]{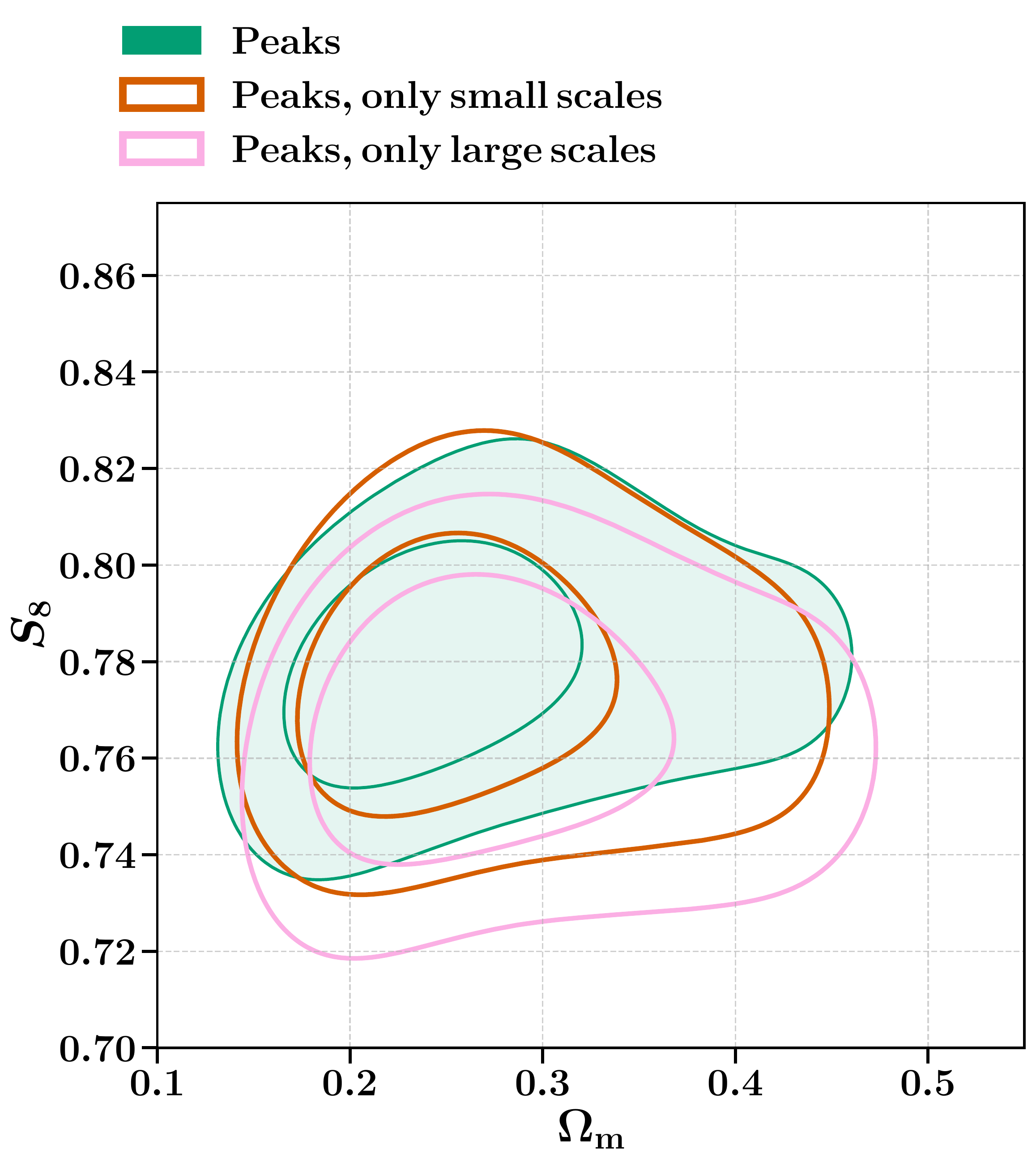} 
\end{center}
\caption{\textbf{Left:} Scale robustness test for the angular power spectra. The shaded contour shows the constraints obtained using all scales. Either all the small ($\ell \in [258, 578]$) or large ($\ell \in [30, 257]$) scales were removed when obtaining the other two constraints. \textbf{Right:} Analogous scale robustness test for the peak counts. The large and small scale samples consist of the scales $\mathrm{FWHM} \in [21.1, 31.6]$ arcmin and $\mathrm{FWHM} \in [7.9, 18.5]$ arcmin, respectively. The contour levels in both plots indicate the 68\% and 95\% confidence regions of the constraints. }
\label{fig:separate_scales}
\end{figure*}

We present how the results for the angular power spectra and peak counts change depending on the scale selection. To do so we split the data vectors into two parts: Large scales ($\ell \in [30, 257]$ and $\mathrm{FWHM} \in [21.1, 31.6]$ arcmin) and small scales ($\ell \in [258, 578]$ and $\mathrm{FWHM} \in [7.9, 18.5]$ arcmin). The resulting change of the $\Omega_{\mathrm{m}} - S_8$ constraints is visualised in Figure~\ref{fig:separate_scales}. Both statistics experience similar shifts.

\bsp	
\label{lastpage}

\bibliography{references}
\bibliographystyle{mn2e_2author_arxiv_amp}

\section*{Affiliations}
$^{1}$ Institute for Particle Physics and Astrophysics, Department of Physics, ETH Z\"uurich, Wolfgang Pauli Strasse 27, 8093 Z\"urich, Switzerland\\
$^{2}$ Department of Physics and Astronomy, University of Pennsylvania, Philadelphia, PA 19104, USA\\
$^{3}$ Department of Physics and Astronomy, University College London, Gower Street, London WC1E 6BT, UK\\
$^{4}$ Department of Astronomy and Astrophysics, University of Chicago, Chicago, IL 60637, USA\\
$^{5}$ Kavli Institute for Cosmological Physics, University of Chicago, Chicago, IL 60637, USA\\
$^{6}$ Department of Physics and Astronomy, Pevensey Building, University of Sussex, Brighton, BN1 9QH, UK\\
$^{7}$ Laboratoire de Physique de l'Ecole Normale Sup\'erieure, ENS, Universit\'e PSL, CNRS, Sorbonne Universit\'e, Universit\'e de Paris, Paris, France\\
$^{8}$ Institute of Cosmology and Gravitation, University of Portsmouth, Portsmouth, PO1 3FX, UK\\
$^{9}$ Argonne National Laboratory, 9700 South Cass Avenue, Lemont, IL 60439, USA\\
$^{10}$ Kavli Institute for Particle Astrophysics \& Cosmology, P. O. Box 2450, Stanford University, Stanford, CA 94305, USA\\
$^{11}$ Physics Department, 2320 Chamberlin Hall, University of Wisconsin-Madison, 1150 University Avenue Madison, WI  53706-1390\\
$^{12}$ Department of Physics, Carnegie Mellon University, Pittsburgh, Pennsylvania 15312, USA\\
$^{13}$ Department of Physics, Duke University Durham, NC 27708, USA\\
$^{14}$ Center for Cosmology and Astro-Particle Physics, The Ohio State University, Columbus, OH 43210, USA\\
$^{15}$ Lawrence Berkeley National Laboratory, 1 Cyclotron Road, Berkeley, CA 94720, USA\\
$^{16}$ Department of Physics, Carnegie Mellon University, Pittsburgh, Pennsylvania 15312, USA\\
$^{17}$ NSF AI Planning Institute for Physics of the Future, Carnegie Mellon University, Pittsburgh, PA 15213, USA\\
$^{18}$ Center for Cosmology and Astro-Particle Physics, The Ohio State University, Columbus, OH 43210, USA\\
$^{19}$ Department of Physics, The Ohio State University, Columbus, OH 43210, USA\\
$^{20}$ Santa Cruz Institute for Particle Physics, Santa Cruz, CA 95064, USA\\
$^{21}$ Jet Propulsion Laboratory, California Institute of Technology, 4800 Oak Grove Dr., Pasadena, CA 91109, USA\\
$^{22}$ Faculty of Physics, Ludwig-Maximilians-Universit\"at, Scheinerstr. 1, 81679 Munich, Germany\\
$^{23}$ Department of Physics, University of Oxford, Denys Wilkinson Building, Keble Road, Oxford OX1 3RH, UK\\
$^{24}$ Jodrell Bank Center for Astrophysics, School of Physics and Astronomy, University of Manchester, Oxford Road, Manchester, M13 9PL, UK\\
$^{25}$ Department of Physics, University of Michigan, Ann Arbor, MI 48109, USA\\
$^{26}$ Department of Applied Mathematics and Theoretical Physics, University of Cambridge, Cambridge CB3 0WA, UK\\
$^{27}$ Perimeter Institute for Theoretical Physics, 31 Caroline St. North, Waterloo, ON N2L 2Y5, Canada\\
$^{28}$ Department of Physics, Stanford University, 382 Via Pueblo Mall, Stanford, CA 94305, USA\\
$^{29}$ SLAC National Accelerator Laboratory, Menlo Park, CA 94025, USA\\
$^{30}$ Instituto de F\'isica Gleb Wataghin, Universidade Estadual de Campinas, 13083-859, Campinas, SP, Brazil\\
$^{31}$ Brookhaven National Laboratory, Bldg 510, Upton, NY 11973, USA\\
$^{32}$ Institut d'Estudis Espacials de Catalunya (IEEC), 08034 Barcelona, Spain\\
$^{33}$ Institute of Space Sciences (ICE, CSIC),  Campus UAB, Carrer de Can Magrans, s/n,  08193 Barcelona, Spain \\
$^{34}$ Laborat\'orio Interinstitucional de e-Astronomia - LIneA, Rua Gal. Jos\'e Cristino 77, Rio de Janeiro, RJ - 20921-400, Brazil \\
$^{35}$ Fermi National Accelerator Laboratory, P. O. Box 500, Batavia, IL 60510, USA \\
$^{36}$ Instituto de F\'{i}sica Te\'orica, Universidade Estadual Paulista, S\~ao Paulo, Brazil \\
$^{37}$ CNRS, UMR 7095, Institut d'Astrophysique de Paris, F-75014, Paris, France \\
$^{38}$ Sorbonne Universit\'es, UPMC Univ Paris 06, UMR 7095, Institut d'Astrophysique de Paris, F-75014, Paris, France \\
$^{39}$ Center for Astrophysical Surveys, National Center for Supercomputing Applications, 1205 West Clark St., Urbana, IL 61801, USA \\
$^{40}$ Department of Astronomy, University of Illinois at Urbana-Champaign, 1002 W. Green Street, Urbana, IL 61801, USA \\
$^{41}$ Institut de F\'{\i}sica d'Altes Energies (IFAE), The Barcelona Institute of Science and Technology, Campus UAB, 08193 Bellaterra (Barcelona) Spain \\
$^{42}$ Jodrell Bank Center for Astrophysics, School of Physics and Astronomy, University of Manchester, Oxford Road, Manchester, M13 9PL, UK \\
$^{43}$ University of Nottingham, School of Physics and Astronomy, Nottingham NG7 2RD, UK \\
$^{44}$ Astronomy Unit, Department of Physics, University of Trieste, via Tiepolo 11, I-34131 Trieste, Italy \\
$^{45}$ INAF-Osservatorio Astronomico di Trieste, via G. B. Tiepolo 11, I-34143 Trieste, Italy \\
$^{46}$ Institute for Fundamental Physics of the Universe, Via Beirut 2, 34014 Trieste, Italy \\
$^{47}$ Observat\'orio Nacional, Rua Gal. Jos\'e Cristino 77, Rio de Janeiro, RJ - 20921-400, Brazil \\
$^{48}$ Hamburger Sternwarte, Universit\"{a}t Hamburg, Gojenbergsweg 112, 21029 Hamburg \\
$^{49}$ School of Mathematics and Physics, University of Queensland,  Brisbane, QLD 4072, Australia \\
$^{50}$ Centro de Investigaciones Energ\'eticas, Medioambientales y Tecnol\'ogicas (CIEMAT), Madrid, Spain \\
$^{51}$ Department of Physics, IIT Hyderabad, Kandi, Telangana 502285, India \\
$^{52}$ Department of Astronomy, University of Michigan, Ann Arbor, MI 48109, USA \\
$^{53}$ Institute of Theoretical Astrophysics, University of Oslo. P.O. Box 1029 Blindern, NO-0315 Oslo \\
$^{54}$ Instituto de Fisica Teorica UAM/CSIC, Universidad Autonoma de Madrid, 28049 Madrid, Spain \\
$^{55}$ Institute of Astronomy, University of Cambridge, Madingley Road, Cambridge CB3 0HA, UK \\
$^{56}$ Kavli Institute for Cosmology, University of Cambridge, Madingley Road, Cambridge CB3 0HA, UK \\
$^{57}$ Center for Astrophysics $\vert$ Harvard \& Smithsonian, 60 Garden Street, Cambridge, MA 02138, USA \\
$^{58}$ Australian Astronomical Optics, Macquarie University, North Ryde, NSW 2113, Australia \\
$^{59}$ Lowell Observatory, 1400 Mars Hill Rd, Flagstaff, AZ 86001, USA \\
$^{60}$ Centre for Gravitational Astrophysics, College of Science, The Australian National University, ACT 2601, Australia \\
$^{61}$ The Research School of Astronomy and Astrophysics, Australian National University, ACT 2601, Australia \\
$^{62}$ Departamento de F\'isica Matem\'atica, Instituto de F\'isica, Universidade de S\~ao Paulo, CP 66318, S\~ao Paulo, SP, 05314-970, Brazil \\
$^{63}$ George P. and Cynthia Woods Mitchell Institute for Fundamental Physics and Astronomy, and Department of Physics and Astronomy, Texas A\&M University, College Station, TX 77843,  USA \\
$^{64}$ Department of Astrophysical Sciences, Princeton University, Peyton Hall, Princeton, NJ 08544, USA \\
$^{65}$ Instituci\'o Catalana de Recerca i Estudis Avan\c{c}ats, E-08010 Barcelona, Spain \\
$^{66}$ Department of Astronomy, University of California, Berkeley,  501 Campbell Hall, Berkeley, CA 94720, USA \\
$^{67}$ School of Physics and Astronomy, University of Southampton,  Southampton, SO17 1BJ, UK \\
$^{68}$ Computer Science and Mathematics Division, Oak Ridge National Laboratory, Oak Ridge, TN 37831 \\
$^{69}$ Max Planck Institute for Extraterrestrial Physics, Giessenbachstrasse, 85748 Garching, Germany \\
$^{70}$ Universit\"ats-Sternwarte, Fakult\"at f\"ur Physik, Ludwig-Maximilians Universit\"at M\"unchen, Scheinerstr. 1, 81679 M\"unchen, Germany \\

\end{document}